\documentclass[sn-mathphys]{sn-jnl}

\usepackage[utf8]{inputenc}
\usepackage{mathrsfs}
\usepackage{amssymb}
\usepackage{amsmath}
\usepackage{mathtools}

\newcommand{\jt}{\vartheta}
\newcommand{\CQ}{\mathcal{Q}}

\newcommand{\CC}{\mathcal{C}}
\newcommand{\CD}{\mathcal{D}}

\newcommand{\CF}{\mathcal{F}}

\newcommand{\CH}{\mathcal{H}}
\newcommand{\CI}{\mathcal{I}}
\newcommand{\CJ}{\mathcal{J}}

\newcommand{\CL}{\mathcal{L}}

\newcommand{\CN}{\mathcal{N}}
\newcommand{\CO}{\mathcal{O}}

\newcommand{\BR}{\mathbb{R}}
\newcommand{\BC}{\mathbb{C}}
\newcommand{\BH}{\mathbb{H}}
\newcommand{\BP}{\mathbb{P}}
\newcommand{\BZ}{\mathbb{Z}}

\newcommand{\BT}{\mathbb{T}}

\newcommand{\bfx}{{\boldsymbol x}}

\newcommand{\SL}{\text{SL}(2,\BZ)}

 \newcommand{\bfmu}{{\boldsymbol \mu}}

\newcommand{\be}{\begin{equation}}
\newcommand{\ee}{\end{equation}} 
\newcommand{\bes}{\begin{equation*}}
\newcommand{\ees}{\end{equation*}}
\DeclareMathOperator{\ima}{\mathring \imath}
\DeclareMathOperator{\sech}{sech}
\newcommand{\slz}{\mathrm{SL}(2,\mathbb Z)}
\newcommand{\sgn}{\mathrm{sgn}}

\jyear{2021}%

\theoremstyle{thmstyleone}%
\theoremstyle{thmstyletwo}%

\theoremstyle{thmstylethree}%

\begin{document}

\title[\small{The $u$-plane integral, mock modularity and enumerative geometry}]{The $u$-plane integral, mock modularity and enumerative geometry}


\author[1,2]{\fnm{Johannes} \sur{Aspman}}\email{aspmanj@maths.tcd.ie}

\author[1,2]{\fnm{Elias} \sur{Furrer}}\email{furrere@maths.tcd.ie}

\author[3,4]{\fnm{Georgios} \sur{Korpas}}\email{georgios.korpas@fel.cvut.cz}

\author[4]{\fnm{Zhi-Cong} \sur{Ong}}\email{ozcong@u.nus.edu}

\author[4]{\fnm{Meng-Chwan} \sur{Tan}}\email{mctan@nus.edu.sg}

\affil[1]{\orgdiv{School of Mathematics, Trinity College Dublin}}

\affil[2]{\orgdiv{Hamilton Mathematics Institute, Trinity College Dublin}}

\affil[3]{\orgdiv{Department of Computer Science, Czech Technical University in Prague}}

\affil[4]{\orgdiv{Department of Physics, National University of Singapore}}


\abstract{We revisit the  low-energy effective $U(1)$ action of  topologically twisted $\mathcal N=2$ SYM theory with gauge group of rank one on a generic oriented smooth 4-manifold $X$ with nontrivial fundamental group. After including a specific new set of $\mathcal Q$-exact operators to the known action, we express the integrand of the path integral of the low-energy $U(1)$ theory as an anti-holomorphic derivative. This allows us to use the theory of mock modular forms and indefinite theta functions for the explicit evaluation of correlation functions of the theory, 
thus facilitating the computations compared to previously used methods. As an explicit check of our results, we compute the path integral for the product ruled surfaces $X=\Sigma_g \times  \mathbb{CP}^1$ for the reduction on either factor and compare the results with existing literature. In the case of reduction on the Riemann surface $\Sigma_g$, via an equivalent topological A-model on $\mathbb{CP}^1$, we will be able to express the generating function of genus zero Gromov-Witten invariants of the moduli space of flat rank one connections over $\Sigma_g$ in terms of an indefinite theta function, whence we would be able to make concrete numerical predictions of these enumerative invariants in terms of modular data, thereby allowing us to derive results in enumerative geometry from number theory.}


 

\maketitle

\section{Introduction, Summary and Plan}

The role of topological quantum field theory in modern physics and mathematics is unambiguously important. One example is Donaldson-Witten (DW) theory, which is a topological formulation of the $\CN=2$ supersymmetric Yang-Mills theory on an oriented smooth four-manifold $X$. Its equivalent IR (long distance) counterpart is an abelian theory where Seiberg-Witten (SW) geometry dictates the physics \cite{Witten:1988ze}. Due to an electric-magnetic duality in the IR, characteristic functions of the theory enjoy powerful modular properties \cite{Seiberg:1994rs}. In the seminal paper \cite{Moore:1997pc}, the solution of the IR theory was derived using the technique of lattice reduction for simply connected $X$. In the same paper, the famous relation
\begin{align}\label{ZDW}
    Z_{\rm DW} = Z_{u} + Z_{\rm SW}
\end{align}
was found, where $Z_{\rm SW}$ denotes the generating function of SW invariants of the four-manifold \cite{Witten:1994cg}, while $Z_u$ denotes the contribution to $Z_{\rm DW}$ from the Coulomb branch of the low-energy effective $U(1)$ theory, the so-called $u$-plane.  The $u$-plane and its contribution to the path integral were studied in detail in \cite{Moore:1997pc,LoNeSha,Lossev:1997bz, Marino:1998rg}. The $u$-plane integral $Z_u$ is of particular interest since it is non-vanishing only for 4-manifolds with $b_2^+(X) \in \{0,1\}$. In turn, such four-manifolds are of particular interest since they are the only candidate topologies that probe the Coulomb branch $\mathcal{B}$ of the theory.

Recently, interest in DW theory, and in particular the $u$-plane integral, was revived due to observations relating the latter for special four-manifolds to the theory of mock theta functions and harmonic Maass forms \cite{Malmendier:2008db,Malmendier:2012zz}. For more generic, but simply connected, compact  four-manifolds, it was later reformulated in terms of the modular completion of a mock modular form \cite{Korpas:2017qdo,Korpas:2019ava,Korpas:2019cwg}. In this series of papers, the possibility of adding $\CQ$-exact operators to the action without affecting the correlation functions was studied in detail. In particular, a specific new $\CQ$-exact operator related to the 2-cycles of the background geometry was added to the action of the low-energy $U(1)$ theory, which makes the connection to mock modular forms apparent and elegant 
\cite{Korpas:2017qdo,Korpas:2019cwg}.
This technique circumvents the cumbersome method of lattice reduction, and allows to evaluate correlation functions efficiently. To the best of our knowledge, all of the recent results relating $Z_u$ and mock modular forms are restricted to the case where the low-energy  $U(1)$ theory is formulated on simply connected 4-manifolds only. 

Taking inspiration from \cite{Marino:1998rg}, we ask the natural question how these recent results carry over to the case when the four-manifold $X$  has a non-trivial fundamental group and non-zero first Betti number $b_1(X)$. When the four-manifold is non-simply connected, the theory is more complicated. This is due to the fact that the manifold now admits more structures, in the form of 1-form fields and 1- and 3-cycles, which are not present in the simply connected case. These cycles give rise to further contact terms in the low-energy $U(1)$ action \cite{LoNeSha,Marino:1998rg}. As a result, we will consider more general $\CQ$-exact operators  related to these cycles. Below, we give a summary of the paper highlighting our results.

\vspace{0.5cm}

\hspace{-0.8cm}{\it{Summary of the Paper}}

\vspace{0.1cm}

In this paper, we present a natural extension of the recent results \cite{Korpas:2017qdo, Korpas:2019cwg} to the case of non-simply connected four-manifolds with $b_2^+=1$. Specifically, we introduce a number of new $\CQ$-exact operators in the low-energy effective $U(1)$ theory that allow us to express the integrand of the $u$-plane integral elegantly as the non-holomorphic completion of a mock modular form. This further allows us to derive a closed-form expression for the $u$-plane integral for any such four-manifold and for arbitrary period point $J$, as is evident from the result of this paper, Equation \eqref{antiderivativeCH}. This solution depends on $H_1(X)$, a fact that is easily seen in the case of product ruled surfaces where it manifests as a genus dependence, while when $H_1(X)$ is trivial, \eqref{antiderivativeCH} reduces to  Equation (4.10) of \cite{Korpas:2017qdo}. As a byproduct of our computations, we present a complete classification of all $\CQ$-exact operators that the theory admits.

DW theory on product ruled surfaces  $X=\BC\BP^1\times \Sigma_g$, with $\Sigma_g$ a genus $g$ Riemann surface, has been argued to be equivalent to a 2d topological A-model on $\BC\BP^1$ in the limit of vanishing volume for $\Sigma_g$ \cite{bershadsky1995topological,Harvey_1995,LoNeSha}. In this paper, we  present a concrete derivation of this equivalence, and in turn show that due to the relation between DW theory and its low-energy $U(1)$ effective theory as given by Eq. \eqref{ZDW}, a connection between Gromov-Witten (GW) theory (realised physically by the A-model) and mock modular forms (appearing in the low-energy effective action) exists, such that one can compute GW invariants using modular data originating from the 4d theory, thereby deriving results in enumerative geometry from number theory. As an example, we make concrete numerical predictions of the genus zero GW invariants of the moduli space of flat $SU(2)$-connections on $\Sigma_2$ via their relation to an indefinite theta function. The GW invariants studied here involve local \textit{and} non-local operators which, as far as we know, have not been studied elsewhere.

\vspace{0.5cm}

\hspace{-0.8cm}{\it{Plan of the Paper}}

\vspace{0.1cm}

The plan of the paper is as follows. In Section \ref{Sec2}, we review the effective DW theory and the low-energy SW geometry. We further discuss the new specific $\CQ$-exact operators that we add to the action. 

In Section \ref{Sec3}, we rederive the $u$-plane integral $Z_u$ by including the $\CQ$-exact operators, and show how $Z_u$ can be written in terms of a mock modular form. 

In Section \ref{Sec4}, we apply our results to the two possible reductions of the product ruled surfaces $X=\mathbb{CP}^1 \times \Sigma_g$ and check them against existing literature \cite{Marino:1998rg, Lozano:1999us}. The second reduction, where the volume of $\Sigma_g$ shrinks, will be related to the genus zero GW invariants realised by the correlation functions of the A-model on $\mathbb{CP}^1$. 

In Section \ref{Sec5}, we discuss the A-model on $\mathbb{CP}^1$, and perform the computations that produce the generating function of GW invariants. Together with the findings of Section \ref{Sec4}, we make concrete numerical predictions of the genus zero GW invariants of the moduli space of flat $SU(2)$-connections on $\Sigma_2$ via modular data. 

In Section \ref{Sec6}, we conclude our paper, where useful appendices follow thereafter.

\section{Effective DW theory}\label{Sec2}

DW theory is the topologically twisted formulation of the pure $\CN=2$ supersymmetric Yang-Mills theory with gauge group $G$ of rank $r_G=1$ on a smooth four-manifold $X$ \cite{Witten:1988ze}. 
In the IR, the theory becomes a $U(1)$ gauge theory that depends on the complexified effective gauge coupling $
\tau=\frac{\theta}{\pi} +\frac{8\pi i}{g^2} \in \BH$, where $\BH$ denotes the Poincar\'e half-plane.
DW theory contains a scalar fermionic BRST operator $\CQ\coloneqq \epsilon^{\dot A\dot B}\overline{\CQ}_{\dot A\dot B}$ that obeys $\CQ^2 =0$.\footnote{Across the literature, this operator is often denoted as $\overline{\CQ}$ instead.} The field content of the theory is a collection of bosonic and fermionic degree $0$, $1$ and $2$ operator valued differential forms on $X$, where the degree of the differential form is equal to the ghost number of the physical operator. In Table \ref{tab:fieldcontent} we summarise the field content of the DW theory.  
\begin{table}[h]
\centering
\begin{tabular}{|c|c|c|} \hline 
   { Bosons} & { Fermions} & { Form degree} \\ \hline 
 $a,\bar{a}$ & $\eta$ &   0 \\ \hline
 $A$ & $\psi$ & 1 \\ \hline
 $D$ & $\chi$ & 2 \\ \hline
\end{tabular}
\caption{Field content of DW theory. The $a,\bar a$ fields originate from the vacuum expectation value of the scalar field of the UV theory. The $D$ field is an auxiliary field.}
 \label{tab:fieldcontent}
 \end{table}
 The BRST transformations on these fields are
\begin{equation}\label{Qtransformations}
	\begin{alignedat}{2}
		[\CQ,A]&=\psi,\qquad &&[\CQ,\psi]=4\sqrt{2}\mathrm da,\\
		[\CQ,a]&=0,\qquad &&[\CQ,\bar a]=\sqrt{2}\ima\eta,\\
		[\CQ,\eta]&=0,\qquad &&[\CQ,\chi]=\ima(F_+-D_+),\\
		[\CQ,D]&=(\mathrm d\psi)_+.
	\end{alignedat}
\end{equation}
 The physical observables of the theory belong to the $\CQ$-cohomology. We are interested in computing the path integral of the theory, the $u$-\emph{plane integral} or \emph{Coulomb branch integral}, when evaluated on a non-simply connected four-manifold. To this end, let us first introduce some notation.
 
Let $b_j\coloneqq b_j(X) = {\rm dim}\, H^j(X)$ be the Betti numbers of the smooth, closed and oriented four-manifold $X$ with $b_2(X)= b_2^+(X) + b_2^{-}(X)$, where the first (second) summand corresponds to the number of positive (negative) eigenvalues of the quadratic form $Q$ of $X$. For $a\in H^i(X)$ and $b\in H^{4-i}(X)$ we define
\begin{align}\label{eq:bilinear}
    B(a,b) = \int_X a\wedge b.
\end{align}
For $a \in H^2(X)$ the quadratic form $Q$ of $X$ corresponds to
\begin{align}\label{eq:quadratic}
    Q(a) := B(a,a).
\end{align}
Furthermore, the signature of $X$ is defined as $\sigma(X)= b_{2}^+(X)-b_{2}^{-}(X)$. Hereafter we consider four-manifolds with $b_2^+(X)=1$. 
By Poincar\'e duality, we have that $b_0=b_4$, $b_1=b_3$. We can assume $b_1$ to be even, since the correlation function of the theory are known to vanish unless $1-b_1+b_2^+$ is even \cite{Moore:1997pc}. 

The Coulomb branch integral is the path integral of the low-energy $U(1)$ theory with the insertion of the observables arising from the descent formalism as well as contact terms and  $\CQ$-exact operators. 
It takes the form
\begin{equation}\label{uplaneint}\begin{aligned}
   Z_u(p,\gamma,S,Y)
   &= \int [\mathcal D\Phi]  \nu(\tau) e^{-\int_X \CL'+I(S,Y)+I_\CO+I_\cap},
\end{aligned}\end{equation}
where $\Phi=\{a,\bar a, A,\eta,\psi,\chi,D\}$ is  the collection of fields of the theory (as in Table \ref{tab:fieldcontent}).

Below, Sections \ref{SWgeometry}---\ref{sec:contacts} are devoted to reviewing and explaining in detail all the ingredients of the $u$-plane integral \eqref{uplaneint}. Finally, in Section \ref{sec:QExact} we introduce the new $\CQ$-exact operators $I_S$ following \cite{Korpas:2017qdo} as well as its generalisation $I(S,Y)$ that will allow us to reformulate $Z_u$ as an integral over a mock modular form.

\subsection{Seiberg-Witten geometry}\label{SWgeometry}
In the seminal papers \cite{Seiberg:1994rs, Seiberg:1994aj}, Seiberg and Witten found the exact low-energy solution of $\CN=2$ supersymmetric Yang-Mills with gauge group $SU(2)$. The $\CN=2$ vector multiplet consists of a gauge field $A$, a scalar $\phi$ and Weyl fermions $\lambda$ and $\psi$, all in the adjoint representation. The potential of the theory is $V(\phi)=\tfrac{1}{g^2}\text{Tr}[\phi,\phi^\dagger]^2$ and we are interested in the moduli space of flat directions. These are found by setting $\phi=a\sigma^3$, with $\sigma^3$ the third Pauli matrix and $a$ a complex parameter. However, note that the Weyl group of $SU(2)$ acts on $a$ by $a\mapsto -a$. We can then construct a gauge invariant parameter as
\begin{equation}
    u=\frac{1}{16\pi^2}\langle \text{Tr}\phi^2\rangle.
\end{equation}
This serves as a good coordinate on the moduli space. 

There are two strong coupling singularities in the gauge theory, located at $u=\pm \Lambda^2$, where a monopole and a dyon become massless, respectively \cite{Seiberg:1994rs}. Here, $\Lambda$ is the dynamical scale of the theory, which is generated by the renormalisation group flow. This will be set equal to one later in the paper. The central charge of a dyonic state with electric and magnetic charges $(n_e, n_m)$ is given by
\begin{equation}
    Z=n_e a+n_m a_D,
\end{equation}
where $a_D$ is the magnetic dual of $a$, $a_D=\frac{\partial\CF}{\partial a}$, with $\CF$ the prepotential of the theory. 

The quantum moduli space of the gauge theory can be described in terms of a certain family of elliptic curves, the so-called Seiberg-Witten (SW) curves,
\begin{equation}\label{curve}
    y^2=x^3-ux^2+\tfrac{1}{4}\Lambda^4 x.
\end{equation}
The complex structure of the curve is identified with the complex coupling $\tau$ of the gauge theory. The fields $a$ and $a_D$ can be determined from the SW differential $\lambda_{\text{SW}}$ as
\begin{equation}
    a=\int_A \lambda_{\text{SW}},\qquad a_D=\int_B \lambda_{\text{SW}},
\end{equation}
where $A$ and $B$ are the canonical basis of homology cycles on the elliptic curve. 

\begin{figure}[!ht]\centering
	\includegraphics[scale=0.8]{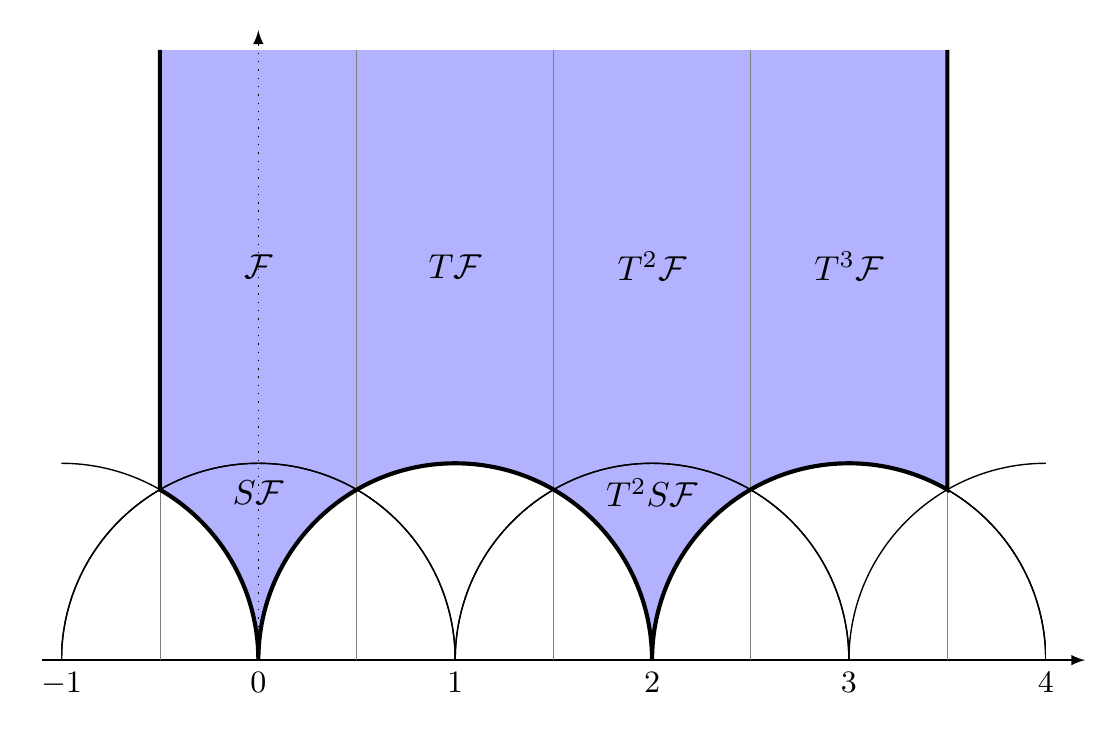} 
	\caption{Fundamental domain $\Gamma^0(4)\backslash \mathbb{H}$
          of the congruence subgroup $\Gamma^0(4)$, the duality group of the pure SW theory. It consists of
          six images of the key-hole fundamental domain $\CF$ of $\SL$. The cusp at $\tau=\ima\infty$ corresponds to weak coupling, while the cusps at $\tau=0$ and $2$ correspond to the monopole and dyon points. }\label{fig:funDom}
\end{figure}

By relating the $j$-invariant of the SW curve \eqref{curve} to that of the Weierstra{\ss} curve, we can solve for $u$ in terms of Jacobi theta functions,
\begin{equation}\label{utau}
    \frac{u(\tau)}{\Lambda^2}=\frac{\jt_2(\tau)^4+\jt_3(\tau)^4}{2\jt_2(\tau)^2\jt_3(\tau)^2}=\tfrac{1}{8}q^{-1/4}+\tfrac 52 q^{1/4}-\tfrac{31}{4}q^{3/4}+\CO(q^{5/4}),
\end{equation}
where $q=e^{2\pi\ima\tau}$.
See Appendix \ref{sec:AutForms} for the definitions of the theta functions. Using this formula it is straightforward to show that $u$ is a modular function for the congruence subgroup $\Gamma^0(4)\subset\slz$. The fundamental domain of this group is shown in Figure \ref{fig:funDom}. The cusp at $\tau=\ima\infty$ corresponds to weak coupling, while the cusps at $\tau=0$ and $\tau=2$ correspond to the monopole and dyon singularities, respectively.

From the curve \eqref{curve} we can directly find other quantities that will be important for the analysis in the paper, such as $\frac{du}{da}$ or $\frac{du}{d\tau}$. See for example \cite{Aspman:2021vhs} for a more detailed discussion on how these quantities can be retrieved from the curve. These expressions and their transformation properties under $\Gamma^0(4)$ are collected in Appendix \ref{app:modtrans}.

The $u$-plane integral can also be formulated for theories other than pure $\CN=2$ SYM with gauge group $SU(2)$. For pure $\CN=2$ SYM with gauge group $SU(N)$ and $N>2$, integrals over the respective Coulomb branches have been performed in \cite{Marino:1998bm}. The modular structure for those theories is however much more involved, as was recently elaborated in \cite{aspman20AHP}. More tractable Coulomb branches are rank one theories with matter hypermultiplets, such as $\CN=2^*$ (with one adjoint hypermultiplet) and $\CN=2$ supersymmetric QCD (with $N_f\leq 4$ fundamental hypermultiplets), both with gauge group $SU(2)$ \cite{Seiberg:1994aj}. Rank 1 Argyres-Douglas theories are  non-Lagrangian examples of $\CN=2$ theories with  1-dimensional Coulomb branches \cite{Argyres:1995jj,Argyres:1995xn}. The partition function has  been studied only for the simplest of those, the $(A_1,A_2)$ theory \cite{Moore:2017cmm}. The extension of such Coulomb branch integrals to manifolds with $b_1(X)>0$ is interesting because the ghost number selection rule admits the possibility of an infinite number of non-vanishing correlation functions, in sharp contrast to the case $b_1(X)=0$.
The $u$-plane integral for $\CN=2^*$ theory has been formulated in \cite{laba1998} and evaluated in \cite{Manschot:2021qqe}, based on the Coulomb branch geometry found in \cite{Huang:2011qx}. The modular structure of  $\CN=2$ QCD on the other hand has been established much more recently \cite{Aspman:2021vhs}. Integration over those Coulomb branches has yet to be completed \cite{AFM:future}.

\subsection{Effective Lagrangian}\label{efflagrangian} The low-energy $U(1)$ effective Lagrangian $\CL$ of the twisted theory is given in \cite[(2.15)]{Moore:1997pc}. The $\mathcal{ Q}$-exact terms as well as the kinetic terms do not contribute since the zero modes are constant in DW theory on a four-manifold $X$ with $b_2^+(X)=1$. For such manifolds there is a useful fact stating that for any $\beta_1,\beta_2, \beta_3,\beta_4\in H^1(X,\BZ)$, we have \cite{Muoz1997WallcrossingFF}
\begin{equation}\label{beta40}
    \beta_1\wedge\beta_2\wedge\beta_3\wedge\beta_4=0.
\end{equation}
We will make extensive use of this fact below.

Let us define $\CL'$ as the part of the zero-mode low-energy $U(1)$ effective Lagrangian that contributes to the $u$-plane integral. It is given by \cite{Moore:1997pc}
\be
\begin{split}\label{lagrangianpi1neq0}
\mathcal{L}'=\, &\pi\ima \bar\tau  k_+^2+\pi\ima\tau k_-^2-\frac{y}{8\pi} D\wedge *D +\frac{\ima\sqrt{2}}{16\pi}\frac{d\bar \tau}{d\bar a}\,\eta\chi \wedge (F_++D)\\
&-\frac{\ima\sqrt{2}}{2^7\pi} \frac{d\tau}{da}\psi\wedge \psi\wedge (F_-+D),
\end{split}
\ee
where  $F_\pm=4\pi k_\pm$ and for any two-form $x$ we abbreviate $B(x,x)=x^2$ as defined in Eq. \eqref{eq:bilinear}.
In $\CL'$, we  disregard any summands of $\CL$  containing $\CQ$-exact terms, exact differential forms and $\wedge$-products of four $1$-forms.
Here and throughout the rest of the paper we use units where the dynamical scale $\Lambda$ of the low-energy effective $U(1)$ theory is equal to one. The gravitational contributions to $\mathcal{L}'$ are described in the following section.

\subsection{Measure factors}\label{sec:Torus}
Assuming $X$ is connected, the (holomorphic) measure factor \cite{Laba05, Moore:1997pc} is
\begin{equation}\label{nu}
\nu(\tau) \coloneqq - \left(2^{7/2}\pi\right)^{\frac{b_1}{2}}\frac{2^{\frac{3\sigma(X)}{4}+1}}{\pi}(u^2-1)^{\frac{\sigma(X)}{8}}\left(\frac{da}{du}\right)^{\frac{\sigma(X)}{2}+b_1-2}.
\end{equation}
Here we used $\chi(X)+\sigma(X)=4-2b_1$ to eliminate the Euler character of $X$, $\chi(X)$. This expression reduces to Eq. (2.9) in \cite{Korpas:2017qdo} if we take $b_1=0$. For the simply connected theory one can use the microscopic definition of the theory to determine the effective gravitational couplings (e.g. by considering expansions of the Nekrasov partition function) \cite{Nakajima2003LecturesOI, Manschot_2020}.

The zero modes of the one-forms $\psi$ live in the tangent space of a $b_1$-dimensional torus $\BT^{b_1}=H^1(X,\BR)/H^1(X,\BZ) = H^1(X, \mathcal{O}^*_X)$ which corresponds to isomorphism classes of invertible sheaves (for $X$ a smooth complex variety that is holomorphic line bundles) on $X$
which are topologically trivial. We can expand $\psi$ in zero-modes as $\psi=\sum_{i=1}^{b_1} c_i\beta_i$ with $\beta_i$ an integral basis of harmonic one-forms, and $c_i$ Grassmann variables. We then have the measure
\begin{equation}\label{measTor}
    \prod_{i=1}^{b_1}\frac{dc_i}{\sqrt y}=y^{-\frac{b_1}{2}}\prod_{i=1}^{b_1}dc_i.
\end{equation}
The photon partition function will also include an integration over $b_1$ zero modes of the gauge field corresponding to flat connections \cite{Marino:1998rg}. These zero modes span the tangent space of $\BT^{b_1}$. As a consequence of this, the photon partition function will have an overall factor of $y^{\frac{1}{2}(b_1-1)}$ \cite{Witten:1995gf}. Combining this with the measure factor \eqref{measTor} we see that in the end there will only be a factor of $y^{-1/2}$ surviving. 

We can also consider the $c_j$ in the expansion of $\psi$ as a basis of one-forms $\beta_j^{\#}\in H^1(\BT^{b_1},\BZ)$, dual to $\beta_j$, such that 
\begin{equation}
    \psi= \sum_{j=1}^{b_1}\beta_j\otimes \beta^{\#}_j.
\end{equation}
A useful fact about four-manifolds with $b_2^+=1$ is that the image of the map
\begin{equation}
    \wedge:\quad H^1(X,\BZ)\otimes H^1(X,\BZ)\to H^2(X,\BZ) 
\end{equation}
is generated by a single rational cohomology class, which we denote as $W$ \cite{Muoz1997WallcrossingFF}.\footnote{This class is denoted $\Sigma$ in \cite{Marino:1998rg} and $\Lambda$ in \cite{Lozano:1999us}. However, since we want to reserve $\Sigma$ for the Riemann surfaces studied below and $\Lambda$ for the dynamical scale of the theory we choose to call the class $W$.} This means that we can write $\beta_i\wedge \beta_j=a_{ij}W$, $i,j=1,\dots,b_1$, where $a_{ij}$ is an anti-symmetric matrix. This further implies that the two-form on $\BT^{b_1}$ can be written as
\begin{equation}
    \Omega= \sum_{i<j}a_{ij}\beta_i^{\#}\wedge \beta_j^{\#},
\end{equation}
where $\beta_i^{\#}\in H^1(\BT^{b_1},\BZ)$, such that 
\begin{equation}
    \text{vol}(\BT^{b_1})=\int_{\BT^{b_1}}\frac{\Omega^{b_1/2}}{(b_1/2)!}.
\end{equation}
Below, we will study four-manifolds of the type $X=\BC\BP^1\times \Sigma_g$ with $\Sigma_g$ a genus $g$ Riemann surface. For these manifolds we have that $W=[\BC\BP^1]$ and $\text{vol}(\BT^{b_1})=1$ \cite{Marino:1998rg}. 

Using the analysis above we can now write $\psi\wedge\psi= 2(W\otimes \Omega)$ \cite{Marino:1998rg}. This will be very useful later on when we want to perform the integral over $\BT^{b_1}$ for the product ruled surfaces.

\subsection{Observables} \label{observables}
$\CQ$-invariant observables can be constructed using the celebrated descent formalism. By starting with the zero-form operator $\CO^{(0)}=2u$, we find all $k$-form valued observables $\CO^{(k)}$ for $k=1,2,3,4$ that are $\CQ$-invariant  modulo exact forms by solving the descent equations 
\begin{equation}\label{descenteq}
    \mathrm d\CQ^{(j)}=\{\CQ,\CO^{(j+1)}\}
\end{equation}
inductively. This ensures that for a $k$-cycle $\Sigma^{(k)}\in H_k(X)$ in $X$, the integrals $\int_{\Sigma^{(k)}}\CO^{(k)}$ are $\CQ$-invariant and only depend on $\Sigma^{(k)}$. Fortunately, there is a canonical solution to the descent equations: Due to the fact that the translation generator is $\CQ$-exact, there is the one-form valued \emph{descent operator} $K$, which satisfies $\mathrm d=\{\CQ,K\}$ \cite{Moore:1997pc}. This implies that \eqref{descenteq} can be solved by  $\mathcal O^{(j)}=K^j \mathcal O^{(0)}$, where the iterated (anti)-commutators are implicit. 
The action of the operator $K$ can be inferred from the BRST transformations \eqref{Qtransformations}  as  \cite{Moore:1997pc}
 \begin{equation}\label{descent}
    \begin{aligned}
   [K,a]=\frac{1}{4\sqrt2}\psi, \quad [K,\bar a]=0, \quad [K,\psi]=-2(F_-+D), \quad 
   [K,A]=-2\ima \chi,\\ [K,\eta]=-\frac{\ima}{\sqrt 2}\mathrm d\bar a, \quad [K,\chi]=-\frac{3\sqrt 2\ima}{4}*\mathrm d\bar a, \quad [K,D]=\frac{3\ima}{4}\left(2\mathrm d\chi-*\mathrm d\eta\right).
    \end{aligned}
\end{equation}

Let us study the insertion of all possible observables. For ease of notation, let us denote $p=\Sigma^{(0)}$ a point class, $\gamma=\Sigma^{(1)}$ a $1$-cycle, $S=\Sigma^{(2)}$ a 2-cycle and  $Y=\Sigma^{(3)}$ a 3-cycle. The cycles $\gamma$, $S$ and $Y$ can be expanded in formal sums as
\begin{equation}
    \gamma=\sum_{i=1}^{b_1}\zeta_i\gamma_i,\qquad S=\sum_{i=1}^{b_2}\lambda_iS_i\qquad Y=\sum_{i=1}^{b_3}\theta_i Y_i,
\end{equation}
where $\gamma_i$, $S_i$ and $Y_i$ are a basis of one-, two- and three-cycles respectively, $\lambda_i$ are  complex numbers, while $\zeta_i$ and $\theta_i$ are Grassmann variables. By the common abuse of notation, we use the same notation for the 3-, 2-, and 1-forms Poincar\'e dual to the cycles, and use the convention
\begin{equation}
    \int_\gamma\omega_1=\int_X \omega_1\wedge \gamma,\qquad \int_S\omega_2=\int_X\omega_2\wedge S, \qquad \int_Y \omega_3=\int_X \omega_3\wedge Y.
\end{equation}
The most general $\CQ$-invariant observable we can add is then 
\begin{equation}\label{observablesIO}
  I_{\CO}=2pu+ a_1\int_\gamma Ku+a_2\int_S K^2u+a_3 \int_Y K^3u,
\end{equation}
where $a_2=\frac{\ima}{\sqrt 2\pi}$ is fixed from matching with the mathematical literature \cite{Moore:1997pc}  and\footnote{We find two small typos in \cite{Marino:1998rg}: $F_+$ should be replaced with $F_-$, and the third term in $K^3u$ misses a factor of $\frac 12$. The contribution from $K F_-$ cancels.} 
\begin{equation}\label{observablesKu}
    \begin{aligned}
    Ku&=\frac{1}{4\sqrt2}\frac{du}{da}\psi,\\
    K^2u&=\frac{1}{32}\frac{d^2u}{da^2}\psi\wedge\psi-\frac{\sqrt2}{4}\frac{du}{da}(F_-+D),\\
    K^3u&=\frac{1}{2^7\sqrt2}\frac{d^3u}{da^3}\psi\wedge\psi\wedge\psi-\frac{3}{16}\frac{d^2u}{da^2}\psi\wedge(F_-+D)-\frac{3\sqrt2\ima}{16}\frac{du}{da}(2\mathrm d\chi-*\mathrm d\eta).
    \end{aligned}
\end{equation}

\subsection{Contact terms}\label{sec:contacts}
The existence of the canonical solution to the descent equations allows to map an observable of the UV theory to the low-energy $U(1)$ effective theory on the $u$-plane. For instance, the operator 
$I(S)=\int_S K^2 u$ of the UV theory 
is mapped to the same observable $\tilde I(S)=\int_S K^2u$ in the IR. This is not quite true for products $I(S_1)I(S_2)\dots I(S_n)$ of such operators for distinct Riemann surfaces $S_i\in H_2(X,\mathbb Z)$. At the intersection of the surfaces, contact terms will appear \cite{Moore:1997pc,LoNeSha}. When mapping a product of surface operators to the IR, the product is corrected by a sum over the intersection points. Due to the $\CQ$-invariance, the inserted operator is holomorphic and the point at which it is inserted is irrelevant. 

Such contact terms appear for all cycles in $X$ that can intersect. They have been classified  and the corresponding contact terms have been found in  \cite[Equations (2.8)-(2.12)]{Marino:1998rg},
\be \label{intersectionterms}
\begin{split}
I_\cap&=   \int_{S\cap S}T + a_{13} \int_{Y \cap \gamma}T+ a_{32} \int_{Y\cap S} K T + a_{33}\int_{Y \cap Y} K^2 T  \\[0.8em]
& + a_{332}\int_{S \cap Y \cap Y} \frac{\partial^3 \CF}{\partial \tau_0^3} + a_{333} \int_{Y \cap Y \cap Y} K \frac{\partial^3 \CF}{\partial \tau_0^3}  + a_{3333}\int_{Y \cap Y \cap Y \cap Y}\frac{\partial^4 \CF}{\partial \tau_0^4}.
\end{split}
\ee
Here $\tau_0$ is the deformation parameter of the prepotential, related to the dynamical scale by  $\Lambda^4=e^{\pi\ima \tau_0}$. The coefficient functions can all be expressed as  quasi-modular functions on the $u$-plane. For instance, the contact term for $S\cap S$ is 
\begin{equation}\label{eq:T}
T=\frac u2-\frac a4\frac{du}{da}=\frac{\jt_2^4+\jt_3^4-E_2}{6 \jt_2^2 \jt_3^2}.
\end{equation}
In terms of the prepotential $\CF$, it is given by  $T(\tau)=\frac{4}{\pi\ima}\frac{\partial^2\CF}{\partial \tau_0^2}$ \cite{mm1998}. It furthermore satisfies the identities \cite{Marino:1998rg}
\begin{equation}\begin{split}\label{dTda}
4\pi\ima \frac{dT}{da}&=\left(\frac{d^2u}{da^2}\right)^2\frac{da}{d\tau}+\pi\ima \frac{du}{da}, \\
\frac{dT}{da}&=\frac14\left(\frac{du}{da}-a\frac{d^2u}{da^2}\right), \\
\frac{d^2T}{da^2}&=-\frac a4 \frac{d^3u}{da^3}, \\
\frac{\partial^3\CF}{\partial \tau_0^3}&=-\frac{\pi^2}{2^4}\left(2T-a\frac{dT}{da}\right).
\end{split}\end{equation}
We can use the action  \eqref{descent} to find 
\begin{equation}\begin{split}\label{KTu}
    KT&= \frac{1}{4\sqrt2}\frac{dT}{da}\psi, \\
    K^2T&=\frac{1}{32}\frac{d^2T}{da^2}\psi\wedge\psi-\frac{1}{2\sqrt2}\frac{dT}{da}(F_-+D).
    \end{split}
\end{equation}
The intersection constants can be obtained from duality invariance \cite{Marino:1998rg}
\begin{equation}\begin{aligned}\label{ajjconst}
&a_1=\pi^{-\frac12}2^{\frac34}e^{-\frac{\pi\ima}{4}}, &&a_3=\pi^{-\frac32}2^{\frac14}e^{\frac{\pi\ima}{4}}/6, \\
&a_{13}=-6\pi^2 a_1a_3,\quad &&a_{32}=-6\sqrt2 \pi\ima a_3, \quad &&a_{33}=-9\pi^2 a_3^2, \\ &a_{332}=-72\sqrt2 \pi \ima a_3^2, \quad
&&a_{333}=36\pi^2 \ima a_3^3, \quad &&a_{3333}=-(6\pi)^3 \ima a_3^4.
\end{aligned}\end{equation}
Due to the identity \eqref{beta40}, the two last terms in \eqref{intersectionterms}  vanish and we can disregard them. Thus, from \eqref{intersectionterms} and \eqref{KTu} we see that all terms in $I_\cap$ except for one are only integrated over $\psi$ and $\tau$, which we we  do in a later step. The remaining term 
\begin{equation}\label{intersectionDetachi}
    -\frac{\sqrt 2a_{33}}{4}\frac{dT}{da}B(F_-+D,Y\wedge Y).
\end{equation}
is to be integrated over $D$, $\chi$ and $\eta$.

\subsection{\texorpdfstring{$\CQ$}{Q}-exact operators}\label{sec:QExact}
As we will later see, the photon path integral combines with the insertion of the surface observable to a Siegel-Narain theta function $\Psi_\mu^J(\tau,z)$. See \eqref{SNdefinition} for the definition. This function can be expressed as a total derivative to a non-holomorphic modular completion of an indefinite theta function, as has been previously shown in the simply connected case \cite{Korpas:2017qdo, Korpas:2019cwg}. To facilitate the calculation further the authors of those papers add the $\CQ$-exact operator $I_S$.\footnote{This term is called $\widetilde{I}_+(\bfx)$ in \cite{Korpas:2017qdo,Korpas:2019ava}. For ease of notation, we remove the tilde from such expressions since we only discuss operators in the IR.} In this section we will generalise this operator insertion to simplify the calculations also in the case of non-simply connected manifolds. This then allows us to evaluate correlation functions efficiently using mock modular forms  \cite{Korpas:2017qdo, Korpas:2019cwg,Korpas:2019ava}.

Since our computations should be valid for any $b_1\geq 0$ and in particular $b_1=0$, this suggests that it is instructive to add the same $\CQ$-exact operator
\cite[(2.11) and (2.12)]{Korpas:2017qdo}
\be\begin{aligned}\label{i+}
I_S &=-\frac{1}{4\pi}\int_S\left\{\CQ,\frac{d\bar u}{d\bar a}\chi\right\}\\&= -\frac{\sqrt{2}\ima}{4\pi} \frac{d^2\bar{u}}{d\bar{a}^2} \int_S \eta \chi - \frac{\ima}{4\pi}\frac{d\bar{u}}{d\bar{a}}\int_S (F_+ - D).
\end{aligned}\ee
The $u$-plane integrand \eqref{uplaneint} with $I_S$ inserted can also in the case where $b_1\neq0$ be written as an anti-holomorphic derivative. 
However, it does not give the same kind of Siegel-Narain theta function as in the simply-connected case. The reason is that the putative elliptic argument $z$ of $\Psi_\mu^J$ does not couple to $H^2_-(X)$ symmetrically to how its conjugate $\bar z$ couples to $H^2_+(X)$. The insertion of $I_S$ in the case $b_1=0$ can be viewed as the unique correction to the path integral that symmetrises the couplings to $H^2_\pm(X)$. Without such an insertion, the resulting theta functions are not symmetric, see for instance  \cite[Equation (3.18)]{Moore:1997pc}.

As we demonstrate below, for $b_1\neq 0$ this issue can be cured by introducing additional $\CQ$-exact operators. More precisely, the new observables and related contact terms require  three new $\CQ$-exact terms. The first two
\begin{equation}\begin{aligned}\label{ic}
    I_Y&=-\frac{3\ima \bar a_3}{16}\int_Y \left[\CQ,\frac{d^2\bar u}{d\bar a^2}\chi\wedge \psi\right]+\frac{\sqrt2}{2^7\pi}\int_X\left\{\CQ,\frac{d\bar \tau}{d\bar a}\chi\wedge\psi\wedge \psi\right\} \\
    &=\frac{3\sqrt2 \bar a_3}{2^4}\frac{d^3\bar u}{d\bar a^3}B(\eta\chi,\psi\wedge Y) +\frac{3\bar a_3}{2^4}\frac{d^2\bar u}{d\bar a^2}B(F_+-D,\psi\wedge Y) \\
    &\quad\,\,\,+\frac{\ima}{2^6\pi}\frac{d^2\bar \tau}{d\bar a^2}B(\eta\chi,\psi\wedge\psi)
   +\frac{\sqrt2\ima}{2^7\pi}\frac{d\bar\tau}{d\bar a}B(F_+-D,\psi\wedge\psi)
\end{aligned}\end{equation}
compensate the observables \eqref{observablesIO}. From the collection of contact terms \eqref{intersectionterms}, only the one from the intersection $Y\cap Y$ gives a term \eqref{intersectionDetachi} that is integrated over $D$, $\eta$ and $\chi$. 
This term requires the addition of the $\CQ$-exact operator
\begin{equation}\begin{split}\label{iYY}
   I_{Y\cap Y}&=-\frac{\sqrt 2\ima \bar a_{33}}{4}\int_{Y\cap Y}\left\{\CQ,\frac{d\bar T}{d\bar a}\chi\right\} \\
   &=\frac{\bar a_{33}}{2}\frac{d^2\bar T}{d\bar a^2}B(\eta\chi,Y\wedge Y)+\frac{\sqrt2 \bar a_{33}}{4}\frac{d\bar T}{d\bar a}B(F_+-D,Y\wedge Y).
\end{split}\end{equation}
We can note that, according to \eqref{ajjconst}, $\bar a_{33}=-a_{33}$. The sum of these additional $\CQ$-exact terms can be compactly written as 
\begin{equation}
   I_Y+  I_{Y\cap Y}   =-\sqrt2 \eta B(\chi, \partial_{\bar a} (y\bar\omega))-y B(F_+-D, \bar\omega),
\end{equation}
where we introduced the $2$-form
\begin{equation}\label{varpi}
\omega\coloneqq \frac{\sqrt{2}\ima}{2^7\pi y}\frac{d\tau}{da}\psi\wedge\psi-\frac{3a_3}{2^4 y}\frac{d^2u}{da^2}\psi\wedge Y-\frac{\sqrt2  a_{33}}{4y}\frac{dT}{da} Y\wedge Y.
\end{equation}
This 2-form has the property that $y\omega$ is holomorphic and thus  $y\bar \omega$ is anti-holomorphic. The form of \eqref{ic} is derived in Appendix \ref{sec:constructionIc}, where we furthermore show that its one-point function evaluates to zero, such that it is safe to include it into the path integral, following the analysis in \cite{Korpas:2019ava, Korpas:2019cwg}. We furthermore find it useful to follow \cite{Korpas:2017qdo} and introduce the notation
\begin{equation}\label{rhob}
\begin{split} 
\rho = \frac{ S}{2\pi} \frac{du}{da},
\qquad b = \frac{{\rm Im}(\rho)}{y}.
\end{split}
\end{equation}
Anticipating the result as a Siegel-Narain theta function, the elliptic variable will turn out to be $z=\rho+2\ima y \omega$, which is a 2-form with holomorphic coefficients. In terms of this variable, the sum of all $\CQ$-exact insertions \eqref{i+}, \eqref{ic} and \eqref{iYY} combine nicely as
\begin{equation}\begin{split}\label{isy}
  I(S,Y)&\coloneqq  I_S+  I_Y+  I_{Y\cap Y}   \\&=-\frac{\ima}{2}\left(\sqrt2 B(\eta\chi,\partial_{\bar a}\bar z)+B(F_+-D,\bar z)\right).
\end{split}\end{equation}
It is clear that this is purely anti-holomorphic. The operator $I(S,Y)$ is then included into the path integral, as in \eqref{uplaneint}.

\section{The \texorpdfstring{$u$}{u}-plane integral for \texorpdfstring{$\pi_1(X) \neq 0$}{pi1(X)!=0}}\label{Sec3}
The $u$-plane integral \eqref{uplaneint} can be expressed as 
\begin{equation}\label{CBI}
    Z_u(p,\gamma,S,Y)=\int[dad\bar a d\eta d\chi dD]\int_{{\rm Pic}(X)}d\psi\, \nu(\tau)\frac{1}{\sqrt{y}} e^{-\int_X\CL'+I_\CO+I_\cap+I(S,Y)},
\end{equation}
where $\int_{{\rm Pic}(X)}$ denotes a sum over isomorphism classes of line bundles, equivalent to a sum over $H^2(X,\BZ)$, followed by an integration over $\BT^{b_1}$.\footnote{${\rm Pic(X)}$ corresponds to the analytic Picard group of $X$. When $X$ is a smooth projective variety, we do not differentiate between the analytic and the algebraic Picard groups since they are isomorphic by GAGA \cite{GAGA}.} The $\psi$ zero modes are tangent to $\rm{Pic}(X)$, so the integral over these modes is understood as the integral of a differential form on $\rm{Pic}(X)$ \cite{Moore:1997pc}. At this point let us make a remark. The $\mathcal{Q}$-exact operator $I(S,Y)$ is not strictly required in order to derive our end result \eqref{antiderivativeCH}. As a matter of fact, as shown in \cite{Korpas:2019cwg} this operator can be added freely as $\alpha I(S,Y)$, with $\alpha$ any number.\footnote{In particular, we can have $\alpha=0$.} However, the case of $\alpha=1$ makes the analysis simpler and more elegant, why we choose to include it. 

Let us perform the integrals above in steps, using an economical notation. We integrate first over the auxiliary field $D$, and then over the fermionic $0$- and $2$-forms, $\eta$ and $\chi$.

\subsection{Integration over \texorpdfstring{$D$}{D}, \texorpdfstring{$\eta$}{eta} and \texorpdfstring{$\chi$}{chi}}\label{sec:intdetachi}
Using \eqref{varpi} and \eqref{rhob}, we can expand the terms in the exponential of \eqref{CBI} that are affected by the integrals over $D$, $\eta$ and $\chi$ as (ignoring the remaining terms for now)
\begin{equation}\begin{aligned}\label{exponent}
&-\int_X (\CL'+a_2K^2u+a_3K^3u)+I(S,Y)-\frac{\sqrt 2a_{33}}{4}\frac{dT}{da}B(F_-+D,Y\wedge Y)\\
&=-\pi \ima \bar\tau k_+^2-\pi\ima \tau k_-^2+\frac{y}{8\pi} D^2-\frac{\sqrt2\ima}{4}\frac{d\bar\tau}{d\bar a}B(\eta\chi,k_+)
-\frac{\sqrt2\ima}{16\pi}\frac{d\bar\tau}{d\bar a}B(\eta\chi,D)\\&-\frac{\ima}{\sqrt2}B(\eta\chi,\frac{d\bar\rho}{d\bar a})-2\pi\ima B(k_-,\rho)-2\pi\ima B(k_+,\bar\rho)+y B(D,b_+)
+\frac{\sqrt2\ima}{2^5} B(\psi\wedge\psi,\tfrac{d\rho}{da})\\&-\sqrt2 \eta B(\chi, \partial_{\bar a} (y\bar\omega))+4\pi y B(k_-,\omega_-)-4\pi yB(k_+,\bar\omega)+y B(D,\omega_+)+y B(D,\bar\omega_+).
\end{aligned}\end{equation}
At any point we discard terms that vanish identically, such as 4-fermion terms or any instance of \eqref{beta40} such as $\psi\wedge\psi\wedge\psi\wedge\psi$, $\psi\wedge\psi\wedge\psi\wedge Y$ or $\omega\wedge\omega$.
The exponential \eqref{exponent} is Gaussian in $D$ with saddle point
\begin{equation}\label{dterm}
D=\frac{\sqrt2\ima}{4y}\frac{d\bar\tau}{d\bar a}\eta \chi-4\pi (b_++\omega_++\bar\omega_+).
\end{equation}
This can be found by differentiating \eqref{exponent} with respect to $D$ and setting it to zero. 
Inserting $D$ in \eqref{exponent} gives \footnote{If we integrate over $D$ instead of inserting the equations of motion we get an additional factor of $2\pi\ima \sqrt{\frac 2y}$ in front of the integral \cite{Korpas:2019ava}. The result of the integration does not change otherwise since both methods agree for Gau{\ss}ian integrals up to an overall factor of $\sqrt{\frac{\pi}{a}}\ima$, if we integrate $e^{aD^2}$ over $D$. It is the same factor as in the simply-connected case because the quadratic $D$-term is the same. According to \cite[p. 68]{Moore:1997pc} the $D$ determinant should be ignored  because it cancels in any case with the fermionic determinants.}
\begin{equation}\begin{aligned}\label{dintegration}
&+\frac{\sqrt2\ima}{2^5} B(\psi\wedge\psi,\tfrac{d\rho}{da})-2\pi y (b_++\omega_++\bar\omega_+)^2-\pi \ima \bar\tau k_+^2-\pi\ima \tau k_-^2\\
&-2\pi\ima B(k_-,\rho)-2\pi\ima B(k_+,\bar\rho)+4\pi y B(k_-,\omega)-4\pi yB(k_+,\bar\omega)\\
&-\frac{\sqrt2\ima}{4}\frac{d\bar \tau}{d\bar a}B(\eta\chi,k_+-b_+-\omega_+-\bar\omega_+)-\frac{\ima}{\sqrt2}B(\eta\chi,\tfrac{d\bar\rho_+}{d\bar a})-\sqrt2 \eta B(\chi, \partial_{\bar a} (y\bar\omega)).
\end{aligned}\end{equation}
The third line are the only terms involving $\eta$ and $\chi$, which we will integrate over next. Before, we can combine those terms in the expression
\begin{equation}
-\frac{\sqrt2\ima}{4}\frac{d\bar \tau}{d\bar a}B\left(\eta\chi,k-b-\omega+\bar\omega-4\ima y\partial_{\bar \tau}\bar\omega+2\partial_{\bar \tau}\bar\rho\right).
\end{equation}
Integrating over $\eta$ and $\chi$, we can rewrite this in a compact way as a total anti-holomorphic derivative times an overall factor that, as we discuss below, cancels with contributions from the rest of the measure,
\begin{equation}\label{taubarder}
\frac{\sqrt2\ima}{4}\frac{d\bar \tau}{d\bar a}B\left(k-b-\omega+\bar\omega-4\ima y\partial_{\bar \tau}\bar\omega+2\partial_{\bar \tau}\bar\rho,\underline J\right)
= \sqrt y \frac{d\bar \tau}{d\bar a}\partial_{\bar\tau} \sqrt{2y}B(k+b+\omega+\bar\omega,\underline J),
\end{equation}
where $\partial_{\bar\tau}$ acts on everything to its right and $\underline J = J/\sqrt{Q(J)}\in H^2_+(X)$ is the normalised self-dual harmonic form on $X$. This result follows directly from the the identities 
\begin{equation}\label{taubarint}
\partial_{\bar\tau} y=\frac\ima2, \quad \partial_{\bar\tau}\sqrt{2y}=\frac{\sqrt 2\ima}{4\sqrt{y}}, \quad \partial_{\bar\tau}\frac1y=\frac{1}{2\ima y^2}, \quad \partial_{\bar\tau}b=\frac{b-\partial_{\bar\tau}\bar \rho}{2\ima  y},\quad \partial_{\bar\tau}\omega=\frac{1}{2\ima y}\omega.
\end{equation}
As previously discussed, the photon path integral together with the measure for the zero modes of $\psi$ contains a sum over all fluxes times a factor of $1/\sqrt y$, and additionally contributes $(-1)^{B(k,K)}$, where $K$ is the canonical class of $X$ \cite{Witten:1995gf}. The $1/\sqrt y$ factor is thus absorbed by the  $\sqrt y$ on the rhs of \eqref{taubarder}.

Using the change of variables
\begin{equation}
u: \Gamma^0(4)\backslash\overline{\mathbb H}\xrightarrow{\sim} \mathbb{CP}^1
\end{equation}
provided by \eqref{utau},
we can further integrate over $d\tau\wedge d\bar \tau$ rather than over $da\wedge d\bar a$. This motivates the definition of the transformed measure
\begin{equation}\label{tnu}
\tilde\nu = \nu \frac{da}{d\tau},
\end{equation} 
such that $da\wedge d\bar a\,\nu=d\tau\wedge d\bar\tau \frac{d\bar a}{d\bar \tau}\,\tilde\nu$. The factor $\frac{d\bar a}{d\bar \tau}$ cancels with the $\frac{d\bar \tau}{d\bar a}$ of \eqref{taubarder}.

\subsection{Siegel-Narain theta function}
Let us demonstrate that the $u$-plane integrand for $\pi_1(X)\neq0$, as in the simply-connected case \cite{Korpas:2017qdo}, evaluates to a Siegel-Narain theta function. To this end, let us define 
\begin{equation}\label{SNdefinition}
\begin{aligned}
\Psi_{\mu}^J(\tau,z)=&\,\,e^{-2\pi y\beta^2_+}   \sum_{k\in L+\mu}\partial_{\bar{\tau}}\left(\sqrt{2y}B( k+\beta,\underline{J})\right)\\
&\,\, \times (-1)^{B(k,K)} q^{-k_-^2/2}\bar q^{k_+^2/2}  e^{-2\pi \ima B(z,k_-)-2\pi \ima B(\bar z, k_+)}
\end{aligned}
\end{equation}
with $q=e^{2\pi\ima\tau}$ and $\beta=\frac{\text{Im}z}{y}\in L\otimes\mathbb R$, where $L=H^2(X,\mathbb Z)$. 

For the elliptic variable $z=\rho+2\ima y \omega$, we have  $\beta=b+\omega+\bar\omega$ (here, we use that $y\omega$ is holomorphic). Both variables appear naturally in \eqref{dintegration} and  \eqref{taubarder}. In fact, we can combine everything to find
\begin{equation}\begin{aligned}\label{Zupsi}
Z_u(p,\gamma,S,Y)=\int\displaylimits_{\Gamma^0(4)\backslash\mathbb H}\!\!\!\!\!d\tau\wedge d\bar\tau \!\!\int\displaylimits_{\BT^{b_1}}\!\![d\psi] \,\tilde \nu \,\Psi_{\mu}^J(\tau,\rho+2\ima y \omega)e^{I_\CO'+I_\cap'}.
\end{aligned}\end{equation}
Here, 
\begin{equation}\label{icapprime}
    I_\cap'=\int_{S\cap S}T+a_{13}\int_{Y\cap\gamma}T+a_{332}\int_{S\cap Y\cap Y}\frac{\partial^3\CF}{\partial\tau_0^3}+\frac{a_{32}}{4\sqrt{2}}\frac{dT}{da}\int_{Y\cap S}\psi
\end{equation}
and 
\begin{equation}\label{ioprime}
I_\CO'=2pu+\frac{\sqrt2a_1}{8}\frac{du}{da}\int_\gamma\psi+\frac{\sqrt2\ima}{2^6\pi}\frac{d^2u}{da^2}\int_S\psi\wedge\psi,
\end{equation}
are the (holomorphic) remainders of the collections of $0,\dots,3$-form observables and their contact terms that has not yet been integrated over, and we eliminated all terms that do not contribute.

Let us check that \eqref{Zupsi} is indeed true from the computations in Section \ref{sec:intdetachi}. Aside from the $\psi\wedge\psi$ term, the exponential of the first two lines in \eqref{dintegration} immediately combine into the definition \eqref{SNdefinition} with said parameters, $z=\rho+2\ima y \omega$ and $\bar z=\bar\rho-2\ima y\bar \omega$. 
Everything not exponentiated is given by the $\bar\tau$ derivative term in \eqref{taubarder}, which precisely gives the derivative term in \eqref{SNdefinition}. This proves \eqref{Zupsi}.


The expression \eqref{Zupsi} generalises the result of the $u$-plane integral \cite[(4.32)]{Korpas:2019cwg} to four-manifolds $X$ with $b_1(X)>0$ by giving a decomposition of the integrand into a  holomorphic and metric-independent measure $\tilde \nu\, e^{I_\CO'+I_\cap'}$ and a metric-dependent, non-holomorphic component $\Psi_{\mu}^J(\tau,z)$. Therefore, the evaluation techniques  of \cite{Korpas:2019cwg} apply. Namely,  we can express the integrand of the $u$-plane  integral as an anti-holomorphic derivative, 
\begin{equation}
    \frac{d}{d\bar\tau}\widehat \CH_\mu^J(\tau,\bar\tau)=\tilde \nu \,\Psi_{\mu}^J(\tau,z)e^{I_\CO'+I_\cap'}.
\end{equation}
The holomorphic exponential $e^{I_\CO'+I_\cap'}$ does not affect the anti-holomorphic derivative, and thus the extension  to $\pi_1(X)\neq 0$ is simply through the elliptic argument $z=\rho+2\ima y\omega$.

Once $\widehat{\CH}_\mu^J(\tau,\bar\tau)$ is found, we can use coset representatives of $\slz / \Gamma^0(4)$ to map the six images of $\CF=\slz\backslash\mathbb H$ back to $\CF$ (see Fig. \ref{fig:funDom}). The regularisation and renormalisation of such integrals originating from insertions of $\CQ$-exact operators has been rigorously established in \cite{Korpas:2019ava}, and we review it in Appendix \ref{sec:regularisation}. 
This then allows to evaluate the partition function as
\begin{equation}\label{antiderivativeCH}
Z_u(p, \gamma,S,Y)=4\,\CI_\mu(\tau)\big\vert_{q^0}+\CI_\mu(-\tfrac1\tau) \big\vert_{q^0} + \CI_\mu\left(\tfrac{2\tau-1}{\tau}\right)\big\vert_{q^0}, 
\end{equation}
where by ${}\vert_{q^0}$ we denote the $q^0$ coefficient of the resulting Fourier expansion, and  the $\tau$-integrand of \eqref{Zupsi} is given by \footnote{One could also contemplate switching the order of integration, and integrate over $\psi$ first. This would however not necessarily result in a similar function to \eqref{antiderivativeCH}, and it might not be possible to use the results of \cite{Korpas:2019cwg}.}
\begin{equation}\label{ciH}
     \CI_\mu(\tau)=\int\displaylimits_{\BT^{b_1}}[d\psi]\widehat \CH_\mu^J(\tau,\bar\tau).
 \end{equation}
 The prefactors in \eqref{antiderivativeCH} can be recognised as the widths of the cusps $\ima\infty$, $0$ and $1$ of the modular curve $\Gamma^0(4)\backslash\mathbb H$.
 
 To derive a suitable anti-derivative $\widehat{\CH}_\mu^J(\tau,\bar\tau)$, it is auxiliary to choose a convenient period point $J$. The $u$-plane integral for a different choice $J'$ is then related to the one for $J$ by a wall-crossing formula, given explicitly in \cite{Marino:1998rg}. 
 It is shown in \cite{Korpas:2019cwg} that for  convenient choices of $J$, $\Psi_\mu^J(\tau,z)$ factors into holomorphic and anti-holomorphic terms, and the anti-derivative $\widehat\CH_\mu^J$ can be found for both $L$ even and odd. Furthermore,  the $u$-plane integral can be evaluated using mock modular forms for point observables $p\in H_0(X)$ and Appell-Lerch sums for surface observables $z\in H_2(X)$ \cite{Korpas:2019cwg}.
 
 In \cite{Korpas:2019ava} it is furthermore shown that in the above mentioned renormalisation, any $\CQ$-exact operator (such as $I(S,Y)$) decouples in DW theory. However, it is clear that the insertion of $I(S,Y)$ crucially changes the \emph{integrand}, making the Siegel-Narain theta function symmetric. Instead of inserting $I(S,Y)$, we can contemplate adding $\alpha I(S,Y)$ for an arbitrary constant $\alpha$. It was noticed in \cite{Korpas:2019cwg} that the Siegel-Narain theta function  $\Psi_{\bfmu,\alpha}^J$ for $b_1=0$ with the insertion $\alpha I_S$ remains finite at weak coupling $(\text{Im} \tau\to\infty)$ if and only if $\alpha=1$. This can be seen from the exponential prefactor in \eqref{SNdefinition}, whose exponent is negative definite if and only if $\bar z$ (which we suppress in the notation) is the complex conjugate of $z$.

\subsection{Single-valuedness of the integrand}
An essential requirement, for the consistency of the theory, is that the path integral \eqref{Zupsi} is single-valued. For this it is  advantageous to first change variables in the $\psi$-integral as 
\begin{equation}\label{psishift}
\psi'=\psi+\frac{12\pi \ima a_3}{\sqrt2}\frac{da}{d\tau}\frac{d^2u}{da^2}Y. 
\end{equation}
This is because the coefficient function of $\psi\wedge\psi$  in $y\omega$ is modular, while the $\psi\wedge Y$ and $Y\wedge Y$ coefficients of $y\omega$ are only quasi-modular. Such shifts \eqref{psishift} leave the measure of $\int [d\psi]$ invariant, as $d\psi=d\psi'$. Due to the order of integration in \eqref{Zupsi}, the change of variables \eqref{psishift} is well-defined. Since $Y$ is also Grassmann-odd, $\psi$ and $Y$ $\wedge$-commute. Using \eqref{ajjconst} and \eqref{dTda},  this gives 
\begin{equation}
\omega=\frac{\sqrt{2}\ima}{2^7\pi y}\frac{d\tau}{da}\psi'\wedge\psi'+\frac{9\sqrt2 \pi^2a_3^2}{16y}\frac{du}{da}Y\wedge Y.
\end{equation}
Let us use the notation of Appendix \ref{app:modtrans}. It is argued in \cite{Marino:1998rg} that $\psi'$ transforms as $(-1,1)^{(1,0)}$. Using \eqref{holomorphicfunctions}, one then finds that $y\omega=(-1,1)^{(-1,0)}$ transforms precisely as $\rho=(-1,1)^{(-1,0)}$, such that $z=\rho+2\ima y \omega=(-1,1)^{(-1,0)}$ is a modular form and transforms exactly as in the $\pi_1(X)=0$ case.

Furthermore, it is auxiliary to define \cite[(2.14)]{Marino:1998rg}
\begin{equation}\label{Sshift}
    S'=S+4\pi\ima y \frac{da}{du}\omega.
\end{equation}
It is well-defined, as $S'=(1,1)^{(0,0)}$ is fully invariant. In contrast to \eqref{psishift}, this is not a change of variables or a redefinition, but rather a substitution to simplify some expressions. For instance, the elliptic variable now reads 
\begin{equation}\label{bfzS'}
    z=\frac{S'}{2\pi}\frac{du}{da},
\end{equation}
which takes the same form \eqref{rhob} as in the simply-connected case.

By incorporating the shift of $\psi\to\psi'$ together with \eqref{Sshift} we find that the contact terms and observables in \eqref{icapprime} and \eqref{ioprime} can be written as
\begin{equation}\label{IcocupMI}\begin{aligned}
    I_{\CO+\cap}&=2pu+S'^2 T+\frac{\sqrt2 a_1}{8}\frac{du}{da}\int_\gamma \psi'-3\pi^2a_1a_3 u\int_\gamma Y+\frac{\sqrt2 }{32}\frac{d\tau}{du}u\int_{S'}\psi'\wedge\psi'\\
    &\quad -\frac{3\pi\ima }{8}a_3\frac{du}{da}\int_Y S'\wedge\psi'+\frac{3\sqrt2}{4}\ima \pi^3a_3^2\, u \int_{S'}Y\wedge Y.
\end{aligned}\end{equation}
All terms but $S'^2T$ are modular functions with trivial multipliers. Due to \eqref{bfzS'}, the quasi-modular shift of $T$ combines precisely with the one of $\Psi(\tau,z)$.

\subsubsection*{Measure factor}
Since $\Delta\propto\frac{\vartheta_4^8}{\vartheta_2^4\vartheta_3^4}$, $\frac{da}{d\tau}=\frac{\pi}{8\ima}\frac{\vartheta_4^8}{\vartheta_2\vartheta_3}$ and $\frac{da}{du}=\frac{1}{2}\vartheta_2\vartheta_3$, from \eqref{nu} we have that $
\nu\propto\frac{\vartheta_4^\sigma}{(\vartheta_2\vartheta_3)^{2-b_1}}$ and therefore 
\begin{equation}
\tilde\nu\propto\frac{\vartheta_4^{8+\sigma}}{(\vartheta_2\vartheta_3)^{3-b_1}}.
\end{equation}
We find that under the generators of $\Gamma^0(4)$, $\tilde\nu=(-1,e^{-\pi\ima\sigma/4})^{(2-\frac{b_2}{2}+b_1,0)}$. For this we have used that $\sigma+b_2=2$ and that $b_1$ is even.

We also need to consider the fermion measure. As we have discussed earlier, this comes with an overall factor of $y^{-\frac{b_1}{2}}$ which gets absorbed by a similar factor coming from the photon partition function. This leaves us with $\prod_{i=1}^{b_1}dc_i$, which has weight $(-b_1,0)$, since $\psi$ has weight $(1,0)$ \cite{Witten:1995gf}. So after the integration over $D$, $\eta$ and $\chi$, and after changing integration variables from $da\wedge d\bar a$ to $d\tau\wedge d\bar\tau$ the measure of the integral will have weight $(-2-b_1,-2)$, and we thus need the rest of the integrand to have weight $(2+b_1,2)$.
Finally, the transformations of the Siegel-Narain theta function $\Psi_\mu^J(\tau,z)$ can be found in Appendix \ref{app:duality}. 

The integrand of the  $u$-plane integral \eqref{Zupsi} reads
\begin{equation}\label{integrand}
   \CJ_\mu^J=d\tau\wedge d\bar\tau \!\!\int\displaylimits_{\BT^{b_1}}\!\![d\psi] \,\tilde \nu \,\Psi_{\mu}^J(\tau,z)e^{I_\CO'+I_\cap'}.
\end{equation}
Since it is integrated over the fundamental domain of $\Gamma^0(4)$, in order to check whether the integral is well-defined  $\CJ_\mu^J$ must transform as a modular function for  $\Gamma^0(4)$ with no phases. In Table \ref{nonsimplytrafo} we collect the phases and weights of the individual factors as discussed above. This shows that the integral is indeed well-defined.

\begin{table}[ht]
\begin{center}
$\begin{tabular}{|c|c|c|c|c|c|c|c|} \hline 
$\text{object}$&
$d\tau\wedge d\bar\tau $&$ \int\displaylimits_{\BT^{b_1}}\![d\psi] $
& $\tilde\nu$&$\Psi_\mu^J(\tau,z)$&$e^{I_\CO'+I_\cap}$&$\CJ_\mu^J$ \\ \hline
\hline
$\text{weight}$&$(-2,-2)$&$(-b_1,0)$&$(2-\frac{b_2}{2}+b_1,0)$&$(\frac{b_2}{2},2)$&$(0,0)$&$(0,0)$\\ 
$T^4$&$1$&$1$&$-1$&$-1$&$1$&$1$\\ 
$S^{-1}T^{-1}S$&$1$&$1$&$e^{-\frac{\pi\ima\sigma}{4}}$&$e^{\frac{\pi\ima\sigma}{4}}e^{-\frac{\pi\ima z^2}{\tau+1}}$&$e^{\frac{\pi \ima z^2}{\tau+1}}$&$1$
\\ \hline
\end{tabular}$
\caption{Modular weights and phases of the $u$-plane integrand \eqref{integrand} under $\Gamma^0(4)$ transformations. This proves that $\CJ_\mu^J(\gamma \tau)=\CJ_\mu^J(\tau)$ for any $\gamma\in\Gamma^0(4)$.}\label{nonsimplytrafo}\end{center}
\end{table}

\section{Computation for product ruled surfaces}\label{Sec4}
As an interesting application of our results we can study the $u$-plane integral for a four-manifold of the type $X=\BC \BP^1\times \Sigma_g$, where $\Sigma_g$ is a genus $g$ Riemann surface. This is a product ruled surface with $b_2^+(X)=1$ (see Appendix \ref{classificationb2+1}).\footnote{One could alternatively consider products $\Sigma_g\times\Sigma_h$ of Riemann surfaces, however those have $b_2^+=1$ if and only if either $g=0$ or $h=0$, such that for $g,h\geq1$ the $u$-plane integral vanishes.}The DW theory for these manifolds was worked out in \cite{Marino:1998rg, Lozano:1999us} and we can use these results as a check of our formula. By shrinking the size of the Riemann surface $\Sigma_g$ we get a topological $\sigma$-model, more specifically the topological A-model, on $\BC\BP^1$  \cite{bershadsky1995topological}.  By calculating certain correlation functions on both sides we will be able to make an indirect connection between mock modular forms and the topological $\sigma$-model on $\BC\BP^1$ in Section \ref{Sec5}. 

The product ruled surfaces that we consider have $b_1=2g$, $b_2=2$, $b_2^+=1$, $K_X=0$, which in turn means that $\sigma=0$ and $\chi=4(1-g)$ \cite{Marino:1998rg}. We consider a general period point
\begin{equation}
    J(\theta)=\frac{1}{\sqrt{2}}\left(e^\theta [\BC\BP^1]+e^{-\theta}[\Sigma_g]\right),
 \end{equation}
where $[\BC\BP^1]$ and $[\Sigma_g]$ are the cohomology classes that generate $H^2(X,\BZ)$ \cite{Marino:1998rg}.\footnote{Sometimes we will be sloppy and write simply $\BC\BP^1$ and $\Sigma_g$ for these classes, and hope that this does not confuse the reader.} For these manifolds we further have that the rational cohomology class W, discussed in Sec. \ref{sec:Torus}, is simply given by $W=[\BC\BP^1]$ \cite{Muoz1997WallcrossingFF}. The intersection matrix is
\begin{equation}\label{PRSintersection}
    Q=\begin{pmatrix}0&1\\1&0\end{pmatrix},
\end{equation}
such that indeed $J(\theta)^2=1$.
Natural representatives of $[\BC\BP^1]$ and $[\Sigma_g]$ are found by choosing coordinates $z\in \BC$ for $[\BC\BP^1]$ and representing $[\Sigma_g]$ (for $g>1$) as a quotient of the Poincar\'{e} disk, $\CD=\{w: \vert w\vert<1\}$ with a Fuchsian group. This gives \cite{Marino:1998rg} 
\begin{equation}
    \begin{aligned}
           [\BC\BP^1]&=\frac{\ima}{2\pi}\frac{dz\wedge d\bar z}{(1+\vert z\vert^2)^2},\\
           [\Sigma_g]&=\frac{\ima}{2\pi(g-1)}\frac{dw\wedge d\bar w}{(1-\vert w\vert^2)^2}.
    \end{aligned}
\end{equation}
The scalar curvature for this metric is $8\pi (e^\theta-e^{-\theta}(g-1))$. We see that this is positive for $e^{2\theta}>g-1$, such that we do not get any contributions from the Seiberg-Witten invariants in these chambers. In particular, this is true when the volume of $\BC\BP^1$ is small, since this has $\theta$ large and positive. 

The connection to the topological $\sigma$-model is made in the chamber where we shrink the volume of $\Sigma_g$ \cite{bershadsky1995topological}. For completeness, we will calculate the $u$-plane integral in both chambers, where either of the factors shrink. The calculations are similar in both cases and we will start with the chamber where the volume of $\BC\BP^1$ is small. 

From Eq. \eqref{nu} we find that the measure factor for these manifolds simplifies to
\begin{equation}
    \tilde\nu =-\frac{2}{\pi}(2^{7/2}\pi)^{g} \left(\frac{da}{du}\right)^{2(g-1)}\frac{da}{d\tau}.
\end{equation}
For these manifolds we also have that the $\Psi_\mu^J$ of \eqref{Zupsi} can be written as a total derivative
\begin{equation}\label{eq:psijj}
\Psi_\mu^J(\tau,z)=\partial_{\bar\tau}\widehat\Theta_\mu^{JJ'},
\end{equation}
of the indefinite theta function \cite{ZwegersThesis} 
\begin{equation}\label{thetahatnonsimply}
\begin{aligned}
\widehat{\Theta}_{\mu}^{JJ'}(\tau,z) =&\sum_{k\in L +\mu} \frac{1}{2}\left[ E(\sqrt{2y}B(k+\beta,\underline{J}))-\text{sgn}( \sqrt{2y}B(k+\beta,J'))\right]\\
&\times (-1)^{B(k,K)}q^{-k^2/2}e^{-2\pi \ima B(z,k)},
\end{aligned}
\end{equation}
where $k^2 = k_+^2+k_-^2$, $J'$ is a reference vector\footnote{The reason for picking $J'$ in the negative cone is to assure that it does not contribute to Eq. \eqref{eq:psijj}. Had we picked $J$ in the positive cone, we would end up with the wall-crossing contributions from the chambers where $J$ ane $J'$ live respectively.} lying in the negative cone such that $Q(J')<0$, and 
\begin{equation}\label{errorf}
E: \mathbb R\to (-1,1), \quad t\mapsto 2\int_0^te^{-\pi x^2}\mathrm dx
\end{equation}
is a reparametrisation of the error function. See also Appendix \ref{Zwegers_theta} for more details on these indefinite theta functions. This means that we can take as $\widehat{\CH}_\mu^J(\tau,\bar\tau)$ in \eqref{ciH}
\begin{equation}\label{HPRS}
    \widehat{\CH}_\mu^J(\tau,\bar\tau)=\tilde\nu \widehat{\Theta}_{\mu}^{JJ'}(\tau,z)e^{I'_\CO+I'_\cap}.
\end{equation}
For the evaluation of the $u$-plane integral using this $\widehat\CH_\mu^J$, one may replace $\widehat\Theta_\mu^{JJ'}$ in \eqref{HPRS} after the modular transformations  as in \eqref{antiderivativeCH} with the mock modular form  $\Theta_\mu^{JJ'}$ defined in Appendix  \ref{Zwegers_theta}. This is also in line with the approach in \cite{Korpas:2017qdo}. 

\subsection{Shrinking \texorpdfstring{$\BC\BP^1$}{CP1}}
Let us start by analysing the chamber where the volume of $\BC\BP^1$ is small. In this chamber we fix the primitive null vector to be $J'=[\BC\BP^1]=W$. Due to \eqref{PRSintersection}, with this choice we have that $B(\psi\wedge\psi,J' )=0$, and in particular $B(S',J')= B(S,W)$. As above, we denote $z=\rho+2\ima y\omega$ and $\beta=b+\omega+\bar\omega$. We can introduce the split $k = m+nW$, with $m$ chosen such that
\begin{equation}
    \frac{B(m+\beta,J)}{B(W,J)}\in [0,1).
\end{equation}
With this split the mock modular form $\Theta_\mu^{JJ'}$ coming from \eqref{HPRS} can be written as
\begin{equation}\footnotesize
\begin{aligned}
    \Theta^{JW}_{\mu}(\tau,z)=&\sum_{n\in\BZ}\sum_{\substack{m\in L+\mu \\ \tfrac{B(m+\beta,J)}{B(W,J)}\in[0,1)}}q^{-\frac{m^2}{2}}e^{-2\pi\ima B(z,m)}q^{-nB(W,m)}e^{-2\pi \ima nB(\rho,W)}\\ &\times\frac{1}{2}\left[\text{sgn}\left(\sqrt{2y}(B(m+\beta,J)+nB(W,J)\right)-\text{sgn}\left(\sqrt{2y}B(m+\beta,W)\right)\right] \\
    =&\sum_{\substack{m\in L+\mu \\ \tfrac{B(m+b,J)}{B(W,J)}\in[0,1)}}\frac{q^{-\frac{m^2}{2}}e^{-2\pi\ima B(z,m)}}{1-q^{-B(W,m)}e^{-2\pi\ima B(\rho, W)}},
\end{aligned}
\end{equation}
where, in the second equality, we performed the sum over $n$. This is an Appell-Lerch sum \cite{ZwegersThesis}. The $u$-plane vanishes in chambers where $w_2(E)\cdot [\BC\BP^1]\neq 0$ \cite{Moore:1997pc}. This means that we only have solutions for $w_2(E)=0$ or $w_2(E)=W$, implying that $B(\mu,W)\in \BZ$. The only solutions for the conditions on $m$ are then $m=0$ for $w_2(E)=0$ and $m=\frac12 W$ for $w_2(E)=W$, this means that the contributions from the theta function are
\begin{equation}
    \begin{aligned}
           \Theta^{JW}_0(\tau,z)=&\frac{1}{1-e^{-2\pi \ima B(\rho,W)}}, \\
           \Theta^{JW}_{W}(\tau,z)=&-\frac{e^{-\pi \ima B(\rho,W)}}{1-e^{-2\pi\ima B(\rho,W)}}. 
    \end{aligned}
\end{equation}
We note that these are independent of $\psi$. The $u$-plane integral in this chamber can now be written as
\begin{equation}
    Z_{u,\mu}(p,\gamma,S,Y)=4\left[\left(\int_{\BT^{b_1}}[d\psi]e^{I'_\CO+I'_\cap}\right)\tilde\nu \Theta^{JW}_\mu(\tau,\rho)\right]_{q^0},
\end{equation}
with $\Theta_\mu^{JW}$ as above. If we only include point and surface observables it is straightforward to do the integral over the torus. The final result is
\begin{equation}
    Z_{u,\mu}(p,S)=\begin{cases}&4\left[\left(\frac{\sqrt{2}\ima}{2^5\pi}\frac{d^2u}{da^2}s\right)^ge^{2pu+2stT}\tilde\nu \frac{1}{1-e^{-\ima\frac{du}{da}s}}\right]_{q^0},\qquad \text{for }\mu=0,\\
    -&4\left[\left(\frac{\sqrt{2}\ima}{2^5\pi}\frac{d^2u}{da^2}s\right)^ge^{2pu+2stT}\tilde\nu \frac{e^{-\frac{\ima}{2}\frac{du}{da}s}}{1-e^{-\ima\frac{du}{da}s}}\right]_{q^0},\qquad \text{for }\mu=W,
    \end{cases}
\end{equation}
where we also defined $S=s[\Sigma_g]+t[\BC\BP^1]$.\footnote{There is a small discrepancy between this result and that of \cite{Marino:1998rg}, namely they differ by an overall phase $\ima^g$. This is most likely due to a known discrepancy in the literature for the normalisation of $\psi$. } 

\subsection{Shrinking \texorpdfstring{$\Sigma_g$}{sigmag} }
We now go on to discuss the chamber where we instead shrink the volume of $\Sigma_g$. For this chamber we pick the primitive null vector to be $J'=[\Sigma_g]$. The procedure is similar to the above. However, note that now $B(\psi\wedge\psi,J')\neq 0$. We start as before by splitting $k=m+n\Sigma_g$ with $m$ chosen such that
\begin{equation}
    \frac{B(m+\beta,J)}{B(\Sigma_g,J)}\in [0,1).
\end{equation}

Let us start by looking at the contribution from infinity. After performing the sum over $n$ we find that the indefinite theta function becomes
\begin{equation}
    \Theta^{J,[\Sigma_g]}(\tau,z)=\sum_{\substack{m\in L+\mu \\ \tfrac{B(m+\beta,J)}{B(\Sigma_g,J)}\in[0,1)}}\frac{q^{-m^2/2}e^{-2\pi\ima B(z,m)}}{1-q^{-B(\Sigma_g,m)}e^{-2\pi\ima B(z,\Sigma_g)}}.
\end{equation}
This is again an Appell-Lerch sum \cite{ZwegersThesis}. Following \cite{Lozano:1999us} we now pick $\omega_2(E)=[\BC\BP^1]+\epsilon[\Sigma_g]$, with $\epsilon=0,1$. For this flux there is no contribution from infinity, as can be seen from the above by realising that there are now no solutions to the conditions on $m$. We therefore turn to the other cusps. 

For the monopole cusp at $\tau=0$ we can use the formulas in the appendix to define the dual indefinite theta function as
\begin{equation}
    \Theta^{J,[\Sigma_g]}_{\mu,D}(\tau_D,z_D)\coloneqq\tau^{-1}e^{\pi\ima\frac{z^2_D}{\tau_D}}\Theta_{\mu}^{J,[\Sigma_g]}(-1/\tau,z/\tau)=\Theta_0^{J,[\Sigma_g]}(\tau_D,z_D-\mu,\bar z-\mu),
\end{equation}
where we used that $K_X=0$ and $b_2(X)=2$ together with the transformation formulas of the appendix. Following the procedure from above, splitting and summing over $n$, and simplifying by only including point and surface observables, we eventually find that
\begin{equation}\small
\begin{aligned}
          \Theta_0^{J,[\Sigma_g]}(\tau_D,z_D-\mu,\bar z_D-\mu)=&\frac{1}{1-e^{-2\pi\ima B(z_D-\mu,\Sigma_g)}}\\
          =&\left(1+\exp\left[-2\pi\ima\left(B(\rho_D,\Sigma_g)-\frac{\sqrt{2}}{2^5\pi}\left(\frac{d\tau}{da}\right)_D\Omega\right)\right]\right)^{-1}.
\end{aligned}
\end{equation}
Here we have used that $B(\mu,\Sigma_g)=\frac12$ and that $\psi\wedge\psi=2W\otimes \Omega$ together with the explicit expressions for $\omega$ when only including points and surfaces as observables. We also continue to denote dual functions with a subscript $D$. The explicit expressions for these are given in the appendix, Eq.\eqref{DualQuantities}.

Next, we want to integrate over the torus. If we only write down the parts that are actually dependent on $\psi$, or equivalently $\Omega$, the integral over the torus is
\begin{equation}\footnotesize\begin{aligned}
\int_{\BT^{b_1}}d\psi \exp\left[\frac{\sqrt{2\ima}}{2^5\pi}\left(\frac{d^2u}{da^2}\right)_DW\wedge S\otimes\Omega\right]  \left(1+\exp\left[-2\pi\ima\left(B(\rho_D,\Sigma_g)-\frac{\sqrt{2}}{2^5\pi}\left(\frac{d\tau}{da}\right)_D\Omega\right)\right]\right)^{-1}.
\end{aligned}\end{equation}
A neat trick we can use is to realise that
\begin{equation}\label{polyLog}
    \frac{1}{1+e^{t+x}}=\frac{1}{1+e^t}+\sum_{n\geq 1}\text{Li}_{-n}(-e^t)\frac{x^n}{n!},
\end{equation}
where $\text{Li}_{n}(y)$ is the polylogarithm \cite{Lozano:1999us}. Using this and again splitting $S=s[\Sigma_g]+t[\BC\BP^1]$ we find that the integral over the torus evaluates to
\begin{equation}\footnotesize
    \begin{aligned}
          \sum_{n=1}^g\binom{g}{n}\text{Li}_{-n}\left(-\exp\left[-it\left(\frac{du}{da}\right)_D\right]\right)\left(\frac{\sqrt{2}\ima}{2^5\pi}\left(\frac{d^2u}{da^2}\right)_Ds\right)^{g-n}\left(\frac{\sqrt{2}\ima}{2^4}\left(\frac{d\tau}{da}\right)_D\right)^n
    \end{aligned}
\end{equation}
where we dropped the first term coming from \eqref{polyLog} since this does not contribute to the $u$-plane integral (it will give a term whose $q$-series starts with a positive exponent). Combining this with the other terms in the $u$-plane integral we find that the contribution from the cusp at $\tau=0$ is given by
\begin{equation}\footnotesize
\begin{aligned}
    Z_{g,\tau=0}^\epsilon=&\Big[\frac{2}{\pi} e^{2pu_D+2st T_D} \sum_{n=1}^g\binom{g}{n}\text{Li}_{-n}\left(-\exp\left[-it\left(\frac{du}{da}\right)_D\right]\right)\left(-\frac{\ima}{2}\left(\frac{da}{du}\right)_D^2\left(\frac{d^2u}{da^2}\right)_Ds\right)^{g-n}\\
    &\times\left(-\ima\pi\left(\frac{da}{du}\right)^2_D\left(\frac{d\tau}{da}\right)_D\right)^{n}\left(\frac{du}{da}\right)^2_D\left(\frac{da}{d\tau}\right)_D \Big]_{q_D^0}.
\end{aligned}
\end{equation}
The contribution from the other cusp is easily calculated using the same procedure. The result is
\begin{equation}\footnotesize
    \begin{aligned}
    Z_{g,\tau=2}^\epsilon=&\Big[\frac{2\ima}{\pi}(-1)^\epsilon e^{-2pu_D-2st T_D} \sum_{n=1}^g\binom{g}{n}\text{Li}_{-n}\left(-\exp\left[-t\left(\frac{du}{da}\right)_D\right]\right)\left(\frac{\ima}{2}\left(\frac{da}{du}\right)_D^2\left(\frac{d^2u}{da^2}\right)_Ds\right)^{g-n}\\
    &\times\left(-\pi\left(\frac{da}{du}\right)^2_D\left(\frac{d\tau}{da}\right)_D\right)^{n}\left(\frac{du}{da}\right)^2_D\left(\frac{da}{d\tau}\right)_D \Big]_{q_D^0}.
    \end{aligned}
\end{equation}
The full $u$-plane integral in this chamber is then the sum of these two terms.\footnote{These expressions again differ from that of the older literature \cite{Lozano:1999us} by an overall phase $(-1)^\epsilon \left(-\ima\right)^g$. }

\subsubsection*{Genus one}
For $g=1$ the Seiberg-Witten contributions vanish and the only contributions comes from the $u$-plane integral \cite{Lozano:1999us}. The above expressions simplifies to
\begin{equation}
Z_1^\epsilon\coloneqq Z_{1,\tau=0}^\epsilon+Z_{1,\tau=2}^\epsilon= 2\ima \left[\frac{e^{\ima tf_D+2stT_D+2pu_D}}{\left(1+e^{\ima tf_D}\right)^2}+(-1)^\epsilon\frac{e^{tf_D-2stT_D-2pu_D}}{\left(1+e^{tf_D}\right)^2}\right]_{q^0},
\end{equation}
where we introduced $f_D=\left(\tfrac{du}{da}\right)_D$ to keep the expressions shorter. We can make various expansions for this. For example, if $s=t=0$ we get
\begin{equation}
    \begin{aligned}
        Z_1^0(p)=&\ima \left(1+2p^2+\frac{2}{3}p^4+\frac{4}{45}p^6+\frac{2}{315}p^8+\CO(p^9)\right), \\
        Z_1^1(p)=&2\ima \left( p+\frac{2}{3}p^3+\frac{2}{15}p^5+\frac{4}{315}p^7+\CO(p^9)\right).
    \end{aligned}
\end{equation}
For $p=0$ we instead find (expanding in small $t$)
\begin{equation} \label{g1compare}\footnotesize
    \begin{aligned}
        Z_1^0(s,t)=&\ima\left(1+\frac{1}{2}s^2t^2-st^3+\frac{1}{24}(16+s^4)t^4+\frac{1}{6}s^3t^5+\frac{1}{720}s^2(240+s^4)t^6+\CO(t^7)\right), \\
        Z_1^1(s,t)=&\ima\left(st-t^2+\frac{1}{6}s^3t^3-\frac{1}{2}s^2t^4+\frac{1}{120}s(80+s^4)t^5-\frac{1}{360}(136+15s^4)t^6+\CO(t^7)\right).
    \end{aligned}
\end{equation}

\subsubsection*{Genus two}
For $g=2$ we find
\begin{equation}
    \begin{aligned}
        Z_2^\epsilon=&\frac{\pi\ima}{2}\Bigg[\left(\frac{d\tau}{da}\right)_D\left(\frac{da}{du}\right)_D^2e^{-2(stT_D+pu_D)} \\
        &\times\Bigg(-e^{4stT_D+4pu_D}\sec^2(tf_D/2)\left(a_Ds-\tan(tf_D/2)\right)\\
        &+(-1)^\epsilon\sech^2(tf_D/2)\left(a_Ds-\tanh(tf_D/2)\right)\Bigg)\Bigg]_{q^0},    \end{aligned}
\end{equation}
where by $a_D$ we actually mean $\frac{\ima}{\pi}\left(\frac{da}{d\tau}\right)_D\left(\frac{d^2u}{da^2}\right)_D$, by use of the relation \eqref{reladadt} \cite{Matone:1995rx}.
For $s=t=0$ we simply get zero, but for $p=0$ we get
\begin{equation}\footnotesize \label{g2compareu}
    \begin{aligned}
        Z_1^0(s,t)=&\frac{1}{8}s^2t-\frac{1}{8}st^2+\frac{4+s^4}{48}t^3-\frac{1}{48}s^3t^4+\frac{s^4-40}{960}s^2t^5+\frac{272-3s^4}{2880}st^6+\CO(t^7), \\
        Z_1^1(s,t)=&\frac{1}{8}s-\frac{1}{8}t+\frac{1}{16}s^3t^2-\frac{1}{16}s^2t^3+\frac{1}{192}s^5t^4-\frac{1}{192}s^4t^5+\frac{s^4-160}{5760}s^3t^6+\CO(t^7).
    \end{aligned}
\end{equation}
For $g=2$ there will also be the Seiberg-Witten contributions given by \cite[Eq.(3.33)]{Lozano:1999us},
\begin{equation}\label{SWcontrg2}
    Z_{\text{SW}}^{g=2}(p,s,t)=\frac{1}{32}(-1)^{\epsilon}\left(e^{-2p-st}\sin(2s-2t)-(-1)^{\epsilon}e^{2p+st}\sinh(2s-2t)\right).
\end{equation}
The first few terms in the expansion for small $s$ and $t$, and $p=0$, are
\begin{equation}\label{g2comparesw1}
\begin{aligned}
        Z_{\text{SW}}^{g=2,\epsilon=0}(s,t)=&\left(-\frac{s^3}{12}-\frac{s^7}{630}+\CO(s^{8})\right)+\left(\frac{s^2}{8}-\frac{s^6}{180}+\CO(s^8)\right)t\\
        &+\left(-\frac{s}{8}+\frac{s^5}{120}+\CO(s^8)\right)t^2+\CO(t^3),
\end{aligned}
\end{equation}
and
\begin{equation}\label{g2comparesw2}
\begin{aligned}
        Z_{\text{SW}}^{g=2,\epsilon=1}(s,t)=&\left(-\frac{s}{8}-\frac{s^5}{60}+\CO(s^{8})\right)+\left(\frac{1}{8}+\CO(s^8)\right)t\\
        &+\left(\frac{s^3}{48}-\frac{s^7}{2520}+\CO(s^8)\right)t^2+\CO(t^3).
\end{aligned}
\end{equation}

\section{Revisiting the A-model computations}\label{Sec5}

The previous discussion is focused on the low-energy $U(1)$ effective action of DW theory on a generic oriented and  non-simply connected four-manifold $X$ ie, the $u$-plane formalism.  There were indications that for $X$  a product ruled surface, the correlation functions calculated in the previous section are related to Gromov-Witten invariants \cite{bershadsky1995topological,Harvey_1995, Lozano:1999us, munoz1999quantum, munoz2002gromov, donaldson1995floer},  that correspond to correlation functions of an $\mathcal{N}=(2,2)$ topological A-model in two dimensions, at the limit where the volume of one of the factors of $X$ vanishes. The aim of this section is to obtain Gromov-Witten invariants, i.e., values of correlation functions of an $\mathcal{N}=(2,2)$ topological A-model in two dimensions, by direct comparison with results obtained from the previous section. There, an explicit calculation was carried out in the case of a product four-manifold, $X=\BC \BP^1\times \Sigma_g$ with $g=2$, where both the $u$-plane and Seiberg-Witten terms contribute, shown in \eqref{g2compareu}, \eqref{g2comparesw1} and \eqref{g2comparesw2}. These expressions are the ones we shall use to obtain Gromov-Witten invariants.


 Using the fact that the twisted $\mathcal{N}=2$ gauge theory is topological, we are free to shrink $\Sigma_g$. We thus obtain an effective 2d theory on $\BC \BP^1$: the $\mathcal{N}=(2,2)$ topological A-model on worldsheet $\BC \BP^1$, with the target space  being the moduli space $\mathcal{M}_{\rm flat}(\Sigma_g)$ of flat $SU(2)$ connections on $\Sigma_g$. As a consequence,  flat $SU(2)$ connections along the directions tangent to $\Sigma_g$ are required to prevent the effective 2d action from blowing up when the limit of small $\Sigma_g$ is taken. This result is rederived in Appendix \ref{reductionaction}, following \cite{bershadsky1995topological}. 
  

 There is however a subtle point about the relation of the 2d A-model to both 4d theories, in that the relation should hold only within the limit of $\Sigma_g \to 0$. More will be said about this further on. Nevertheless, what we can achieve with this relation are predictions for Gromov-Witten invariants via coefficients from the expansion of a 4d low-energy $U(1)$ effective theory. This also offers an alternative approach in the calculation of Gromov-Witten invariants from 4d theories via physical principles, as illustrated in Figure 2.

\begin{figure}[!ht]\centering
	\includegraphics[scale=0.9]{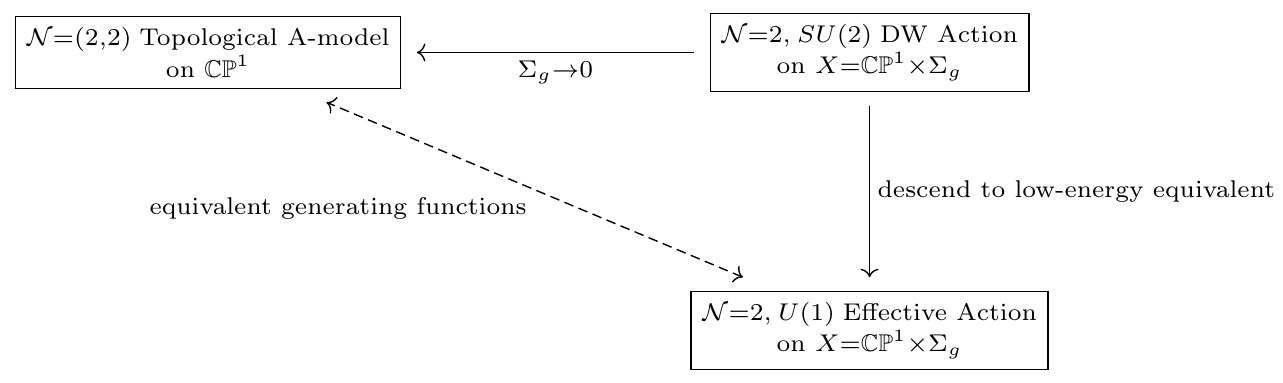} 
	\caption{Schematic diagram of the relation between the 4d and 2d theories.}
\end{figure}

We will now proceed to obtain Gromov-Witten invariants in a similar fashion to the steps in the previous sections in obtaining the $u$-plane integral. The $\mathcal{N}=(2,2)$ topologically twisted A-model has the action 
\begin{equation}\label{action_A}
    \begin{aligned}
        S &= \frac{1}{e^2}\int_{ \BC \BP^1} \Bigg( d^{2}z G_{i \bar{j}}\Big(\frac{1}{2}\partial_{z}\varphi^{i} \partial_{\bar{z}}\varphi^{\bar{j}} + \frac{1}{2}\partial_{\bar{z}}\varphi^{i} \partial_{z}\varphi^{\bar{j}} + \ima\rho^{\bar{j}}_{z}\nabla_{\bar{z}}\chi^{i} + \ima\rho^{i}_{\bar{z}}\nabla_{z}\chi^{\bar{j}}\Big)\\
        & \qquad - R_{i\bar{j}k\bar{l}}\rho^{i}_{\bar{z}} \rho^{\bar{j}}_{z}\chi^{k}\chi^{\bar{l}}\Bigg) + \ima \theta\int_{ \BC \BP^1} \varphi^{*}\omega,
    \end{aligned}
\end{equation}
with $z$, $\bar{z}$ as worldsheet coordinates and $i, \bar{i}$ etc, are coordinates on $\mathcal{M}_{\rm flat}(\Sigma_g)$ for the map $\varphi: \BC \BP^1\to \mathcal{M}_{\rm flat}(\Sigma_g)$. In Eq. \eqref{action_A}, $e$ denotes the gauge coupling and $\theta$ the instanton theta angle. The last term is a pullback of a K\"{a}hler form $\omega$ that directly descends from the instanton term of $\int F \wedge F$ in 4d. The bosonic field is the worldsheet scalar $\varphi$, and fermionic fields are the scalar $\chi$ and 1-form $\rho$. The covariant derivative on the worldsheet is defined as  $\nabla_{z}\chi^{\bar{i}}=\partial_{z}\chi^{\bar{i}}+\chi^{\bar{j}}\Gamma_{\bar{j}\bar{k}}^{\bar{i}}\partial_{z}\varphi^{\bar{k}}$ and $ R_{i\bar{j}k\bar{l}}$ is the Riemann curvature tensor on $\mathcal{M}_{\rm flat}(\Sigma_g)$. The BPS condition of the A-model localizes \eqref{action_A} to a moduli space of holomorphic maps
\begin{equation}\label{homoclass}
    \mathcal{M}_{\text{maps}}(\BC\BP^1, \beta) = \{ \varphi: \BC\BP^1 \to \mathcal{M}_{\text{flat}}(\Sigma_g) \mid \partial_{\bar{z}}\varphi ^{i}=0 \},
\end{equation}
with the condition coming directly from that of the 4d theory, namely $F^{+}=0$. Here, $\beta$ is the homology class of the map $\varphi$ into the moduli space of flat connections on $\Sigma_g$:
\begin{equation}
    \beta = \varphi_{*}[\mathbb{CP}^1] \in H_2\big(\mathcal{M}_{\rm flat}(\Sigma_g),\mathbb{Z}\big).
\end{equation}
The class $\beta$ can be further indexed as $\beta_I$ according to the dimension of each stratum of $\mathcal{M}_{\rm flat}$, that is 
\begin{equation}
    \beta_I = \varphi_{*}[\mathbb{CP}^1] \in H_2\big(\mathcal{M}_{\rm flat}^{(I)}(\Sigma_g),\mathbb{Z}\big),
\end{equation}
where $I = \mathrm{dim}(\mathcal{M}_{\rm flat})$.

Since we wish to obtain the Gromov-Witten invariants by comparison with coefficients from \eqref{g2compareu}, \eqref{g2comparesw1} and \eqref{g2comparesw2} as obtained via calculations in the $u$-plane from the previous sections, only the dimensionally reduced version of the 4d surface operators on $\Sigma_g \times \BC\BP^1$ as $S=s[\Sigma_g]+t[\BC\BP^1]$ will be considered in this section. Note that surface operator insertions on $\Sigma_g$ are non-local operators in 4d (on $\Sigma_g$), but get reduced to local point operators in 2d on $\BC \BP^1$ as $\Sigma_g \to 0$.

The surface operator inserted is Tr$\int_{s} (\psi_{\mu}\psi_{\nu}+\phi F_{\mu\nu})dx^{\mu}\wedge dx^{\nu}$. When inserted on $S \in \Sigma_g$, it becomes a point operator on $\BC \BP^1$, as mentioned:
\begin{equation}
\begin{aligned}
    \mathcal{O}^{(0)}&=\text{Tr}\int_{S}\bigg(\psi_{w}\psi_{\bar{w}}+\phi F_{w\bar{w}} \bigg)dw \wedge d\bar w\\
    &=  \omega_{i\bar{j}}\chi^{i}\chi^{\bar{j}} \in H^{0}(\BC \BP^1)\otimes\Omega^{2}(\mathcal{M}_{\text{maps}}).  
\end{aligned}
\end{equation}
Here $F_{w\bar{w}}=0$ was used, where $w$, $\bar{w}$ are complex coordinates on $\Sigma_g$,  and $\omega_{ij}$ is the K\"{a}hler form (see appendix \ref{reductionaction}). Note that $\mathcal{O}^{(0)}$ is consistent with the features required for being a point operator on the worldsheet, since it does not contain worldsheet indices. The presence of worldsheet indices require a contraction with the worldsheet metric $g_{z\bar{z}}$, thereby making the operator $\CQ$-exact. Hence this operator is indeed $\CQ$-closed. The ghost number of $\mathcal{O}^{(0)}$ is two, reflecting the fact that it is a 2-form on $\mathcal{M}_{\text{maps}}$.

On the other hand, when inserted into $S' \in \BC\BP^1$, it remains as a surface operator:
\begin{equation}
\begin{aligned}
   \mathcal{O}^{(2)}&= \text{Tr}\int_{S'}\bigg(\psi_{z}\psi_{\bar{z}}+\phi F_{z\bar{z}} \bigg)dz\wedge d\bar z\\
   &= \text{Tr}\int_{S'}\bigg(\chi^{i}\chi^{\bar{j}}\Phi_{i\bar{j}}F_{z\bar{z}} \bigg)dz\wedge d\bar z\in H^{2}(\BC \BP^1)\otimes\Omega^{2}(\mathcal{M}_{\text{maps}}).
\end{aligned}
\end{equation}
Here the $\psi_{z}, \psi_{\bar{z}}$ drop out since they do not survive the reduction, while $F_{z\bar{z}}$ contains the components of the gauge field $A_{z}$, $A_{\bar{z}}$ on $\BC\BP^1$ which are auxiliary fields (see Eq. \eqref{Aonsigma}). The fermionic parts of  $A_{z}$, $A_{\bar{z}}$ do not contribute since there are no $\rho$ zero modes, to be explained below). Using $F_{\mu\nu}=\partial_{\mu}A_{\nu}-\partial_{\nu}A_{\mu}+[A_{\mu},A_{\nu}]$, $F_{z\bar{z}}$ can then be written in terms of $\varphi^{i}, \varphi^{\bar{i}}$. 

Next, we have to look at which fermionic zero-modes exists in the 2d theory after dimensional reduction. The self-dual 2-form fermions $\chi_{\mu\nu}$ descend down to $\rho^{i}_{\bar{z}}$, $\rho^{\bar{i}}_{z}$, which are one-forms over $\BC \BP^1$. This dimensional reduction is coming from the high energy theory, in which the gauge group ${ SU}(2)$ is still left unbroken. Connections in this regime then remain irreducible and regular \cite{Witten:1988ze}, and hence we should \textit{not} expect $\rho$ zero-modes. The only other fermionic zero-modes are from the $\chi$ fields from the vertex operators, which can be absorbed by the  measure of the fermionic zero-modes in the path integral. The path integral that generates the GW invariants takes the form 
\begin{equation}\label{gwpath}
Z_{\rm GW}=\sum_{\beta \in H_2(\mathcal{M}_{\rm flat})} \int_{\beta} [\mathcal{D}\Phi]\; e^{-S}e^{\mathcal{O}^{(0)}+\mathcal{O}^{(2)}},
\end{equation}
where $\Phi$ represents all fields we integrate over and we also perform a discrete sum over $\beta$. This is equivalent to summing over the 4d instanton numbers $k$.  

In the evaluation of Eq. \eqref{gwpath}, all fields will have both zero- and fluctuating modes.
However, due to the independence of \eqref{gwpath} on the coupling, we are free to take the  weak coupling limit. The functional integral over the fluctuating modes in the action then  equals $\pm 1$ and zero-modes for both bosons and fermions are annihilated in the kinetic terms in the action when we take the quadratic approximation. With this approximation, fluctuating modes in the operators $\mathcal{O}^{(0)}$ and $\mathcal{O}^{(2)}$  can be suppressed, and we are then left with fields in terms of zero-modes only. These shall henceforth be labelled as $\mathcal{O}^{(0)}$ and $\mathcal{O}^{(2)}$ as well. It is necessary to have fields only in terms of zero-modes since both bosonic and fermionic zero-modes correspond to tangent vectors in $\mathcal{M}_{\text{maps}}$.

The resulting fields should then correspond to differential forms in this moduli space which, when combined together to obtain the correct index for the absorption of fermionic zero-modes give us a top form on $ \mathcal{M}_{\text{maps}}$ to be integrated over. In particular, recall that the ghost numbers of $\mathcal{O}^{(0)}$ and $\mathcal{O}^{(2)}$ are two, reflecting their degree as differential forms in $\mathcal{M}_{\text{maps}}$. To achieve that, we expand the vertex operators in powers of $\mathcal{O}^{(i)}$ from the vertex to soak up the extra zero-modes of $\chi^{i}$ and $\chi^{\bar{i}}$ in the measure. We then obtain correlation functions, for each $\beta$, that localizes on $\mathcal{M}_{\text{maps}}$. Only the terms which allow for the correct absorption of fermionic zero-modes give a non-zero contribution to Eq. \eqref{gwpath}. In the end, we obtain the usual Gromov-Witten invariants associated with a 2d topological A-model.  From the relation in dotted lines in Figure 2, we are able to conclude, \textit{in the limit of small $\Sigma_g$}:
\begin{equation}\label{thebox}
    Z_{\rm GW}= Z_{u}+Z_{\rm SW}
\end{equation}

On both sides of Eq. \eqref{thebox}, we have generating functions which contain terms that are graded by the instanton numbers. This is because the instanton term in the 4d action also descends down to a corresponding term in the 2d effective action. One can thus identify terms on both sides of \eqref{thebox} and we can then see that modular forms appearing in the $u$-plane integral can play an indirect (computation facilitating) role in Gromov-Witten invariants for holomorphic maps to the moduli space of flat connections on a Riemann surface.
 
With regards to wall-crossing, despite the condition of $b_2^+=1$, where wall-crossing phenomena are expected in $Z_{u}+Z_{\rm SW}$, we should not expect to see wall-crossing behaviour for $Z_{\rm GW}$. This is due to the fact that in shrinking $\Sigma_g$, we are restricting ourselves to the chamber of small $\Sigma_g$ and we should not expect any walls within a chamber, by definition. Hence the relation \eqref{thebox} should only be understood to hold within this particular chamber.

We can make a further comparison of \eqref{thebox} in another way: looking at how the operators $\mathcal{O}^{(0)}$ and $\mathcal{O}^{(2)}$ in 2d were derived, we see that they come from dimensional reduction of the operators in the four-dimensional high-energy theory. As mentioned, we consider only surface operators on $\Sigma_g \times \BC\BP^1$ as $S=s[\Sigma_g]+t[\BC\BP^1]$. The Gromov-Witten generating functional Eq. \eqref{gwpath} with only (4d) surface operators inserted will then be 
\begin{equation}\label{gwpathg1}
Z_{\rm GW}=\sum_{\beta} \int_{\mathcal{M}_{\text{maps}}} [\mathcal{D}\Phi]\; e^{-S_{0}}\;e^{-B(\omega,\beta)}e^{s\mathcal{O}^{(0)}+ t\mathcal{O}^{(2)}},
\end{equation}
where $S=S_{0}+B(\omega,\beta)$, and $B(\omega,\beta)$ as the instanton contribution to the action.

 Expanding the generating function \eqref{gwpathg1}, we can then compare Gromov-Witten invariants with $ Z_{u}+Z_{\rm SW}$  for different powers of $s$ and $t$. In the genus two case \footnote{Gauge equivalent classes of flat $G$-connections on a manifold $M$ correspond to equivalence classes of homomorphisms $f: \pi_{1}(M) \to G$ for a gauge group $G$, up to conjugation. The number of homology cycles on $M$ minus the number of restrictions and redundancy of conjugations determine the number of solutions for a linearization of the flat connection equations, which then determine the dimension of $\mathcal{M}_{\text{flat}}$. For $g=0$ we require at least 3 punctures on $\Sigma_0$ for $\mathcal{M}_{\text{flat}}(\Sigma_0)$ to be well-defined. For $g=1$, though $\mathcal{M}_{\text{flat}}(\Sigma_1)$ can be well-defined with 1 puncture, it is still possible to define a $\mathcal{M}_{\text{flat}}(\Sigma_1)$ without punctures, albeit with certain complications that we wish to avoid. Since there are no punctures being considered on $\Sigma_g$, we shall only consider cases of $g \geq 2$. The formula for dim$\mathcal{M}_{\text{flat}}(\Sigma_g)$ is dim$\mathcal{M}_{\text{flat}}(\Sigma_g)=\text{rank}(G)(2g-2)$ and rank$(G)=3$ for gauge group $G=SU(2)$.} where we have $\varphi: \BC\BP^1 \to \mathcal{M}_{\text{flat}}(\Sigma_2)$ and we have to include both the $u$-plane and Seiberg-Witten contributions, we can compare \eqref{gwpathg1} with \eqref{g2compareu}, \eqref{g2comparesw1} and \eqref{g2comparesw2}. Performing the procedure mentioned above of taking the weak coupling limit and integrating out fluctuating modes in the action, we are left with  $\mathcal{O}^{(0)}$ and $\mathcal{O}^{(2)}$ insertions in the path integral. The operators in the vertex will just be expanded and collected to match the different index numbers for the absorption of the correct number of fermionic zero-modes.
 For example, for a map of index 0, where $\mathcal{M_{\text{maps}}}=\mathcal{M}_{\text{flat}}(\Sigma_{2})$, we can have an invariant with the usual point operators that are inserted at $x_i\in \mathbb{CP}^1$. These are identified with the pullback of $\omega_i\in H^{*}(\mathcal{M}_{\text{flat}})$ by the evaluation map $\text{ev}_i: \mathcal{M}_{\text{maps}} \to \mathcal{M}_{\text{flat}}$ at $x_i$. 
 
In the $s^3$ term, we have 
 \begin{equation}\label{s3}
    \begin{aligned}
      -\frac{1}{12}
        &= \int_{\mathcal{M_{\text{maps}}}}[\mathcal{D}\chi \mathcal{D}\varphi] \, (\mathcal{O}^{(0)})^{3}e^{-B(\omega,\beta_{6})}   \\
         &= \int_{\mathcal{M_{\text{maps}}}}[\mathcal{D}\chi \mathcal{D}\varphi] ( \omega_{i\bar{j}})^{3}\chi^{6}e^{-B(\omega,\beta_{6})}\\
         &= e^{-B(\omega,\beta_{6})}\int_{\mathcal{M_{\text{maps}}}}\text{ev}_{1}^{*}\omega_{1}\wedge\text{ev}_{2}^{*}\omega_{2}\wedge\text{ev}_{3}^{*}\omega_{3} .
    \end{aligned}
  \end{equation}
We can also look at the less commonly studied non-local surface operators. Collecting the $s^{3}t^2$ terms, we require terms from \eqref{gwpathg1} to have a total of index 10:
 \begin{equation}\label{leftside}
    \begin{aligned}
      \frac{1}{12}
        &= \int_{\mathcal{M_{\text{maps}}}}[\mathcal{D}\chi \mathcal{D}\varphi] \, (\mathcal{O}^{(0)})^{3}(\mathcal{O}^{(2)})^{2}e^{-B(\omega,\beta_{10})}   \\
         &= \int_{\mathcal{M_{\text{maps}}}} [\mathcal{D}\chi \mathcal{D}\varphi] \, \bigg[( \omega_{i\bar{j}})^{3}\bigg(\int_{S'}\Phi_{k\bar{l}} F_{z\bar{z}} dz\wedge d\bar{z}\bigg)^{2}\bigg]\chi^{10}e^{-B(\omega,\beta_{10})} .
    \end{aligned}
  \end{equation}
  In \eqref{s3} and \eqref{leftside}, $\beta_{6}$ and $\beta_{10}$ are the homology classes for terms of index 6 and 10, respectively. We can identify $\beta_I$ with $d$, the degree of the map. From \cite{munoz1999quantum, munoz2002gromov, Hori:2003ic}, the given formula for the index $I=\text{dim}(\mathcal{M}_{\text{maps}})$ and degree $d\geq 0$ of $\varphi: \BC\BP^1 \to \mathcal{M}_{\text{maps}}(\Sigma_2)$ is $I=6+4d$. This formula relates $\text{dim}(\mathcal{M}_{\text{maps}})$ to the instanton number (degree of the map). The example  in \eqref{leftside} then corresponds to an invariant $(H^0_{\BC\BP^1})^{\otimes3}\otimes(H^2_{\BC\BP^1})^{\otimes2}\otimes\Omega^{10}_{\mathcal{M}_{\text{maps}}}$ of a degree 1 map. The prescription for comparison is thus simple: since $\mathcal{O}^{(0)}$ and $\mathcal{O}^{(2)}$ are labelled by $s$ and $t$, respectively, we just have to insert the relevant number of $\mathcal{O}^{(0)}$'s and $\mathcal{O}^{(2)}$'s based on the corresponding powers in the polynomial.

  Hence, for an index 14 term, we can have 
  \begin{equation}\label{index14}
    \begin{aligned}
      -\frac{1}{48}
        &= \int_{\mathcal{M_{\text{maps}}}} [\mathcal{D}\chi \mathcal{D}\varphi]\, (\mathcal{O}^{(0)})^{3}(\mathcal{O}^{(2)})^{4}e^{-B(\omega,\beta_{14})},   \\
    \end{aligned}
  \end{equation}
  of $(H^0_{\BC\BP^1})^{\otimes 3}\otimes(H^2_{\BC\BP^1})^{ \otimes 4}\otimes\Omega^{14}_{\mathcal{M}_{\text{maps}}} $ of a degree 2 map. And an index 18 term as
  \begin{equation}\label{index18}
    \begin{aligned}
      \frac{1}{192}
        &= \int_{\mathcal{M_{\text{maps}}}} [\mathcal{D}\chi \mathcal{D}\varphi]\,  (\mathcal{O}^{(0)})^{5}(\mathcal{O}^{(2)})^{4}e^{-B(\omega,\beta_{18})}   \\
    \end{aligned}
  \end{equation}
  for $(H^0_{\BC\BP^1})^{\otimes 5}\otimes (H^2_{\BC\BP^1})^{\otimes 4}\otimes\Omega^{18}_{\mathcal{M}_{\text{maps}}} $ of a degree 3 map.
  At first glance, the existence of a negative sign in \eqref{index14} might be a surprise, since these numbers actually represent values of correlation functions between operators, i.e. scattering amplitudes. The A-model considered, however, is non-unitary \cite{MR958805}, implying the existence of negative norm states. 
  
  As a consistency check, we can see that the lowest dimension of  $\mathcal{M_{\text{maps}}}$ is 6, which agrees with \eqref{g2compareu}, \eqref{g2comparesw1} and \eqref{g2comparesw2} since the lowest combined power of $s$ and $t$ is cubic, which have an index of $I=6$. This is because  terms with $I< 6$ (equivalently, maps of negative degree $d<0$) vanish. In fact, a quick examination of the combined powers of $s$ and $t$ of various terms in \eqref{g2compareu}, \eqref{g2comparesw1} and \eqref{g2comparesw2} show that the index of all terms obey the formula. This thus provides further evidence of having a direct correspondence between values obtained via computations in the 4d low-energy regime from previous sections and that of the 2d A-model. 

\section{Conclusions}\label{Sec6}
In this paper, we studied the low-energy $U(1)$ path integral of  DW theory evaluated on non-simply connected four-manifolds. Following the analysis of \cite{Korpas:2017qdo,Korpas:2019ava}, we derived the full solution for the correlation functions of the theory in terms of the modular completion of a mock modular form. The result can be readily extended to the case with surface defects \cite{Korpas:2018dag} or theories with matter hypermultiplets \cite{Seiberg:1994aj,Aspman:2021vhs,Manschot:2021qqe,Kanno:1998qj,Labastida:1998sk,AFM:future,AFMM:future,Moore:1997pc,LoNeSha} and even class $\mathcal{S}$ theories \cite{Gaiotto:2009we,Gaiotto:2009hg}, although we leave this for future work. 

We also presented a concrete reduction of  the theory on $\mathbb{CP}^1 \times \Sigma_g$ over $\Sigma_g$, whence we obtained a topological A-model on $\mathbb{CP}^1$, thereby demonstrating a novel connection between mock modular forms and  genus zero Gromov-Witten invariants.

We can also consider 4-manifolds of the form $X=M_3\times M_1$ with suitable topological numbers that allow probing the Coulomb branch. We expect that the mock modular form reformulation of $Z_u$ can be applied in this case too, whence relations between mock modular forms and topological invariants of 3-manifolds $M_3$ can be precisely formulated, thereby allowing us to derive results in geometric topology from  number theory. We will leave this for future work as well. 


\vspace{5pt}
\noindent\textbf{Acknowledgements. } 
We are happy to thank Robin Karlsson, Jan Manschot and Gregory Moore for correspondence and discussions, and the two referees for their helpful comments and remarks. JA is funded by the Irish Research Council under award number GOIPG/2020/910. GK acknowledges support of the OP RDE funded project CZ.02.1.01/0.0/0.0/16\_019/0000765 ``Research Center for Informatics''. EF is supported by the TCD Provost’s PhD Project Award. MC-Tan is supported by the MOE Tier 2 grant R-144-000-396-112.

\section*{Declarations}
\textbf{Conflict of interest.} On behalf of all authors, the corresponding author states that there is no conflict of interest.

\appendix

\section{Automorphic forms}\label{sec:AutForms}
In this Appendix, we collect some important aspects of modular and automorphic forms. 

\subsection{Modular forms}
The Jacobi theta functions $\vartheta_j:\mathbb{H}\to \mathbb{C}$,
$j=2,3,4$, are defined as
\be
\label{Jacobitheta}
\begin{split}
\vartheta_2(\tau)= \sum_{r\in
  \mathbb{Z}+\frac12}q^{r^2/2},\quad 
\vartheta_3(\tau)= \sum_{n\in
  \mathbb{Z}}q^{n^2/2},\quad
\vartheta_4(\tau)= \sum_{n\in 
  \mathbb{Z}} (-1)^nq^{n^2/2},
\end{split}
\ee
with $q=e^{2\pi i\tau}$. 
They  transform under $\slz$ in the following way:
\begin{alignat}{3}\small\nonumber
S:\quad& \vartheta_2(-1/\tau)=\sqrt{-\ima\tau}\vartheta_4(\tau),\quad&&\vartheta_3(-1/\tau)=\sqrt{-\ima\tau}\vartheta_3(\tau),\quad&&\vartheta_4(-1/\tau)=\sqrt{-\ima\tau}\vartheta_2(\tau)\\
T:\quad&\vartheta_2(\tau+1)=e^{\frac{\pi \ima}{4}}\vartheta_2(\tau),\quad &&\vartheta_3(\tau+1)=\vartheta_4(\tau),&&\vartheta_4(\tau+1)=\vartheta_3(\tau).
\end{alignat}
Under the generators $T^4$, $ST^{-1}S$ of $\Gamma^0(4)$ they transform as
\be
\label{Jacobitheta_trafos}
\begin{split}
&\vartheta_2(\tau+4)=-\vartheta_2(\tau),\qquad
\vartheta_2\!\left(\frac{\tau}{\tau+1}\right)=\sqrt{\tau+1}\,\vartheta_3(\tau),  \\
&\vartheta_3(\tau+4)=\vartheta_3(\tau),\qquad
\vartheta_3\!\left(\frac{\tau}{\tau+1}\right)=\sqrt{\tau+1}\,\vartheta_2(\tau),  \\
&\vartheta_4(\tau+4)=\vartheta_4(\tau),\qquad
\vartheta_4\!\left(\frac{\tau}{\tau+1}\right)=e^{-\frac{\pi
    \ima}{4}}\sqrt{\tau+1}\,\vartheta_4(\tau).  \\
\end{split}
\ee

The Eisenstein series $E_k:\mathbb{H}\to \mathbb{C}$ for even $k\geq 2$ are defined as the $q$-series 
\be
\label{Ek}
E_{k}(\tau)=1-\frac{2k}{B_k}\sum_{n=1}^\infty \sigma_{k-1}(n)\,q^n,
\ee
with $\sigma_k(n)=\sum_{d\vert n} d^k$ the divisor sum. For $k\geq 4$, $E_{k}$ is a modular form of
$\operatorname{SL}(2,\mathbb{Z})$ of weight $k$. On the other hand $E_2$ is a quasi-modular form, which means that the $\operatorname{SL}(2,\mathbb{Z})$ transformation of $E_2$ includes a shift in addition to the weight,
\be 
\label{E2trafo}
E_2\!\left(\frac{a\tau+b}{c\tau+d}\right) =(c\tau+d)^2E_2(\tau)-\frac{6\ima}{\pi}c(c\tau+d).
\ee

\subsection{Ingredients of the \texorpdfstring{$u$}{u}-plane integrand}\label{app:modtrans}
In this section, we give explicit modular expressions for the ingredients of the $u$-plane integrand. The integrand transforms under the duality group $\Gamma^0(4)$, which is generated by $T^4$ and $S^{-1}T^{-1}S$. Let us introduce the shorthand $f=(\phi_1,\phi_2)^{(k,l)}$ if a function $f$ is a non-holomorphic modular form of weight $(k,l)$ for $\Gamma^0(4)$ with multipliers, i.e. transforms as
\begin{equation}\begin{aligned}
f(\tau+4,\bar\tau+4)&=\phi_1 f(\tau,\bar\tau), \\ f\left(\frac{\tau}{\tau+1},
\frac{\bar\tau}{\bar\tau+1}\right)&=\phi_2(\tau+1)^k (\bar\tau+1)^l f(\tau,\bar\tau).
\end{aligned}\end{equation}
It is clear that 
\begin{equation}\begin{aligned}
(\phi_1,\phi_2)^{(k_1,l_1)}(\varphi_1,\varphi_2)^{(k_2,l_2)}&=(\phi_1\varphi_1,\phi_2\varphi_2)^{(k_1+k_2,l_1+l_2)},\\
\overline{(\phi_1,\phi_2)^{(k,l)}}&=(\bar\phi_1,\bar\phi_2)^{(l,k)},\\
\frac{1}{(\phi_1,\phi_2)^{(k,l)}}&=(\bar\phi_1,\bar\phi_2)^{(-k,-l)},
\end{aligned}\end{equation}
since $\vert\phi_i\vert=1$.
The functions
\begin{equation}\begin{split}\label{holTrans}
\frac{d^2u}{da^2}=4\frac{2E_2+\vartheta_2^4+\vartheta_3^4}{3\vartheta_4^8}, \quad 
\frac{u}{\Lambda^2}=\frac{\vartheta_2^4+\vartheta_3^4}{2\vartheta_2^2 \vartheta_3^2}, \quad 
\frac{a}{\Lambda}=\frac{2E_2+\vartheta_2^4+\vartheta_3^4}{6\vartheta_2 \vartheta_3}, \\ 
\frac{du}{d\tau}=\frac{\pi \Lambda^2}{4\ima} \frac{\vartheta_4^8}{\vartheta_2^2 \vartheta_3^2}, \qquad
\frac{da}{du}=\frac{1}{2\Lambda}\vartheta_2 \vartheta_3, \qquad
\frac{d\tau}{da}=\frac{8\ima}{\pi \Lambda}\frac{\vartheta_2 \vartheta_3}{\vartheta_4^8}
\end{split}\end{equation} 
transform as 
\begin{equation}\begin{split}\label{holomorphicfunctions}
u&=(1,1)^{(0,0)}, \quad \frac{du}{d\tau}=(1,1)^{(2,0)}, \quad \frac{da}{du}=(-1,1)^{(1,0)},\\ \rho&=(-1,1)^{(-1,0)}, \quad
\frac{d\tau}{da}=(-1,1)^{(-3,0)},\quad  y=(1,1)^{(-1,-1)}
\end{split}\end{equation} 
The derivative $\frac{d^2u}{da^2}$ is invariant under $T^4$ but quasi-modular,
\begin{equation}
    \frac{d^2u}{da^2}\left(\frac{\tau}{\tau+1}\right)=\frac{1}{(\tau+1)^2}\frac{d^2u}{da^2}-\frac{16\ima}{\pi}\frac{1}{(\tau+1)^3\vartheta_4^8}
\end{equation}
We also have that \cite{Matone:1995rx,Matone:1995jr}
\begin{equation}\label{reladadt}
    a=\frac{\ima}{\pi}\frac{d^2u}{da^2}\frac{da}{d\tau}.
\end{equation}

We will also need what we call the dual expressions for the above quantities. These are the expressions for the dual variable $\tau_D=-1/\tau$, disregarding the modular weights. They are
\begin{equation}\label{DualQuantities}
    \begin{split}
        u_D=&\frac{\jt_3^4+\jt_4^4}{2\jt_3^2\jt_4^2},\qquad \left(\frac{da}{du}\right)_D=\frac{1}{2\ima}\jt_3\jt_4,\\
        \left(\frac{da}{d\tau}\right)_D=&\frac{\pi}{8}\frac{\jt_2^8}{\jt_3\jt_4},\qquad \left(\frac{d^2u}{da^2}\right)_D=\frac{4}{3}\frac{2E_2-\jt_3^4-\jt_4^4}{\jt_2^8},
    \end{split}
\end{equation}
as well as
\begin{equation}
T_D=-\frac{1}{24}\left(E_2\left(\frac{du}{da}\right)_D^2-8u_D\right).
\end{equation}
\subsection{Siegel-Narain theta function}\label{app:duality}

Siegel-Narain theta functions form a large class of theta functions of
which the Jacobi theta functions are a special case (see \cite{Korpas:2017qdo}) . We restrict here
to a specific Siegel-Narain theta function for which the associated lattice $L$
is a uni-modular lattice of signature
$(1,n-1)$ (or a Lorentzian lattice). We denote the bilinear form by
$B(x,y)$ and the quadratic form $B(x,x)\equiv Q(x)\equiv x^2 $.
Let $K$ be a characteristic vector of $L$, such that
$ Q(k) + B(k, K) \in 2\mathbb{Z}$ for each $k\in L$.   

Given an element $J\in L\otimes \mathbb{R}$ with $Q(J)>0$, we
may decompose the space $L\otimes \mathbb{R}$ in a positive
definite subspace $L_+$ spanned by $J$, and a negative definite
subspace $L_-$, orthogonal to $L_+$. Let  $\underline J=J/\sqrt{Q(J)}$
be the normalization of $J$. The projections of a
vector $k\in L$ to $L_+$ and $L_-$ are then given by
\be
\label{k+k-}
k_+=B(k,\underline J)\, \underline J, \qquad \qquad k_{-} = k-k_+.
\ee

Given this notation, we can introduce the Siegel-Narain theta function
of our interest $\Psi^J_\mu:\mathbb{H}\times \mathbb{C}\to \BC$. Let $J$ be
as discussed above (\ref{k+k-}) and $\mu\in L\otimes
\mathbb{R}$. Then $\Psi^J_\mu$ is defined by\footnote{For brevity we
list in $\Psi^J_\mu$ only the holomorphic arguments $\tau$ and $z$,
even though the function does also depend on $\bar \tau$ and $\bar z$.}
\be 
\label{PsiJ} 
\begin{split} 
\Psi^J_\mu(\tau,z)&=e^{-2\pi y b_+^2} \sum_{k\in
  L + \mu} \partial_{\bar \tau} (\sqrt{2y}B(k+b,
\underline J))\,(-1)^{B(k, K)} q^{-k_-^2/2} \bar q^{k_+^2/2} \\
&\quad \times e^{-2\pi i B(z, k_-)-2\pi i B(\bar z,k_+)},
\end{split}
\ee
where $b=\mathrm{Im}(z)/y\in L \otimes \mathbb{R}$. 

One finds for the modular transformations of $\Psi^J_\mu$ under the generators of
SL$(2,\mathbb{Z})$ the following identities
\be
\label{Psi_trafos}
\begin{split}
\Psi^J_{\mu+K/2}(\tau+1,z)=&e^{\pi i (\mu^2-K^2/4)}\,\Psi^J_{\mu+K/2}(\tau,z+\mu), \\
\Psi^J_{\mu+K/2}(-1/\tau,z/\tau)=& 
-i(-i\tau)^{\frac{n}{2}}(i\bar \tau)^{2} \exp\!\left(-\pi
  i z^2/\tau+\pi i K^2/2\right) (-1)^{B(\mu,K)}\\
&\times  \Psi^J_{K/2}(\tau,z-\mu).
\end{split}
\ee
For  $\mu\in L/2$ one can show that $\Psi^J_\mu$ is a
modular form of the congruence subgroup $\Gamma^0(4)$. The  action of
the generators of $\Gamma^0(4)$ on $\Psi^J_\mu$ with $\mu\in L/2$ is given by
\begin{eqnarray} 
&&\label{Psi-1} \Psi^J_\mu(\tau,-z)=-e^{2\pi i B(\mu,K)}\,\Psi^J_\mu(\tau,z), \\
&&\label{PsiS}\Psi^J_\mu\!\left( \tfrac{\tau}{\tau+1}, \tfrac{z}{\tau+1} \right)=(\tau+1)^{\frac{n}{2}}(\bar \tau+1)^2\exp\!\left(-\tfrac{\pi iz^2}{\tau+1} +\tfrac{\pi i}{4}K^2\right) \Psi^J_{\mu}(\tau,z),\\
&&\label{PsiT4}\Psi^J_\mu(\tau+4,z)=e^{2\pi i B(\mu,K)}\,\Psi_\mu(\tau,z).
\end{eqnarray}

Notice that for $\ell=4$ we can use that $2\mu^2+B(\mu, K)\in \mathbb Z$.

\subsection{Indefinite theta functions}
\label{Zwegers_theta} 
In this appendix we present various aspects of indefinite theta
functions and their modular completions. We assume that the associated
lattice $L$ is unimodular and of signature $(1,n-1)$. 
 
To define the indefinite theta function, we choose two positive
definite vectors $J$ and $J'\in L\otimes \mathbb{R}$ with
$B(J,J')>0$, such that they both lie in the same positive cone of
$L$. Let $\underline J$ and $\underline J'$ be their
normalizations as before.
The arguments of theta function are $\tau\in \BH$, $z\in L \otimes \mathbb{C}$ and $\mu\in L\otimes \mathbb{R}$. We let
$b=\mathrm{Im}(z)/y\in L \otimes \mathbb{R}$. In terms of this data, the indefinite theta function $\Theta_\mu^{JJ'}$ is defined as 
\be 
\label{indeftheta} 
\begin{split}
\Theta^{JJ'}_{\mu}\!(\tau,z)=&\sum_{k\in L+\mu} 
\tfrac{1}{2}\left( \sgn(B(k+b,J))-\sgn(B(k+b,J'))\right)\\
&(-1)^{B(k,K)} q^{-k^2/2}
e^{-2\pi i B(z,k)}.
\end{split} 
\ee  
It is possible to show that the sum over $L$ is convergent \cite{ZwegersThesis}. However, $\Theta^{JJ'}_{\mu}$ does only transform as a modular form after the addition of certain non-holomorphic terms. References \cite{ZwegersThesis, MR2605321} explain that the modular completion $\widehat \Theta^{JJ'}_\mu$ of $\Theta^{JJ'}_\mu$ is obtained by substituting (rescaled) error functions for the sgn-functions in (\ref{indeftheta}). The completion $\widehat \Theta^{JJ'}_\mu$ then transforms as a modular form of weight $n/2$, and is explicitly given by
\be 
\label{hatTheta}
\begin{split}   
\widehat \Theta^{JJ'}_{\mu}\!(\tau,z)=\sum_{k\in L+\mu} &
\tfrac{1}{2}\left( E(\sqrt{2y}\,B(k+b, \underline
  J))-E(\sqrt{2y}\,B(k+b, \underline J'))\right)\\
& \times (-1)^{B(k,K)} q^{-k^2/2} e^{-2\pi i B(z,k)},
\end{split}  
\ee   
where $E$ is a reparametrization of the error function \eqref{errorf}.
Note that in the limit $y\to \infty$, $E$ in (\ref{hatTheta})
approaches the original $\sgn$-function of (\ref{indeftheta}),
$$\lim_{y\to \infty} E\left(\sqrt{2y}\,u\right)=\sgn(u).$$
If we analytically continue $E$ to a function with complex argument,
then this limit is only convergent for
$-\frac{\pi}{4}<\mathrm{Arg}(u)<\frac{\pi}{4}$. 

The transformation properties under SL$(2,\BZ)$ follow from chapter 2 of Zwegers'
thesis \cite{ZwegersThesis} or Vign\'eras \cite{Vigneras:1977}. One
finds for the action of the generators on $\widehat
\Theta^{JJ'}_{\mu+K/2}(\tau,z)$ 
\be
\label{theta_comp_mod}
\begin{split}
&\widehat \Theta^{JJ'}_{\mu+K/2}(\tau+1,z)=e^{\pi i (\mu^2-K^2/4)}\, \widehat\Theta^{JJ'}_{\mu+K/2}(\tau,z+\mu),\\
&\widehat\Theta^{JJ'}_{\mu+K/2}(-1/\tau,z/\tau)=i(-i\tau)^{n/2} \exp\!\left(-\pi
  i z^2/\tau+\pi i K^2/2\right) \widehat\Theta^{JJ'}_{K/2}(\tau,z-\mu).
\end{split}
\ee

For our application, the  $\bar \tau$-derivative of $\widehat \Theta^{JJ'}_\mu$ is of particular
interest. This gives the ``shadow'' of
$\Theta^{JJ'}_\mu$, whose modular properties are easier to
determine than those of $\Theta^{JJ'}_\mu$. We obtain here 
\be
\label{shadow}
\begin{split} 
\partial_{\bar \tau} \widehat
\Theta^{JJ'}_\mu(\tau,z)=& \Psi^J_\mu(\tau,z)-\Psi^{J'}_\mu(\tau,z),
\end{split}
\ee
with $\Psi^J_\mu$ defined in (\ref{PsiJ}). The modular properties of $\Psi^J_\mu$ are  given in (\ref{Psi_trafos}), and can be obtained using standard Poisson resummation. 

The completion (\ref{hatTheta}) may simplify if the lattice $L$ contains vectors $k_0\in L$ with norm $k_0^2=0$. For such lattices $J$ and/or $J'$ can be chosen to equal such a vector, and careful analysis of the limit shows that the error function reduces to the original sgn-function \cite{ZwegersThesis}.  We assume now that $J'\in L$ such that $(J')^2=0$.  To ensure convergence of the sum, one needs to require furthermore that $B(k+b,J')\neq 0$ for any $k\in L+K/2+\mu$, except if one also has $B(k+b,J)=0$. Then the completion $\widehat \Theta^{JJ'}_\mu$ is given by
\be  
\label{hat2Theta} 
\begin{split}   
\widehat \Theta^{JJ'}_{\mu}\!(\tau,z)=\sum_{k\in L+K/2+\mu} &
\tfrac{1}{2}\left( E(\sqrt{2y}B(k+b, \underline J))-\sgn(B(k+b, J'))\right)\\
& \times (-1)^{B(k, K)} q^{-k^2/2} e^{-2\pi i B(z,k)},
\end{split}  
\ee  
with shadow 
\be
\label{shadow2}
\partial_{\bar \tau}\widehat\Theta^{JJ'}_\mu(\tau,z)=\Psi^J_\mu(\tau,z).
\ee

\section{Regularising the \texorpdfstring{$u$}{u}-plane integral}\label{sec:regularisation}
The $u$-plane integrand can diverge at the cusps. We therefore need to be careful about regularising the integral. This procedure was worked out in detail in \cite{Korpas:2019ava} for the simply connected case, but the analysis goes through without alteration when allowing for $\pi_1(X)\neq 0$. For completeness, we summarise the important steps, and refer the reader to \cite{Korpas:2019ava} for more details. 

We are interested in evaluating integrals on the form \eqref{antiderivativeCH} or \eqref{operTotDeriv}. To make the analysis simpler, we map these integrals over the fundamental domain $\BH/\Gamma^0(4)$ to integrals over the ordinary key-hole domain $\CF$ of $\SL$ by mapping the six copies of $\CF$ in $\BH/\Gamma^0(4)$ back to $\CF$ (see Fig. \ref{fig:funDom}). This means that the integrals now take the form
\begin{equation}
	\int_{\CF} d\tau\wedge d\bar\tau y^{-s}f(\tau,\bar\tau),
\end{equation}
where $f(\tau,\bar\tau)$ is a non-holomorphic modular form for $\SL$ of weight $(2-s,2-s)$, and it corresponds to the sum of the six transformations of the corresponding integrands by the elements of $\SL/\Gamma^0(4)$. It has a Fourier expansion
\begin{equation}
	f(\tau,\bar\tau)=\sum_{m,n\gg-\infty}c(m,n)q^m\bar q^n,
\end{equation} 
with $c(m,n)$ only non-zero for $m-n\in\BZ$. These coefficients are bounded
\begin{equation}
	c(m,n)< e^{\sqrt{Km}+\sqrt{Kn}},
\end{equation}
for some constant $K>0$ and the sum over $m$ and $n$ is therefore absolute convergent for $y<\infty$. The integral does however diverge when $m+n\leq 0$. Which is the domain of our integrals when considering addition of the $\CQ$-exact operators. To deal with this we start by compactifying $\CF$ to a domain $\CF_Y$ by introducing a cut-off $Y\geq y$ for $\tau\to\ima\infty$ and some $Y\gg1$ and defining the integral
\begin{equation}
	\CI_f(Y)=\int_{\CF_Y}d\tau\wedge d\bar\tau y^{-s}f(\tau,\bar\tau).
\end{equation}
We can now regularise our integrals by using the generalised exponential integral $E_l(z)$, defined by
\begin{equation}
	E_l(z)=\begin{cases}
		z^{l-1}\int_z^\infty e^{-t}t^{-l}dt,\quad &\text{for }z\in \BC^*, \\
		\frac{1}{l-1}, \quad &\text{for }z=0,l\neq 0,\\
		0,\quad &\text{for }z=0,l=1,
	\end{cases}
\end{equation}
where for non-integral $l$, the branch of $t^{-l}$ is fixed by restricting the argument of any complex number $\rho\in\BC^*$ to lie in the domain $(-\pi,\pi]$. The regularised integral is
\begin{equation}\label{regI}
	\CI_f^r=\lim_{Y\to\infty}\left[\CI_f(Y)-2\ima\sum_{m\gg-\infty}c(m,m)Y^{1-s}E_s(4\pi mY)\right].
\end{equation}

Now, to evaluate the integrals we can make use of Stokes theorem. Since in our case we have $\partial_{\bar\tau}\hat h=y^{-s}f$ for some $\hat h$ a modular form of weight 2 and $f$ as in the above. This can be integrated using $E_l(z)$ to give
\begin{equation}\label{hhat}
	\hat h(\tau,\bar\tau)=h(\tau)+2\ima y^{1-s}\sum_{m,n\gg-\infty} c(m,n)q^{m-n}E_s(4\pi ny),
\end{equation}
for $s\neq 1$ and when $s=1$ we need to change the terms with $n=0$ in the sum to 
\begin{equation}
	-2\ima \log(y)\sum_{m\gg-\infty}c(m,0)q^m.
\end{equation}
Here, $h$ is a (weakly) holomorphic function with Fourier expansion
\begin{equation}
	h(\tau)=\sum_{m\gg -\infty}d(m)q^m,
\end{equation}
and $h(\tau)$ is uniquely determined by the coefficients $d(m)$ with $m<0$. We can now use Stokes theorem and the integral over $\CF_Y$ becomes
\begin{equation}
	d(0)+2\ima \lim_{Y\to\infty}\sum_{m\gg-\infty} Y^{1-s}c(m,m)E_s(4\pi mY),
\end{equation} 
and regularising this as in \eqref{regI} we find that the only contribution of the regularised integral is the constant term of $h(\tau)$,
\begin{equation}
	\CI^r_f=d(0).
\end{equation}

In Appendix \ref{sec:IcTotDeriv} we demonstrate that the correlation function of the $\CQ$-exact operator $I(S,Y)$ that we add can be written as
\begin{equation}
	\langle I(S,Y)\rangle = \int_{\BH/\Gamma^0(4)}d\tau\wedge d\bar\tau \partial_{\bar\tau} F_I, 
\end{equation}
where 
\begin{equation}
	F_I(\tau,\bar \tau)=y^{-s}\sum_{m,n}c(m,n)q^m\bar q^n.
\end{equation}
We then have that 
\begin{equation}
	\partial_{\bar\tau}F_I=-\ima y^{-s}\sum_{m,n} c(m,n)(2\pi n+\tfrac{1}{2}sy^{-1})q^m\bar q^n,
\end{equation}
which in the above prescription allow us to identify $F_I$ with some $\hat{h}_1+\hat h_2$. Where both $\hat h_1$ and $\hat h_2$ are of the form \eqref{hhat} but with $s$ replaced by $s+1$ in the later case. But, since $F_I$ is a modular form of weight 2 the sum $h_1+h_2$ should vanish, as there is no such holomorphic modular form, and in particular the sum of the constant terms $d_1(0)+d_2(0)=0$, which in term shows us that $\langle I(S,Y)\rangle = 0$. We can thus safely add this operator to the action.

\section{Construction of \texorpdfstring{$I(S,Y)$}{ISY}}\label{sec:constructionIc}
In this appendix, we explain the construction of the $\CQ$-exact operator $I(S,Y)$ in \eqref{isy}, which aids the evaluation of the $u$-plane integral using mock modular forms. 
A constructive approach is to classify all $\CQ$-exact operators in DW theory, add all of them to the path integral, evaluate the path integral and solve for all coefficient functions that lead to the desired properties. For two reasons, this is fortunately not necessary. First, it is convenient that most such operators do not even alter the $u$-plane integrand after integrating out the fermions and the auxiliary field. Second, the path integral can be performed without insertions of any additional operators, or with the insertion of just $I_S$ as was done in the case that $b_1(X)=0$ \cite{Korpas:2017qdo}. Such calculations lead to 
integrands that do not contain the Siegel-Narain theta function $\Psi_\mu^J(\tau,z)$ for any $z$, however only a few terms are missing with an educated guess of $z$. Only very specific $\CQ$-exact operators can provide the necessary terms for the new integrands to complete into $\Psi_\mu^J(\tau,z)$. 

In Section \ref{qexactops}, we classify all possible $\CQ$-exact operators that contribute to the $u$-plane \emph{integrand}. In Section \ref{solutioniy} we demonstrate how the correct $\CQ$-exact operators can be selected, for the simplified example where the intersection $Y\cap Y$ is empty (such that there is no intersection term for $Y\cap Y$). In Section \ref{sec:IcTotDeriv} we finally show that the $\CQ$-exact operators we add do not alter the $u$-plane \emph{integral}.

\subsection{\texorpdfstring{$\CQ$}{Q}-exact observables}\label{qexactops}
Let us complete the result of \cite{Korpas:2019ava} by computing the all $\CQ$-exact observables on a four-manifold $X$ with $\pi_1(X)\neq 0$. Let us first collect
    \begin{alignat*}{2}
    \CC_1&=\{&&\psi\}, \\\
    \CC_2&=\{&&D,F_{\pm},\chi,\psi\wedge\psi\}, \\
    \CC_3&=\{&&\psi\wedge D,\psi \wedge\chi, \psi\wedge F_\pm, \psi\wedge\psi\wedge\psi \}, \\
    \CC_4&=\{&&\psi\wedge\psi\wedge D,\psi\wedge\psi\wedge\chi,\psi\wedge\psi\wedge F_\pm, \psi\wedge\psi\wedge\psi\wedge\psi, \\ \, & &&D\wedge D,D\wedge F_+,D\wedge\chi,F_\pm\wedge F_\pm,F_+\wedge\chi\}.
    \end{alignat*}
These are all $1\dots 4$-forms that can be constructed out of the field content in Table  \ref{tab:fieldcontent}. Since any operator must be gauge invariant, we do not use the $1$-form $A$ to construct operators but only $F= \mathrm dA$. Furthermore, some operators are identically zero due to fermion saturation. The sets $\CC_k$ are then generating sets for the spaces of $k$-form  observables \cite{Korpas:2019ava},
\begin{equation}\label{cok}
    \CO_k=\sum_{j=0}^1\sum_{X\in \CC_k}f_{X,j}(a,\bar a)\eta^j X.
\end{equation}
Here, $f_{X,j}(a,\bar a)$ are real-analytic functions without singularities away from strong and weak coupling. The most generic $0$-form observable is $\CO_0=f_0(a,\bar a)+ f_1(a,\bar a)\eta$.
Let us restrict now to the $\CQ$-exact $k$-observables $[\CQ,\CO_k\}^{\int}$ that survive integration. These do in particular either contain $\eta\chi$ or neither, since otherwise they would not survive the fermionic integration, and they do not contain any derivative term $\mathrm{d} X$, as we consider $b_{2}^{+}(X)=1$ and thus their zero modes vanish. By the notation $[\CQ,\CO\}$ we furthermore mean either $\{\CQ,\CO\}$ or $[\CQ,\CO]$, depending on whether $\CO$ is Grassmann odd or even.

Recall the action \eqref{Qtransformations} of the supersymmetry generator $\CQ$.
It follows that $[\CQ,F_\pm]=(\mathrm d\psi)_\pm$. The identities 
\begin{equation}\label{abcidentities}
  \begin{aligned}
    [A,BC] &= +B[A,C] + [A,B]C \\
    [A,BC]&=-B\{A,C\}+\{A,B\}C \\
\{A,BC\} &= - B[A,C]+\{A,B\}C 
\end{aligned}
\end{equation}
are helpful when computing (anti-)commutators. The action of $\CQ$ on functions $f(a,\bar a)$ is given by
\begin{equation}
    [\CQ ,f(a,\bar a)]=\partial_{\bar a}f(a,\bar a)[\CQ,\bar a]=\sqrt2\ima  \partial_{\bar a}f(a,\bar a)\eta.
\end{equation}

The general $\CQ$-exact observable $[\CQ,\CO_k\}$ from \eqref{cok} is very tedious to compute, luckily in $[\CQ,\CO_k\}^{\int}$ generally not many terms survive. Furthermore, \eqref{cok} has $2\vert\CC_k\vert$  terms, however due to \eqref{Qtransformations} and  \eqref{abcidentities} we have
$[\CQ,\eta \CO\}=-\eta[\CQ,\CO\}$ and for the terms with $j=1$ it remains to multiply the $j=0$ term by $-\eta$. Then only one of $[\CQ,\CO\}$ and $-\eta [\CQ,\CO\}$ is Grassmann even in the variables $\eta$ and $\chi$, such that only one of those can contribute to $[\CQ,\CO\}^{\int}$. 
Lastly, if   $\CO=\prod_l \tilde \CO_l$ is a composite operator, its $\CQ$-action $[\CQ, \CO\}$ is an alternating sum $\sim\sum_{l} [\CQ,\tilde \CO_l\}\prod_{k\neq l}\tilde \CO_k$, and so if all such summands  do not contribute (which can be easily checked) then the whole commutator does not either. With this, it is now slightly less work to extract those summands $\CO_k^{\int}$ of \eqref{cok} that contribute, $[\CQ,\CO_k\}^{\int}=[\CQ,\CO_k^{\int}\}$. They are
\begin{equation}
    \begin{aligned}
   \CO_2^{\int}&=f_1 \chi, \\
   \CO_3^{\int}&=f_2\psi\wedge\chi, \\
   \CO_4^{\int}&=f_3 \psi\wedge\psi\wedge\chi+f_4  D\wedge\chi+f_5F_+\wedge\chi,
    \end{aligned}
\end{equation}
while $[\CQ,\CO_0\}^{\int}=[\CQ,\CO_1\}^{\int}=0$.
Their $\CQ$-commutators give
\begin{equation}
    \begin{aligned}
   \{\CQ,\CO_2^{\int}\}&=+\sqrt2\ima \partial_{\bar a}f_1\eta\chi+\ima f_1(F_+-D), \\
    [\CQ,\CO_3^{\int}]&=-\sqrt2\ima \partial_{\bar a}f_2\eta\chi\wedge\psi-\ima f_2(F_+-D)\wedge\psi, \\
    \{\CQ,\CO_4^{\int}\}&=+\sqrt 2\ima\partial_{\bar a}f_3 \eta \chi\wedge \psi\wedge\psi+\ima f_3(F_+-D)\wedge \psi\wedge\psi \\
       &\quad\, +\sqrt2\ima \partial_{\bar a}f_4\eta\chi\wedge D+\ima f_4 D\wedge (F_+-D) \\
       &\quad\,  +\sqrt2\ima \partial_{\bar a}f_5\eta\chi\wedge F_++\ima f_5 F_+\wedge(F_+-D).
    \end{aligned}
\end{equation}
These are \emph{all} $\CQ$-exact operators in  DW theory. 
The following $\CQ$-exact terms can then be added to the action
\begin{equation}
    \begin{aligned}
   I_2=\int_S\{\CQ,\CO_2^{\int}\}, \quad
   I_3=\int_Y[\CQ,\CO_3^{\int}], \quad
   I_4=\int_X\{\CQ,\CO_4^{\int}\}.
    \end{aligned}
\end{equation}

\subsection{Solution for \texorpdfstring{$I_Y$}{Ic}}\label{solutioniy}
By adding only $I_S$ as suggested in \cite{Korpas:2017qdo,Korpas:2018dag,Korpas:2019ava}, the $u$-plane integrand can be written as a total derivative, however it does not complete to a Siegel-Narain theta function. Let us construct the operator $I_Y$ such that this becomes true.  For simplicity, we ignore the contact terms $I_\cap$. This is possible since all contact terms other than the $Y\cap Y$  are integrated over $\psi$ and $\tau$ only and therefore do not affect the path integral calculation. For simplicity and only in this section, we take the intersection $Y \cap Y$ to be empty. 

We therefore aim to find the functions $f_1,\dots f_5$.
In the case $\pi_1(X)=0$, the total integrand must go back to \eqref{CBI}. If $f_4$ and $f_5$ are nonzero, this is not the case since they alter the integral. \footnote{This is certainly true if $f_4$ and $f_5$ can be varied. It is possible in principle that for specific functions $f_4$ and $f_5$ the $\pi_1(X)=0$ \emph{integral} does not change.} We therefore set $f_4=f_5=0$. Thus, in the simply connected case, we have $I_2= I_S$, which implies  $f_1=-\frac{1}{4\pi}\frac{d\bar u}{d\bar a}$.
We shall therefore consider adding the correction
\begin{equation}
\begin{aligned}
   I_Y&=-   \sqrt2\ima \partial_{\bar a}f_2\eta B(\chi,\psi\wedge Y)-\ima f_2B(F_+-D,\psi\wedge Y) \\
   &\quad +\sqrt 2\ima\partial_{\bar a}f_3 \eta B(\chi,\psi\wedge\psi)+\ima f_3B(F_+-D,\psi\wedge\psi).
\end{aligned}
\end{equation}
to the exponential in \eqref{CBI}.
The terms $\psi\wedge Y$ and $\psi\wedge\psi$ are precisely the terms that lead to the problems if only $I_S$ is added.  We can organise $h\coloneqq f_3\psi\wedge\psi-f_2\psi\wedge Y$,
such that 
\begin{equation}
    I_Y=\sqrt2\ima \eta B(\chi, \partial_{\bar a} h)+\ima B(F_+-D, h).
\end{equation}
Inserting it into the path integral we find
\begin{equation}
D=\frac{\sqrt2\ima}{4y}\frac{d\bar\tau}{d\bar a}\eta \chi-4\pi (b_++\omega_+)+\frac{4\pi\ima}{y}h_+.
\end{equation}
After integrating out $D$, this produces new terms
\begin{equation}
    4\pi\ima B(k_++b_+,h)+\frac{\sqrt2}{4y}\frac{d\bar\tau}{d\bar a}\eta B(\chi,h)+\sqrt{2}\ima \eta B(\chi,\partial_{\bar a} h)
\end{equation}
to \eqref{dintegration} (notice that $\omega\wedge h=h\wedge h=0$). The first term is only integrated over $\psi$ and $\tau$, so it will not play a role immediately. The second and third term yield after the fermionic integration,
\begin{equation}\label{hbar_der}
    \frac{d\bar \tau}{d\bar a}\left(-\frac{\sqrt2}{4y}h-\sqrt2\ima \partial_{\bar \tau} h\right).
\end{equation}
In view of \eqref{taubarder} and the above discussion, we can aim this new contribution to give the missing factor
\begin{equation}\label{omegabar}
\sqrt y \frac{d\bar \tau}{d\bar a}\partial_{\bar \tau} \sqrt{2y}\,  \bar\omega=\frac{d\bar\tau}{d\bar a}\left(\frac{\sqrt2\ima}{4}\bar\omega+\sqrt2 y\partial_{\bar\tau}\bar\omega\right),
\end{equation}
such that the Siegel-Narain theta function has an elliptic variable $z=\rho+2\ima y \omega$ and $\beta=b+\omega+\bar\omega$. \footnote{Another possibility would be to chose $h$ to be holomorphic and cancelling the $\omega$ inside the derivative. This is possible, however, the $\omega$ dependence does not drop from the SN theta function.}
Motivated by the computation \eqref{taubarint}, we make the ansatz $h=\ima c y \bar\omega$, with $c\in\mathbb C$ some number. From this it follows that $y\partial_{\bar\tau}\bar\omega=-\frac{\ima}{c}\partial_{\bar \tau}h-\frac\ima2\bar\omega$. Notice that $h$ is purely anti-holomorphic, while $\bar\omega$ is not. We find that \eqref{hbar_der} equals 
\eqref{omegabar} precisely for $c=1$. From this, it is easy to find
\begin{equation}
    f_2=\frac{3\ima a_3}{16}\frac{d^2\bar u}{d\bar a^2}, \qquad f_3=\frac{\sqrt2}{2^7\pi}\frac{d\bar \tau}{d\bar a}.
\end{equation}
 In the simply connected case, the correction $I_2=I_S$ is necessary in order for the surface observable $\tilde I_-(S)$ to combine into a Siegel-Narain function such that the $u$-plane integral is a total derivative. In the case $\pi_1(X)\neq0$, an analogous procedure is required for the $3$-cycle $Y$, which combines to a $2$-form as $\psi\wedge Y$. In the $\pi_1(X)\neq0$ Lagrangian \eqref{lagrangianpi1neq0} there is a new term $\psi\wedge\psi$ that is integrated over $\eta$, $\chi$ and $D$, such that the $u$-plane integral \emph{is} a total derivative but does not contain a SN theta function. After the insertion of an anti-holomorphic $\CQ$-exact $4$-form operator, the integrand indeed becomes a Siegel-Narain theta function.

\subsection{Ward-Takahashi identity for \texorpdfstring{$ I_Y$}{Ic}}\label{sec:IcTotDeriv}
In \cite{Korpas:2019ava} it was shown that the vacuum expectation value of any $Q$-exact operators vanish in the simply connected case. This then allows one to safely add such operators to the $u$-plane integral. In this appendix we only consider the operator insertion $I_Y$, the analyses is similar for the other factors in $I(S,Y)$. To this end, we will demonstrate that $\langle I_Y\rangle$ can be written as an integral where the integrand can be written as a total derivative. We can then use the renormalisation procedure of \cite{Korpas:2019ava}, as discussed in Sec. \ref{sec:regularisation}, to show that $\langle  I_Y\rangle=0$, such that the insertion of this operator does not change the $u$-plane integral.  

To see that the one-point function can be written as a total derivative we start from
\begin{equation}
    \langle I_Y\rangle = \int [dad\eta d\chi dD]\int_{{\rm Pic}(X)}d\psi\nu(\tau)\sum_{k\in L+\mu} I_Y e^{-\int_X \CL'},
\end{equation}
with $\CL'$ as in \eqref{lagrangianpi1neq0} and $\nu(\tau)$ as in \eqref{nu}. 

The integration over $D$ yields
\begin{equation}\footnotesize
\begin{aligned}
       \langle  I_Y\rangle = 2\pi i\int [dad\eta d\chi]\int_{{\rm Pic}(X)}d\psi\nu(\tau)\sum_{k\in L+\mu} \sqrt{\frac{2}{y}}\Big(&-\sqrt{2}\eta\chi\wedge\partial_{\bar a}(y\bar\omega)-4\pi yk_+\wedge\bar\omega \\
    &+\frac{i\sqrt{2}}{4}\left(\frac{d\bar\tau}{d\bar a}\eta\chi-\frac{1}{8}\frac{d\tau}{da}\psi\wedge\psi\right)\wedge \bar\omega\Big)  e^{-\int \CL''},
\end{aligned}
\end{equation}
where now 
\begin{equation}
\begin{aligned}
    \CL''=&\pi i\bar\tau k_+^2+\pi i\tau k_-^2+\frac{i\sqrt{2}}{4}\frac{d\bar \tau}{d\bar a}\eta\chi\wedge k_+-\frac{i\sqrt{2}}{2^5}\frac{d\tau}{da}\psi\wedge\psi\wedge k_- \\
    &+\frac{1}{2^{11}}\left(\frac{i}{3}\frac{d^2\tau}{da^2}-\frac{1}{2\pi y}\left(\frac{d\tau}{da}\right)^2\right)\psi\wedge\psi\wedge\psi\wedge\psi\\
    &-\frac{1}{2^6\pi y}\left(\frac{d\bar\tau}{d\bar a}\right)^2\eta\chi\wedge \eta\chi+\frac{1}{8\pi y}\frac{d\tau}{da}\frac{d\bar\tau}{d\bar a}\psi\wedge\psi\wedge\eta\chi.
\end{aligned}
\end{equation}
Next, we integrate over $\eta$ and $\chi$. This yields 
\begin{equation}\footnotesize
\begin{aligned}
       \langle  I_Y\rangle =\int[da]&\int_{{\rm Pic}(X)} d\psi\nu(\tau) \sum_{k\in L+\mu}\frac{ 4\pi i}{\sqrt{y}}e^{-\int  \CL'''} \\
       &\times B\left(y\partial_{\bar a}\bar\omega+\frac{i}{4}\frac{d\bar\tau}{d\bar a}\bar \omega -y\left(\pi i\frac{d\bar\tau}{d\bar a}k_+^2+\frac{i}{2^{13}\pi y^2}\left(\frac{d\tau}{da}\right)^2\frac{d\bar\tau}{d\bar a}\psi^4\right) \bar\omega,\underline J\right),
\end{aligned}
\end{equation}
where by $k_+^2$ and $\psi^4$ we mean $B(k_+,k_+)$ and $B(\psi\wedge\psi,\psi\wedge\psi)$ and we now have
\begin{equation}
    \CL'''=\pi i\bar \tau k_+^2+\pi i\tau k_-^2-\frac{i\sqrt{2}}{2^5}\frac{d\tau}{da}\psi\wedge\psi\wedge k_-+\left(\frac{i}{3\cdot 2^{11}}\frac{d^2\tau}{da^2}-\frac{1}{2^{12}\pi y}\left(\frac{d\tau}{da}\right)^2\right)\psi^4,
\end{equation}
and its now straightforward to recognise that we indeed can write the integrand as a total derivative
\begin{equation}\label{operTotDeriv}
    \langle I_Y\rangle =4\pi i\int d\tau\wedge d\bar \tau\int_{{\rm Pic}(X)} d\psi \partial_{\bar \tau}\sum_{k\in L+\mu}\left(\tilde\nu(\tau)B(\sqrt{y}\bar\omega,\underline J) e^{-\int \CL'''}\right).
\end{equation}

We can now use the renormalisation procedure of \cite{Korpas:2019ava} to show that $\langle  I_Y\rangle$ in fact vanishes. The broad strokes are summarised in Sec. \ref{sec:regularisation}, and further details can be found in \cite[Sec. 5]{Korpas:2019ava}.

\section{Classical topological invariants}
Let $X$ and $Y$ be topological spaces, and consider the homology over a field $\mathbb K$. The K\"unneth theorem states that there exists an isomorphism
\begin{equation}\label{knn}
H_n(X\times Y, \mathbb K) \cong \bigoplus_{k=0}^n H_k(X, \mathbb K)\otimes H_{n-k}(Y,\mathbb K).
\end{equation}
Written in terms of Poincar\'e  polynomials $
p_X(z)\coloneqq \sum_{k=0}^\infty b_k(X)z^k$,
that is,  generating functions of Betti numbers $b_k(X) \coloneqq \text{rank} \, H_k(X)$, we have
\begin{equation}\label{pxy}
p_{X\times Y}(z) = p_X (z)p_Y(z).
\end{equation}
The Betti numbers are then related as
\begin{equation}\label{bnxy}
b_n(X\times Y) = \sum_{k=0}^n b_k(X)b_{n-k}(Y).
\end{equation}
Poincar\'e duality states that $
b_k(M)= b_{n-k}(M)$
for any oriented closed $n$-manifold $M$. The manifolds $M$ under consideration satisfy $b_2^+(M)=1$ and vanish unless $b_1(M)$ is even. The topological invariants are then related as $\sigma+b_2=2$ and $\chi+\sigma=4-2b_1$.

\emph{Example.$\quad$} Note that $
p_X(-1) =\chi(X)$
gives the Euler characteristic of $X$. 

\emph{Example.$\quad$}
The circle $\mathbb S^1$ has Betti numbers $b_0(\mathbb S^1)=b_1(\mathbb S^1)=1$ and all other zero. We therefore have $
p_{\,\mathbb S^1}(z)=1+z$ (more generally, $p_{\,\mathbb S^n}(z)=1+z^n$).
For the torus $\mathbb T^n \coloneqq (\mathbb S^1)^{{\times n}}$, it follows from \eqref{pxy} that 
\begin{equation}
 p_{\,\mathbb T^n}(z) =p_{\,\mathbb S^1}(z)^n= (1+z)^n = \sum_{k=0}^n \binom{n}{k} z^k,
\end{equation}
that is, $
b_k(\mathbb T^n)=\binom{n}{k}$.

\subsection{Classification of  product 4-manifolds with \texorpdfstring{$b_2^+=1$}{b2+=1}}\label{classificationb2+1}
Let us study all decompositions of $M$ into Cartesian products of smooth, closed, oriented, connected $1,2,3$-manifolds such that $b_2^+(M)=1$. The dimension 4 has five partitions, namely $4$, $3+1$, $2+2$, $2+1+1$ and $1+1+1+1$.
First, note that if $M=X\times Y$, then $\sigma(M)=0$.\footnote{Sketch of the proof: From \eqref{knn} we have that \begin{equation}
H_2(X\times Y)=H_0(X)\otimes H_2(Y)\oplus H_1(X)\otimes H_1(Y)\oplus H_2(X)\otimes H_0(Y).
\end{equation}
Since $H_2(X\times Y)$ is symmetric in $X$ and $Y$, the intersection form $H^2(X\times Y,\mathbb R)\times H^2(X\times Y,\mathbb R)\to \mathbb R$ has equally many positive as negative eigenvalues, $b_2^+(X\times Y)=b_2^-(X\times Y)$. Therefore, $\sigma(M)=b_2^+(M)-b_2^-(M)=0$.} The same is true for all decompositions of $4$. Since $\sigma(M)+b_2(M)=2b_2^+(M)$, we have that $b_2^+(M)=\frac12 b_2(M)$ whenever $M$ is a product.

Let us begin with $4=1+1+1+1$. If $X$ is a $1$-manifold, then $b_0(X)=1$, and by Poincar\'e duality also $b_1(X)=1$, such that $X\cong \mathbb S^1$. One easily  computes $b_2^+(\mathbb S^1\times \mathbb S^1\times\mathbb S^1\times\mathbb S^1)=3$, and thus the $u$-plane integral vanishes.

For $4=2+1+1$ the most general manifold is $M=\Sigma_g\times \mathbb S^1\times\mathbb S^1$ with $\Sigma_g$ a genus $g$ Riemann surface. One finds that $b_2^+(M)=1+2g$, such that only at genus $g=0$ the $u$-plane integral is nonzero. This gives the manifold
\begin{equation}\label{s2s1s1}
\mathbb{CP}^1\times\mathbb S^1\times\mathbb S^1
\end{equation}
with $\chi=\sigma=0$ and $b_1=2$.

For $4=2+2$ the most generic $4$-manifold is $M=\Sigma_h\times \Sigma_g$, a product of genus $g$ and $h$ Riemann surfaces. It has $b_2^+=1+2gh$, and w.l.o.g. we can take $h=0$, such that $\Sigma_h$ is a 2-sphere. Then the product ruled surface
\begin{equation}\label{fgs2}
\mathbb{CP}^1\times \Sigma_g , \quad g\in\mathbb N_0
\end{equation}
has $\sigma=0$, $\chi=4(1-g)$ and $b_1=2g$.

When $4=3+1$ we have $M= Y\times \mathbb S^1$, where $b_1(Y)=b_2(Y)$ by Poincar\'e duality.  It has $b_2^+(M)=b_1(Y)$, such that for $b_2^+(M)=1$ we have $p_Y(z)=1+z+z^2+z^3=p_{\,\mathbb S^2}(z)p_{\,\mathbb S^1}(z)$ and therefore $M\cong\mathbb S^2\times\mathbb S^1\times\mathbb S^1$, just as in \eqref{s2s1s1}.

Now \eqref{fgs2} in fact includes \eqref{s2s1s1} because $\Sigma_1\cong \mathbb T^2=\mathbb S^1\times\mathbb S^1$. This proves that the Betti numbers of any smooth, closed, oriented and  connected  product $4$-manifold with $b_2^+=1$ are given by those of \eqref{fgs2}.  The simplest examples are at genus $0$ and $1$, which are $\mathbb{CP}^1\times \mathbb{CP}^1$ and $\mathbb{CP}^1\times \mathbb S^1\times \mathbb S^1$.

\subsection{Gromov-Witten invariants as \texorpdfstring{$A$}{A}-model correlation functions}\label{gwappendix}
For a smooth projective variety $X$, GW invariants
essentially count, in a refined way, algebraic curves with certain incidence conditions. Physically, GW invariants are given in terms of specific correlation functions in the topological A-model and count specific holomorphic curves. This construction is well known \cite{Hori:2003ic} but we will provide a small review in this section. 

The action of the A-model is written explicitly in \eqref{action_A} for our case of interest in the main body of the paper. For the purposes of this subsection, we can recast it as
\begin{equation}\label{action_AV}
    S \sim \int_{\Sigma} \{ \mathcal{Q}, V\} + \int_\Sigma \varphi^*\omega,
\end{equation}
with $V = g_{i\bar j}( \rho_z^{\bar i}\partial_{\bar z}\varphi^j + \partial_z \varphi^{\bar i} \rho_{\bar z}^j )$ and $\varphi: \Sigma \to X$ where $\Sigma$ is the worldsheet and $X$ is the target space. The second summand corresponds to the pullback of the K\"ahler form of X and it only depends on the cohomology
class of $\omega$ as well as the homotopy type of $\varphi$ making it invariant under continuous deformations
of $g=g(\Sigma)$. As discussed in the main body of the paper, the correlation functions are given in the form:
\begin{align}
    \langle \CO_1,\ldots, \CO_n\rangle = \int [\mathcal{D}\Phi]\,\CO_1 \ldots \CO_n e^{-S}
\end{align}
The fixed loci of the supersymmetry transformations (see \cite{Hori:2003ic}) are given as
\begin{align}
    \partial_{\bar i} \varphi^j & = 0,\\
    \partial_i \varphi^{\bar j} & = 0,
\end{align}
and $\varphi$ that satisfy these equations are called worldsheet instantons. For an image of $\varphi(\Sigma)$ in $X$, there is a class $\beta \in H_2(X,\mathbb{Z})$ whose basis is given by $\{c_1,\ldots,c_m \}$, $m={\rm dim}(H_2(X,\mathbb{Z}))=b_2(X)$. Restricting to the bosonic part $S_{\rm bos}$ of \eqref{action_AV} for worldsheet instantons, it is possible to show that $S_{\rm bos} = B(\omega,\beta)$ which corresponds to the volume of the image of the worldsheet, where $\omega$ is the K\"{a}hler form. Then
\begin{align}
    B(\omega,\beta) &= -\ima\theta\bigg(\int_{\beta} \omega \bigg)\\
                &= -\ima\theta\bigg(\sum_{i=1}^{m}n_i\int_{c_i} \omega\bigg)\\
                &= -\ima\theta\bigg(\sum_i n_i t_i\bigg)
\end{align}
for weights $n_i$ such that
\begin{align}\label{restrictions}
    t_i := B(\omega,c_i).
\end{align}
Correlation functions vanish unless
\begin{align}
    \sum_i^n p_i &= \sum_i^n q_n \\
                 & = {\rm dim}(X)(1-g)+ B(c_1(X),\beta),
\end{align}
where $p_i$ and $q_i$ are the holomorphic (chiral) and anti-holomorphic (anti-chiral) degrees of the operator $\CO_i$. For example, a three-point function gives
\begin{align}
    \langle \CO_1 \CO_2 \CO_3 \rangle = \# (D_1\cap D_2 \cap D_3 )e^{B(\omega,\beta)}
\end{align}
where $\exp(B(\omega,\beta))$ is the instanton contribution to the correlation function.

The correlation function essentially contains the information on the counts the number of holomorphic maps of genus $g$ at $n$ intersection points to the class $\beta$ such that the operator insertions $\CO_i$ are mapped into divisors $D_i$ of $X$. 

\section{Reduction of DW theory to 2d}\label{reductionaction}
Using $A, B$ and $a, b$ to denote indices on the large and small Riemann surfaces, respectively, for a  four-manifold $M_4$ of the form $\Sigma \times C$, where each factor is a real 2d surface, the metric can be written in a block diagonal form
\begin{equation}\label{metricdeform}
    ds^2 = (G_{\Sigma})_{AB}dx^{A}dx^{B} + \epsilon(G_{C})_{ab}dx^{a}dx^{b}.
\end{equation}
Eventually, we want to let $\epsilon\rightarrow 0$ in order to shrink $C$. The high energy DW action (after twisting) can be written as 
\begin{equation}
    \begin{aligned}\label{highedwaction}
        S_{DW} &= \frac{1}{e^2} \int_{M_4} \sqrt{G_{M_4}} {\rm Tr} \Bigg(-\frac{1}{4}F^{\mu\nu}F_{\mu\nu} + D_{\mu}\phi D^{\mu}\phi^{\dagger} + \ima\eta D^{\mu}\psi_{\mu} \\&- \ima\chi_{\dot{\alpha}\dot{\beta}}\left(\bar{\sigma}_{\mu\nu} \right)^{\dot{\alpha}\dot{\beta}}D^{\mu}\psi^{\nu} +\frac{1}{\sqrt{2}}\chi^{\dot{\alpha}\dot{\beta}} [\chi_{\dot{\alpha}\dot{\beta}}, \phi] - \frac{1}{2\sqrt{2}}\psi_{\mu}[\psi^{\mu}, \phi^{\dagger}] \\
        &+ \ima\sqrt{2}\eta[\phi, \eta] -\frac{\ima}{2}[\phi, \phi^{\dagger}]^2 \Bigg). 
    \end{aligned}
\end{equation}
Using \eqref{metricdeform}, we have $\sqrt{G_{C}}\rightarrow \epsilon\sqrt{G_{C}}$, $g^{ab}\rightarrow \epsilon^{-1} g^{ab}$. Thus for the inner product of two 2-form terms $O_{(2)}$, we have
\begin{equation}
    \begin{aligned}\label{2formdeform}
        &\int_{M_4} O_{(2)}\wedge* O_{(2)}\\ &=  \int_{M_4}  \sqrt{G_{M_4}}O_{(2)}^{\mu\nu}O_{{(2)\mu\nu}} \\
   &= \epsilon \int_{M_4} \sqrt{G_{\Sigma}}\sqrt{G_{C}} g^{\mu\rho}g^{\nu\lambda}O_{{(2)\mu\nu}}O_{{(2)\rho\lambda}}\\
   &= \epsilon \int_{M_4}  \sqrt{G_{\Sigma}}\sqrt{G_{C}} \Bigg(g^{AC}g^{BD}O_{{(2)AB}}O_{{(2)CD}} + \epsilon^{-2}g^{ac}g^{bd}O_{{(2)ab}}O_{{(2)cd}}\\
   &\qquad+ 2\epsilon^{-1}g^{AC}g^{bd}O_{{(2)Ab}}O_{{(2)Cd}} \Bigg).
    \end{aligned}
\end{equation}
 Taking $\epsilon\rightarrow 0$, we see that only the terms $O_{{(2)Ab}}$ with mixed indices survive. Repeating the same process for the inner product of 1-forms gives us $O_{{(1)a}}$, where only 1-forms with a small index survives. Note also that the scalar interaction terms $[\phi, \phi^{\dagger}]^2$ and $\eta[\phi, \eta]$ in \eqref{highedwaction} do not survive when the metric is deformed. 

 Upon reduction on $C$ for $M_4=\Sigma \times C$, the fields $\phi$, $\phi^{\dagger}$, $\eta$ and $A_{\Sigma}$ have no derivatives on $\Sigma$, and become auxiliary fields which can be integrated out.  What is left of the fermionic fields are $\psi_{C}$ and $\chi_{\Sigma C}$ which can be interpreted as a 0-form, 1-form on $\Sigma$, respectively. The bosonic field left will be the gauge field $A_{C}$. From \eqref{2formdeform}, we see that deformation of the metric forces us to have $O_{{(2)ab}}=0$. For the field strength $F_{\mu\nu}$, we thus have 
\begin{equation}
    F_{ab} = 0 .
\end{equation}
Since $A_{C}$ is the only leftover bosonic field on $\Sigma$, we must have configurations of $A_C$ that gives us $SU(2)$ flat connections on $C$. We can thus take $A_C$ to be a map $A_{C} : \Sigma \to \mathcal{M}_C$, where $\mathcal{M}_C$ is the moduli space of flat connections on $C$. We can express variations of $A_C$ about configurations that give flat connections on $C$ in terms of basis cotangent vectors $\alpha_{IC}$ on $\mathcal{M}_C$ by 
\begin{equation}
    \frac{\partial A_{\bar{w}}}{\partial \varphi^i} = \alpha_{i\bar{w}} - D_{\bar{w}}E_{i},
\end{equation}
where $E_I$ are connections on $\mathcal{M}_C$, and the indices $i, \bar{i}$ etc, are for collective complex coordinates on $\mathcal{M}_C$, $z, \bar{z}$ are complex coordinates on $\Sigma$, and $w, \bar{w}$ are complex coordinates on $C$. The connection $E_i$ helps define a covariant derivative on $\mathcal{M}_C$,
\begin{equation}
    \nabla_i = \partial_{i} - \ima E_i.
\end{equation}
 The remaining fermionic fields $\psi_{C}$ and $\chi_{\Sigma C}$, being cotangent vectors on $\mathcal{M}_C$, can be expressed as linear combinations of $\alpha_{i\bar{w}}$, $\alpha_{\bar{i}w}$ :
\begin{equation}
     \psi_{\bar{w}} = \chi^{i}\alpha_{i\bar{w}}, \quad \psi_{w} = \chi^{\bar{i}}\alpha_{\bar{i}w}\quad \chi_{zw}= \rho_{z}^{\bar{i}}\alpha_{\bar{i}w}, \quad \chi_{\bar{z}\bar{w}}= \rho_{\bar{z}}^{i}\alpha_{i\bar{w}}.
\end{equation}
Next, we integrate out the fields in 4d that do not depend on derivatives on $\Sigma$, namely $\phi$, $\phi^{\dagger}$, $\eta$ and $A_{z}$. The equation of motion for $A_{z}$ gives \cite{bershadsky1995topological, harvey1993string}
\begin{equation}\label{Aonsigma}
    A_{z} =  \partial_{z}\varphi^{i}E_i+\partial_{z}\varphi^{\bar{i}}E_{\bar{i}} - \Phi_{\bar{i}j}\rho_{z}^{\bar{i}}\chi^{j},
\end{equation}
(and corresponding terms for $A_{\bar{z}}$) where $\Phi_{i\bar{j}}$ is the curvature on $\mathcal{M}_C$, defined by
\begin{equation}
    \Phi_{i\bar{j}} =  \ima\big[\nabla_i , \nabla_{\bar{j}} \big].
\end{equation}
For $\phi$ and $\phi^{\dagger}$, we obtain 
\begin{equation}
    \begin{aligned}\label{DDphi}
         D_{w} D_{\bar{w}} \phi &=  \ima\big\{ \psi_{w},\psi_{\bar{w}} \big\}\quad, \quad   D_{w} D_{\bar{w}} \phi^{\dagger} =  \ima g^{z\bar{z}}\big\{ \chi_{zw},\chi^{\bar{z}\bar{w}} \big\}.
    \end{aligned}
\end{equation}
Lastly, we obtain, for $\eta$, the constraints $D_w \psi_{\bar{w}} = D_{\bar{w}} \psi_{w} = D_{w} \chi_{\bar{z}\bar{w}} = D_{\bar{w}} \chi_{zw} = 0$.

Since the moduli spaces of flat connections on Riemann surfaces are K\"{a}hler manifolds, we introduce a metric (symmetric in its indices) and symplectic form on $\mathcal{M}_C$ as
\begin{equation}
    \begin{aligned}
         G_{i\bar{j}} &= \int_{C} \; {\rm Tr}(\alpha_{i}\wedge\star\; \alpha_{\bar{j}}),\\
    \omega_{i\bar{j}}&= \int_{C}  \; {\rm Tr}(\alpha_{i}\wedge\alpha_{\bar{j}}),
    \end{aligned}
\end{equation}
respectively. Given these two objects, we can also obtain the connection coefficients and the Riemann tensor on  $\mathcal{M}_C$
\begin{equation}\label{riemanngamma}
    \begin{aligned}
        \Gamma_{ij}^{k} &= \partial_{j} G_{i\bar{k}}= \int_{C} d^2 w \; {\rm Tr}(\alpha_{\bar{k}w}\nabla_{j}\alpha_{i\bar{w}}) \quad,\\
    R_{i\bar{j}k\bar{l}} &= \partial_{\bar{l}}\Gamma_{ik}^{j}=\int_{C} d^2 w \; {\rm Tr}(\nabla_{i}\alpha_{\bar{j}w}\nabla_{\bar{l}}\alpha_{k\bar{w}}+\nabla_{k}\alpha_{\bar{j}w}\nabla_{\bar{l}}\alpha_{i\bar{w}}).
    \end{aligned}
\end{equation}
These (and the conjugates with barred indices) are the only components of the connection and the Riemann tensor due to the fact that $\mathcal{M}_C$ is a K\"{a}hler manifold.

 We now use \eqref{Aonsigma} and the constraints from integrating $\eta$ out to obtain the field strength as 
\begin{equation}
    \begin{aligned}\label{mixedf}
        F_{zw} &=  \partial_{z}A_{w}-D_{w}A_{z}\\
    &= \partial_{z}\varphi^{\bar{i}}\alpha_{\bar{i}w}+\nabla_{j}\alpha_{\bar{i}w}\rho_{z}^{\bar{i}}\chi^{j}
    \end{aligned}
\end{equation}
where the relation $D_{\bar{w}}\Phi_{i\bar{j}}=\nabla_{\bar{j}}\alpha_{i\bar{w}}$ has been used. Next, with the identity $[\alpha_{\bar{w}i}, \alpha_{w\bar{k}}]=\ima D_{\bar{w}}D_{w}\Phi_{i\bar{k}}$ and \eqref{DDphi}, we obtain
\begin{equation}\label{phiphis}
    \phi = \chi^{i}\chi^{\bar{j}}\Phi_{i\bar{j}}, \quad \phi^{\dagger} = g^{z\bar{z}}\rho^{\bar{j}}_{z}\rho^{i}_{\bar{z}}\Phi_{i\bar{j}} .
\end{equation}
Dealing with the kinetic terms in the 4d theory, we can substitute \eqref{phiphis} into $D_{w}\phi D_{\bar{w}}\phi^{\dagger}$, and together with the square of the second term in \eqref{mixedf} that comes from $F\wedge \star F$, we will obtain the Riemann tensor of $\mathcal{M}_C$ as shown in \eqref{riemanngamma}. Similarly, the square of the first term in \eqref{mixedf} with the K\"{a}hler metric will give the kinetic term for the bosonic field in the 2d A-model. 

The only fermionic kinetic term left in 4d after reduction on $C$ is $\chi_{zw}D_{\bar{z}}\psi_{\bar{w}}$ and $\chi_{\bar{z}\bar{w}}D_{z}\psi_{w}$. Since there is a covariant derivative on $\Sigma$, we make use of the expression for $A_{z}$ in \eqref{Aonsigma}, and together with the cross terms of \eqref{mixedf} from $F\wedge \star F$, we obtain the kinetic terms for the fermion fields $\chi^{i}$, $\chi^{\bar{i}}$, $\rho_{z}^{\bar{i}}$ and $\rho_{\bar{z}}^i$. In particular, we obtain the covariant derivative on $\Sigma$ as $\nabla_{\bar{z}}\chi^{i}=\partial_{\bar{z}}\chi^{i}+\chi^{j}\Gamma_{jk}^{i}\partial_{\bar{z}}\varphi^{i}$.

One can also add an instanton term $\frac{\ima\theta}{8\pi^2}\int_{M_4}  {\rm Tr}\left(F\wedge F \right)$ to the DW action in \eqref{highedwaction}, which then translates to the pullback of the K\"{a}hler form in the A-model\cite{Ashwinkumar:2019owj}.

At the end, we obtain the A-model, whose action is 
\begin{equation}
    \begin{aligned}
        S &= \frac{1}{e^2}\int_{\Sigma} \Bigg( d^{2}z G_{i \bar{j}}\Big(\frac{1}{2}\partial_{z}\varphi^{i} \partial_{\bar{z}}\varphi^{\bar{j}} + \frac{1}{2}\partial_{\bar{z}}\varphi^{i} \partial_{z}\varphi^{\bar{j}} + \ima\rho^{\bar{j}}_{z}\nabla_{\bar{z}}\chi^{i} + \ima\rho^{i}_{\bar{z}}\nabla_{z}\chi^{\bar{j}}\Big)\\
        &\quad  - R_{i\bar{j}k\bar{l}}\rho^{i}_{\bar{z}} \rho^{\bar{j}}_{z}\chi^{k}\chi^{\bar{l}}\Bigg) + \ima\theta\int_{\Sigma} \varphi^{*}\omega.
    \end{aligned}
\end{equation}
Explicitly, the pullback of the K\"{a}hler form goes as (with $I, J$ as real coordinates on $\mathcal{M}_{C}$)
\begin{equation}
     \ima\theta\int_{\Sigma} \varphi^{*}\omega = \frac{\ima\theta}{2\pi^2} \int_{\Sigma} g^{z\bar{z}} \omega_{IJ}\partial_{z}\varphi^I\partial_{\bar{z}}\varphi^J .
\end{equation}
This term is the same as the corresponding one in the purely bosonic nonlinear sigma model, even after adding supersymmetry. There are no fermions despite it being supersymmetric. This is due to the fact that it is a topological term which does not change under small continuous variations of the fields. Hence, there are neither bosonic nor fermionic degrees of freedom which allows it to remain supersymmetric. 

We can compare the fields from both theories in table \ref{fieldmatchtable}. Fields that become auxiliary after reduction are left out from the 2d side of the table. On the 2d side, the bosonic field is the map $X : \Sigma \to \mathcal{M}_C $ and from the ghost charge, can be identified with the leftover gauge field $A_C$. Similarly, we can identify the corresponding fermionic fields as well. 
\begin{table}[h]
\begin{center}
$\begin{tabular}{|c|c|c|c|} 
\hline
4d $\mathcal{N}=2$ DW & U(1) Ghost Charge & 2d A-model & U(1) Ghost Charge\\
\hline
$A_{\mu}$ & 0 & $\varphi^{i}, \varphi^{\bar{i}}$ & 0\\
$\psi_{\mu}$ & 1 & $\chi^{i},\chi^{\bar{i}} $ & 1 \\ 
$\chi_{\mu\nu}$ & -1 & $\rho^{\bar{i}}_{z}$, $\rho^{i}_{\bar{z}}$ & -1 \\ 
$\eta$ & -1 & - & -\\
$\phi$ & 2 & - & -\\ 
$\phi^{\dagger}$ & -2 & - &- \\ 
D & 0 & - & - \\
\hline
\end{tabular}$
\caption{Fields and ghost charges of the 4d and 2d actions.}\label{fieldmatchtable}
\end{center}
\end{table}


\bibliography{DWns}


\begin{thebibliography}{51}
\ifx \bisbn   \undefined \def \bisbn  #1{ISBN #1}\fi
\ifx \binits  \undefined \def \binits#1{#1}\fi
\ifx \bauthor  \undefined \def \bauthor#1{#1}\fi
\ifx \batitle  \undefined \def \batitle#1{#1}\fi
\ifx \bjtitle  \undefined \def \bjtitle#1{#1}\fi
\ifx \bvolume  \undefined \def \bvolume#1{\textbf{#1}}\fi
\ifx \byear  \undefined \def \byear#1{#1}\fi
\ifx \bissue  \undefined \def \bissue#1{#1}\fi
\ifx \bfpage  \undefined \def \bfpage#1{#1}\fi
\ifx \blpage  \undefined \def \blpage #1{#1}\fi
\ifx \burl  \undefined \def \burl#1{\textsf{#1}}\fi
\ifx \doiurl  \undefined \def \doiurl#1{\url{https://doi.org/#1}}\fi
\ifx \betal  \undefined \def \betal{\textit{et al.}}\fi
\ifx \binstitute  \undefined \def \binstitute#1{#1}\fi
\ifx \binstitutionaled  \undefined \def \binstitutionaled#1{#1}\fi
\ifx \bctitle  \undefined \def \bctitle#1{#1}\fi
\ifx \beditor  \undefined \def \beditor#1{#1}\fi
\ifx \bpublisher  \undefined \def \bpublisher#1{#1}\fi
\ifx \bbtitle  \undefined \def \bbtitle#1{#1}\fi
\ifx \bedition  \undefined \def \bedition#1{#1}\fi
\ifx \bseriesno  \undefined \def \bseriesno#1{#1}\fi
\ifx \blocation  \undefined \def \blocation#1{#1}\fi
\ifx \bsertitle  \undefined \def \bsertitle#1{#1}\fi
\ifx \bsnm \undefined \def \bsnm#1{#1}\fi
\ifx \bsuffix \undefined \def \bsuffix#1{#1}\fi
\ifx \bparticle \undefined \def \bparticle#1{#1}\fi
\ifx \barticle \undefined \def \barticle#1{#1}\fi
\bibcommenthead
\ifx \bconfdate \undefined \def \bconfdate #1{#1}\fi
\ifx \botherref \undefined \def \botherref #1{#1}\fi
\ifx \url \undefined \def \url#1{\textsf{#1}}\fi
\ifx \bchapter \undefined \def \bchapter#1{#1}\fi
\ifx \bbook \undefined \def \bbook#1{#1}\fi
\ifx \bcomment \undefined \def \bcomment#1{#1}\fi
\ifx \oauthor \undefined \def \oauthor#1{#1}\fi
\ifx \citeauthoryear \undefined \def \citeauthoryear#1{#1}\fi
\ifx \endbibitem  \undefined \def \endbibitem {}\fi
\ifx \bconflocation  \undefined \def \bconflocation#1{#1}\fi
\ifx \arxivurl  \undefined \def \arxivurl#1{\textsf{#1}}\fi
\csname PreBibitemsHook\endcsname

\bibitem{Witten:1988ze}
\begin{barticle}
\bauthor{\bsnm{Witten}, \binits{E.}}:
\batitle{{Topological Quantum Field Theory}}.
\bjtitle{Commun. Math. Phys.}
\bvolume{117},
\bfpage{353}
(\byear{1988}).
\doiurl{10.1007/BF01223371}
\end{barticle}
\endbibitem

\bibitem{Seiberg:1994rs}
\begin{barticle}
\bauthor{\bsnm{Seiberg}, \binits{N.}},
\bauthor{\bsnm{Witten}, \binits{E.}}:
\batitle{{Electric - magnetic duality, monopole condensation, and confinement
  in N=2 supersymmetric Yang-Mills theory}}.
\bjtitle{Nucl. Phys.}
\bvolume{B426},
\bfpage{19}--\blpage{52}
(\byear{1994})
{\href{https://arxiv.org/abs/hep-th/9407087}{{arXiv:hep-th/9407087}}}
{[hep-th]}.
\doiurl{10.1016/0550-3213(94)90124-4, 10.1016/0550-3213(94)00449-8}.
\bcomment{[Erratum: Nucl. Phys.B430,485(1994)]}
\end{barticle}
\endbibitem

\bibitem{Moore:1997pc}
\begin{barticle}
\bauthor{\bsnm{Moore}, \binits{G.W.}},
\bauthor{\bsnm{Witten}, \binits{E.}}:
\batitle{{Integration over the u plane in Donaldson theory}}.
\bjtitle{Adv. Theor. Math. Phys.}
\bvolume{1},
\bfpage{298}--\blpage{387}
(\byear{1997})
{\href{https://arxiv.org/abs/hep-th/9709193}{{arXiv:hep-th/9709193}}}
{[hep-th]}
\end{barticle}
\endbibitem

\bibitem{Witten:1994cg}
\begin{barticle}
\bauthor{\bsnm{Witten}, \binits{E.}}:
\batitle{{Monopoles and four manifolds}}.
\bjtitle{Math. Res. Lett.}
\bvolume{1},
\bfpage{769}--\blpage{796}
(\byear{1994})
{\href{https://arxiv.org/abs/hep-th/9411102}{{arXiv:hep-th/9411102}}}
{[hep-th]}.
\doiurl{10.4310/MRL.1994.v1.n6.a13}
\end{barticle}
\endbibitem

\bibitem{LoNeSha}
\begin{barticle}
\bauthor{\bsnm{Losev}, \binits{A.}},
\bauthor{\bsnm{Nekrasov}, \binits{N.}},
\bauthor{\bsnm{Shatashvili}, \binits{S.L.}}:
\batitle{{Issues in topological gauge theory}}.
\bjtitle{Nucl. Phys.}
\bvolume{B534},
\bfpage{549}--\blpage{611}
(\byear{1998})
{\href{https://arxiv.org/abs/hep-th/9711108}{{arXiv:hep-th/9711108}}}
{[hep-th]}.
\doiurl{10.1016/S0550-3213(98)00628-2}
\end{barticle}
\endbibitem

\bibitem{Lossev:1997bz}
\begin{barticle}
\bauthor{\bsnm{Lossev}, \binits{A.}},
\bauthor{\bsnm{Nekrasov}, \binits{N.}},
\bauthor{\bsnm{Shatashvili}, \binits{S.L.}}:
\batitle{{Testing Seiberg-Witten solution}}.
\bjtitle{NATO Sci. Ser. C}
\bvolume{520},
\bfpage{359}--\blpage{372}
(\byear{1999})
{\href{https://arxiv.org/abs/hep-th/9801061}{{arXiv:hep-th/9801061}}}
\end{barticle}
\endbibitem

\bibitem{Marino:1998rg}
\begin{barticle}
\bauthor{\bsnm{Marino}, \binits{M.}},
\bauthor{\bsnm{Moore}, \binits{G.W.}}:
\batitle{{Donaldson invariants for nonsimply connected manifolds}}.
\bjtitle{Commun. Math. Phys.}
\bvolume{203},
\bfpage{249}
(\byear{1999})
{\href{https://arxiv.org/abs/hep-th/9804104}{{arXiv:hep-th/9804104}}}
{[hep-th]}.
\doiurl{10.1007/s002200050611}
\end{barticle}
\endbibitem

\bibitem{Malmendier:2008db}
\begin{barticle}
\bauthor{\bsnm{Malmendier}, \binits{A.}},
\bauthor{\bsnm{Ono}, \binits{K.}}:
\batitle{{SO(3)-Donaldson invariants of $\mathbb{P}^2$ and Mock Theta
  Functions}}.
\bjtitle{Geom. Topol.}
\bvolume{16},
\bfpage{1767}--\blpage{1833}
(\byear{2012})
{\href{https://arxiv.org/abs/0808.1442}{{arXiv:0808.1442}}}
{[math.DG]}.
\doiurl{10.2140/gt.2012.16.1767}
\end{barticle}
\endbibitem

\bibitem{Malmendier:2012zz}
\begin{botherref}
\oauthor{\bsnm{Malmendier}, \binits{A.}},
\oauthor{\bsnm{Ono}, \binits{K.}}:
{Moonshine and Donaldson invariants of $\mathbb{P}^2$}
(2012)
{\href{https://arxiv.org/abs/1207.5139}{{arXiv:1207.5139}}}
{[math.DG]}
\end{botherref}
\endbibitem

\bibitem{Korpas:2017qdo}
\begin{barticle}
\bauthor{\bsnm{Korpas}, \binits{G.}},
\bauthor{\bsnm{Manschot}, \binits{J.}}:
\batitle{{Donaldson-Witten theory and indefinite theta functions}}.
\bjtitle{JHEP}
\bvolume{11},
\bfpage{083}
(\byear{2017})
{\href{https://arxiv.org/abs/1707.06235}{{arXiv:1707.06235}}}
{[hep-th]}.
\doiurl{10.1007/JHEP11(2017)083}
\end{barticle}
\endbibitem

\bibitem{Korpas:2019ava}
\begin{barticle}
\bauthor{\bsnm{Korpas}, \binits{G.}},
\bauthor{\bsnm{Manschot}, \binits{J.}},
\bauthor{\bsnm{Moore}, \binits{G.}},
\bauthor{\bsnm{Nidaiev}, \binits{I.}}:
\batitle{{Renormalization and BRST Symmetry in Donaldson--Witten Theory}}.
\bjtitle{Annales Henri Poincare}
\bvolume{20}(\bissue{10}),
\bfpage{3229}--\blpage{3264}
(\byear{2019})
{\href{https://arxiv.org/abs/1901.03540}{{arXiv:1901.03540}}}
{[hep-th]}.
\doiurl{10.1007/s00023-019-00835-x}
\end{barticle}
\endbibitem

\bibitem{Korpas:2019cwg}
\begin{botherref}
\oauthor{\bsnm{Korpas}, \binits{G.}},
\oauthor{\bsnm{Manschot}, \binits{J.}},
\oauthor{\bsnm{Moore}, \binits{G.W.}},
\oauthor{\bsnm{Nidaiev}, \binits{I.}}:
{Mocking the $u$-plane integral}
(2019)
{\href{https://arxiv.org/abs/1910.13410}{{arXiv:1910.13410}}}
{[hep-th]}
\end{botherref}
\endbibitem

\bibitem{bershadsky1995topological}
\begin{barticle}
\bauthor{\bsnm{Bershadsky}, \binits{M.}},
\bauthor{\bsnm{Johansen}, \binits{A.}},
\bauthor{\bsnm{Sadov}, \binits{V.}},
\bauthor{\bsnm{Vafa}, \binits{C.}}:
\batitle{Topological reduction of 4d sym to 2d $\sigma$-models}.
\bjtitle{Nuclear Physics B}
\bvolume{448}(\bissue{1-2}),
\bfpage{166}--\blpage{186}
(\byear{1995})
\end{barticle}
\endbibitem

\bibitem{Harvey_1995}
\begin{barticle}
\bauthor{\bsnm{Harvey}, \binits{J.A.}},
\bauthor{\bsnm{Moore}, \binits{G.W.}},
\bauthor{\bsnm{Strominger}, \binits{A.}}:
\batitle{{Reducing S-duality to T-duality}}.
\bjtitle{Phys. Rev. D}
\bvolume{52},
\bfpage{7161}--\blpage{7167}
(\byear{1995})
{\href{https://arxiv.org/abs/hep-th/9501022}{{arXiv:hep-th/9501022}}}.
\doiurl{10.1103/PhysRevD.52.7161}
\end{barticle}
\endbibitem

\bibitem{Lozano:1999us}
\begin{barticle}
\bauthor{\bsnm{Lozano}, \binits{C.}},
\bauthor{\bsnm{Marino}, \binits{M.}}:
\batitle{{Donaldson invariants of product ruled surfaces and two-dimensional
  gauge theories}}.
\bjtitle{Commun. Math. Phys.}
\bvolume{220},
\bfpage{231}--\blpage{261}
(\byear{2001})
{\href{https://arxiv.org/abs/hep-th/9907165}{{arXiv:hep-th/9907165}}}
{[hep-th]}.
\doiurl{10.1007/s002200100442}
\end{barticle}
\endbibitem

\bibitem{Seiberg:1994aj}
\begin{barticle}
\bauthor{\bsnm{Seiberg}, \binits{N.}},
\bauthor{\bsnm{Witten}, \binits{E.}}:
\batitle{{Monopoles, duality and chiral symmetry breaking in N=2 supersymmetric
  QCD}}.
\bjtitle{Nucl. Phys.}
\bvolume{B431},
\bfpage{484}--\blpage{550}
(\byear{1994})
{\href{https://arxiv.org/abs/hep-th/9408099}{{arXiv:hep-th/9408099}}}
{[hep-th]}.
\doiurl{10.1016/0550-3213(94)90214-3}
\end{barticle}
\endbibitem

\bibitem{Aspman:2021vhs}
\begin{botherref}
\oauthor{\bsnm{Aspman}, \binits{J.}},
\oauthor{\bsnm{Furrer}, \binits{E.}},
\oauthor{\bsnm{Manschot}, \binits{J.}}:
{Cutting and gluing with running couplings in $\mathcal{N}=2$ QCD}
(2021)
{\href{https://arxiv.org/abs/2107.04600}{{arXiv:2107.04600}}}
{[hep-th]}
\end{botherref}
\endbibitem

\bibitem{Marino:1998bm}
\begin{barticle}
\bauthor{\bsnm{Marino}, \binits{M.}},
\bauthor{\bsnm{Moore}, \binits{G.W.}}:
\batitle{{The Donaldson-Witten function for gauge groups of rank larger than
  one}}.
\bjtitle{Commun. Math. Phys.}
\bvolume{199},
\bfpage{25}--\blpage{69}
(\byear{1998})
{\href{https://arxiv.org/abs/hep-th/9802185}{{arXiv:hep-th/9802185}}}
{[hep-th]}.
\doiurl{10.1007/s002200050494}
\end{barticle}
\endbibitem

\bibitem{aspman20AHP}
\begin{barticle}
\bauthor{\bsnm{Aspman}, \binits{J.}},
\bauthor{\bsnm{Furrer}, \binits{E.}},
\bauthor{\bsnm{Manschot}, \binits{J.}}:
\batitle{{Elliptic Loci of SU(3) Vacua}}.
\bjtitle{Annales Henri Poincar{\'e}}
\bvolume{22}(\bissue{8}),
\bfpage{2775}--\blpage{2830}
(\byear{2021})
{\href{https://arxiv.org/abs/2010.06598}{{{arXiv}:2010.06598}}}
{[hep-th]}.
\doiurl{10.1007/s00023-021-01040-5}
\end{barticle}
\endbibitem

\bibitem{Argyres:1995jj}
\begin{barticle}
\bauthor{\bsnm{Argyres}, \binits{P.C.}},
\bauthor{\bsnm{Douglas}, \binits{M.R.}}:
\batitle{{New phenomena in SU(3) supersymmetric gauge theory}}.
\bjtitle{Nucl. Phys.}
\bvolume{B448},
\bfpage{93}--\blpage{126}
(\byear{1995})
{\href{https://arxiv.org/abs/hep-th/9505062}{{arXiv:hep-th/9505062}}}
{[hep-th]}.
\doiurl{10.1016/0550-3213(95)00281-V}
\end{barticle}
\endbibitem

\bibitem{Argyres:1995xn}
\begin{barticle}
\bauthor{\bsnm{Argyres}, \binits{P.C.}},
\bauthor{\bsnm{Plesser}, \binits{M.R.}},
\bauthor{\bsnm{Seiberg}, \binits{N.}},
\bauthor{\bsnm{Witten}, \binits{E.}}:
\batitle{{New N=2 superconformal field theories in four-dimensions}}.
\bjtitle{Nucl. Phys.}
\bvolume{B461},
\bfpage{71}--\blpage{84}
(\byear{1996})
{\href{https://arxiv.org/abs/hep-th/9511154}{{arXiv:hep-th/9511154}}}
{[hep-th]}.
\doiurl{10.1016/0550-3213(95)00671-0}
\end{barticle}
\endbibitem

\bibitem{Moore:2017cmm}
\begin{botherref}
\oauthor{\bsnm{Moore}, \binits{G.W.}},
\oauthor{\bsnm{Nidaiev}, \binits{I.}}:
{The Partition Function Of Argyres-Douglas Theory On A Four-Manifold}
(2017)
{\href{https://arxiv.org/abs/1711.09257}{{arXiv:1711.09257}}}
{[hep-th]}
\end{botherref}
\endbibitem

\bibitem{laba1998}
\begin{barticle}
\bauthor{\bsnm{Labastida}, \binits{J.}},
\bauthor{\bsnm{Lozano}, \binits{C.}}:
\batitle{Mass perturbations in twisted n = 4 supersymmetric gauge theories}.
\bjtitle{Nuclear Physics B}
\bvolume{518},
\bfpage{37}--\blpage{58}
(\byear{1998}).
\doiurl{10.1016/S0550-3213(98)00135-7}
\end{barticle}
\endbibitem

\bibitem{Manschot:2021qqe}
\begin{botherref}
\oauthor{\bsnm{Manschot}, \binits{J.}},
\oauthor{\bsnm{Moore}, \binits{G.W.}}:
{Topological correlators of $SU(2)$, $\mathcal{N}=2^*$ SYM on four-manifolds}
(2021)
{\href{https://arxiv.org/abs/2104.06492}{{arXiv:2104.06492}}}
{[hep-th]}
\end{botherref}
\endbibitem

\bibitem{Huang:2011qx}
\begin{barticle}
\bauthor{\bsnm{Huang}, \binits{M.-x.}},
\bauthor{\bsnm{Kashani-Poor}, \binits{A.-K.}},
\bauthor{\bsnm{Klemm}, \binits{A.}}:
\batitle{{The $\Omega$ deformed B-model for rigid $\mathcal{N}=2$ theories}}.
\bjtitle{Annales Henri Poincare}
\bvolume{14},
\bfpage{425}--\blpage{497}
(\byear{2013})
{\href{https://arxiv.org/abs/1109.5728}{{arXiv:1109.5728}}}
{[hep-th]}.
\doiurl{10.1007/s00023-012-0192-x}
\end{barticle}
\endbibitem

\bibitem{AFM:future}
\begin{botherref}
\oauthor{\bsnm{Aspman}, \binits{J.}},
\oauthor{\bsnm{Furrer}, \binits{E.}},
\oauthor{\bsnm{Manschot}, \binits{J.}}:
To appear
(2021)
\end{botherref}
\endbibitem

\bibitem{Muoz1997WallcrossingFF}
\begin{botherref}
\oauthor{\bsnm{{Mu{\~n}oz}}, \binits{V.}}:
{Wall-crossing formulae for algebraic surfaces with $q>0$}.
arXiv e-prints
(1997)
{\href{https://arxiv.org/abs/alg-geom/9709002}{{arXiv:alg-geom/9709002}}}
{[math.AG]}
\end{botherref}
\endbibitem

\bibitem{Laba05}
\begin{bbook}
\bauthor{\bsnm{Labastida}, \binits{J.}},
\bauthor{\bsnm{Marino}, \binits{M.}}:
\bbtitle{Topological Quantum Field Theory and Four Manifolds}.
\bpublisher{Springer}, \blocation{???}
(\byear{2005}).
\burl{http://www.springer.com/physics/book/978-1-4020-3058-1}
\end{bbook}
\endbibitem

\bibitem{Nakajima2003LecturesOI}
\begin{bchapter}
\bauthor{\bsnm{Nakajima}, \binits{H.}},
\bauthor{\bsnm{Yoshioka}, \binits{K.}}:
\bctitle{Lectures on instanton counting}.
(\byear{2003})
\end{bchapter}
\endbibitem

\bibitem{Manschot_2020}
\begin{botherref}
\oauthor{\bsnm{Manschot}, \binits{J.}},
\oauthor{\bsnm{Moore}, \binits{G.W.}},
\oauthor{\bsnm{Zhang}, \binits{X.}}:
Effective gravitational couplings of four-dimensional $ \mathcal{N} = 2$
  supersymmetric gauge theories.
Journal of High Energy Physics
\textbf{2020}(6)
(2020).
\doiurl{10.1007/jhep06(2020)150}
\end{botherref}
\endbibitem

\bibitem{Witten:1995gf}
\begin{barticle}
\bauthor{\bsnm{Witten}, \binits{E.}}:
\batitle{{On S duality in Abelian gauge theory}}.
\bjtitle{Selecta Math.}
\bvolume{1},
\bfpage{383}
(\byear{1995})
{\href{https://arxiv.org/abs/hep-th/9505186}{{arXiv:hep-th/9505186}}}
{[hep-th]}.
\doiurl{10.1007/BF01671570}
\end{barticle}
\endbibitem

\bibitem{mm1998}
\begin{barticle}
\bauthor{\bsnm{Marino}, \binits{M.}},
\bauthor{\bsnm{Moore}, \binits{G.}}:
\batitle{{Integrating over the Coulomb branch in $\mathcal N = 2$ gauge
  theory}}.
\bjtitle{Nuclear Physics B - Proceedings Supplements}
\bvolume{68}(\bissue{1}),
\bfpage{336}--\blpage{347}
(\byear{1998}).
\doiurl{10.1016/S0920-5632(98)00168-6}.
\bcomment{Strings '97}
\end{barticle}
\endbibitem

\bibitem{GAGA}
\begin{barticle}
\bauthor{\bsnm{Serre}, \binits{J.-P.}}:
\batitle{G\'eom\'etrie alg\'ebrique et g\'eom\'etrie analytique}.
\bjtitle{Annales de l'Institut Fourier}
\bvolume{6},
\bfpage{1}--\blpage{42}
(\byear{1956}).
\doiurl{10.5802/aif.59}
\end{barticle}
\endbibitem

\bibitem{ZwegersThesis}
\begin{botherref}
\oauthor{\bsnm{{Zwegers}}, \binits{S.}}:
{Mock Theta Functions}.
arXiv e-prints,
0807--4834
(2008)
{\href{https://arxiv.org/abs/0807.4834}{{arXiv:0807.4834}}}
{[math.NT]}
\end{botherref}
\endbibitem

\bibitem{Matone:1995rx}
\begin{barticle}
\bauthor{\bsnm{Matone}, \binits{M.}}:
\batitle{{Instantons and recursion relations in N=2 SUSY gauge theory}}.
\bjtitle{Phys. Lett.}
\bvolume{B357},
\bfpage{342}--\blpage{348}
(\byear{1995})
{\href{https://arxiv.org/abs/hep-th/9506102}{{arXiv:hep-th/9506102}}}
{[hep-th]}.
\doiurl{10.1016/0370-2693(95)00920-G}
\end{barticle}
\endbibitem

\bibitem{munoz1999quantum}
\begin{barticle}
\bauthor{\bsnm{Mu{\~n}oz}, \binits{V.}}:
\batitle{Quantum cohomology of the moduli space of stable bundles over a
  riemann surface}.
\bjtitle{Duke mathematical journal}
\bvolume{98}(\bissue{3}),
\bfpage{525}--\blpage{540}
(\byear{1999})
\end{barticle}
\endbibitem

\bibitem{munoz2002gromov}
\begin{botherref}
\oauthor{\bsnm{Mu{\~n}oz}, \binits{V.}}:
On the gromov-witten invariants of the moduli of bundles on a surface
(2002)
\end{botherref}
\endbibitem

\bibitem{donaldson1995floer}
\begin{botherref}
\oauthor{\bsnm{Donaldson}, \binits{S.}}:
Floer homology and algebraic geometry
(1995)
\end{botherref}
\endbibitem

\bibitem{Hori:2003ic}
\begin{bbook}
\bauthor{\bsnm{Hori}, \binits{K.}},
\bauthor{\bsnm{Katz}, \binits{S.}},
\bauthor{\bsnm{Klemm}, \binits{A.}},
\bauthor{\bsnm{Pandharipande}, \binits{R.}},
\bauthor{\bsnm{Thomas}, \binits{R.}},
\bauthor{\bsnm{Vafa}, \binits{C.}},
\bauthor{\bsnm{Vakil}, \binits{R.}},
\bauthor{\bsnm{Zaslow}, \binits{E.}}:
\bbtitle{Mirror Symmetry}.
\bsertitle{Clay mathematics monographs},
vol. \bseriesno{1}.
\bpublisher{AMS},
\blocation{Providence, USA}
(\byear{2003})
\end{bbook}
\endbibitem

\bibitem{MR958805}
\begin{barticle}
\bauthor{\bsnm{Witten}, \binits{E.}}:
\batitle{Topological sigma models}.
\bjtitle{Comm. Math. Phys.}
\bvolume{118}(\bissue{3}),
\bfpage{411}--\blpage{449}
(\byear{1988})
\end{barticle}
\endbibitem

\bibitem{Korpas:2018dag}
\begin{botherref}
\oauthor{\bsnm{Korpas}, \binits{G.}}:
{Donaldson-Witten theory, surface operators and mock modular forms}
(2018)
{\href{https://arxiv.org/abs/1810.07057}{{arXiv:1810.07057}}}
{[hep-th]}
\end{botherref}
\endbibitem

\bibitem{Kanno:1998qj}
\begin{barticle}
\bauthor{\bsnm{Kanno}, \binits{H.}},
\bauthor{\bsnm{Yang}, \binits{S.-K.}}:
\batitle{{Donaldson-Witten functions of massless N=2 supersymmetric QCD}}.
\bjtitle{Nucl. Phys. B}
\bvolume{535},
\bfpage{512}--\blpage{530}
(\byear{1998})
{\href{https://arxiv.org/abs/hep-th/9806015}{{arXiv:hep-th/9806015}}}.
\doiurl{10.1016/S0550-3213(98)00560-4}
\end{barticle}
\endbibitem

\bibitem{Labastida:1998sk}
\begin{barticle}
\bauthor{\bsnm{Labastida}, \binits{J.M.F.}},
\bauthor{\bsnm{Lozano}, \binits{C.}}:
\batitle{{Duality in twisted N=4 supersymmetric gauge theories in
  four-dimensions}}.
\bjtitle{Nucl. Phys.}
\bvolume{B537},
\bfpage{203}--\blpage{242}
(\byear{1999})
{\href{https://arxiv.org/abs/hep-th/9806032}{{arXiv:hep-th/9806032}}}
{[hep-th]}.
\doiurl{10.1016/S0550-3213(98)00653-1}
\end{barticle}
\endbibitem

\bibitem{AFMM:future}
\begin{botherref}
\oauthor{\bsnm{Aspman}, \binits{J.}},
\oauthor{\bsnm{Furrer}, \binits{E.}},
\oauthor{\bsnm{Manschot}, \binits{J.}}:
To appear
(2021)
\end{botherref}
\endbibitem

\bibitem{Gaiotto:2009we}
\begin{barticle}
\bauthor{\bsnm{Gaiotto}, \binits{D.}}:
\batitle{{N=2 dualities}}.
\bjtitle{JHEP}
\bvolume{08},
\bfpage{034}
(\byear{2012})
{\href{https://arxiv.org/abs/0904.2715}{{arXiv:0904.2715}}}
{[hep-th]}.
\doiurl{10.1007/JHEP08(2012)034}
\end{barticle}
\endbibitem

\bibitem{Gaiotto:2009hg}
\begin{botherref}
\oauthor{\bsnm{Gaiotto}, \binits{D.}},
\oauthor{\bsnm{Moore}, \binits{G.W.}},
\oauthor{\bsnm{Neitzke}, \binits{A.}}:
{Wall-crossing, Hitchin Systems, and the WKB Approximation}
(2009)
{\href{https://arxiv.org/abs/0907.3987}{{arXiv:0907.3987}}}
{[hep-th]}
\end{botherref}
\endbibitem

\bibitem{Matone:1995jr}
\begin{barticle}
\bauthor{\bsnm{Matone}, \binits{M.}}:
\batitle{{Koebe 1/4 theorem and inequalities in N=2 supersymmetric QCD}}.
\bjtitle{Phys. Rev. D}
\bvolume{53},
\bfpage{7354}--\blpage{7358}
(\byear{1996})
{\href{https://arxiv.org/abs/hep-th/9506181}{{arXiv:hep-th/9506181}}}.
\doiurl{10.1103/PhysRevD.53.7354}
\end{barticle}
\endbibitem

\bibitem{MR2605321}
\begin{botherref}
\oauthor{\bsnm{Zagier}, \binits{D.}}:
Ramanujan's mock theta functions and their applications (after {Z}wegers and
  {O}no-{B}ringmann).
Ast\'erisque
(326),
986--1431642010
(2009).
S{\'e}minaire Bourbaki. Vol. 2007/2008
\end{botherref}
\endbibitem

\bibitem{Vigneras:1977}
\begin{barticle}
\bauthor{\bsnm{Vign\'eras}, \binits{M.-F.}}:
\batitle{{S\'eries th\^eta des formes quadratiques ind\'efinies}}.
\bjtitle{Springer Lecture Notes}
\bvolume{627},
\bfpage{227}--\blpage{239}
(\byear{1977})
\end{barticle}
\endbibitem

\bibitem{harvey1993string}
\begin{barticle}
\bauthor{\bsnm{Harvey}, \binits{J.A.}},
\bauthor{\bsnm{Strominger}, \binits{A.}}:
\batitle{{String theory and the Donaldson polynomial}}.
\bjtitle{Commun. Math. Phys.}
\bvolume{151},
\bfpage{221}--\blpage{232}
(\byear{1993})
{\href{https://arxiv.org/abs/hep-th/9108020}{{arXiv:hep-th/9108020}}}.
\doiurl{10.1007/BF02096766}
\end{barticle}
\endbibitem

\bibitem{Ashwinkumar:2019owj}
\begin{barticle}
\bauthor{\bsnm{Ashwinkumar}, \binits{M.}},
\bauthor{\bsnm{Png}, \binits{K.-S.}},
\bauthor{\bsnm{Tan}, \binits{M.-C.}}:
\batitle{{Boundary N=2 Theory, Floer Homologies, Affine Algebras, and the
  Verlinde Formula}}.
\bjtitle{Adv. Theor. Math. Phys.}
\bvolume{25},
\bfpage{1}--\blpage{58}
(\byear{2021})
{\href{https://arxiv.org/abs/1909.04058}{{arXiv:1909.04058}}}
{[hep-th]}
\end{barticle}
\endbibitem

\end{thebibliography}


\end{document}